\documentclass[12pt]{article}
\usepackage{enumerate}
\usepackage{amsfonts}
\usepackage{amsmath}
\usepackage{amssymb}
\usepackage{amsthm}
\usepackage{latexsym}
\usepackage{color}

\def\H{\mathcal{H}}

\def\M{\mathbb{M}}
\def\P{\mathcal{P}}
\def\S{\mathfrak{S}}
\def\C{\mathfrak{C}}

\def\F{\mathfrak{F}}
\def\T{\mathfrak{T}}
\def\B{\mathfrak{B}}

\newcommand{\supp}{\mathrm{supp}}
\newcommand{\rank}{\mathrm{rank}}
\newcommand{\id}{\mathrm{Id}}
\newcommand{\Tr}{\mathrm{Tr}}

\newcommand{\shs}{\hspace{1pt}}
\newcounter{defin}  \newcounter{lemma}  \newcounter{theorem}
\newcounter{property} \newcounter{corol}  \newcounter{remark} \newcounter{example}
\newenvironment{lemma}{\par\refstepcounter{lemma}     \textbf{Lemma \thelemma.} }{\rm\par}

\newenvironment{property}{\par\refstepcounter{property}     \textbf{Proposition \theproperty.}\ }{\rm\par}
\newenvironment{corollary}{\par\refstepcounter{corol}     \textbf{Corollary \thecorol.} }{\rm\par}

\newenvironment{remark}{\par\refstepcounter{remark}     \textbf{Remark \theremark.}}{\rm\par}
\newenvironment{example}{\par\refstepcounter{example}     \textbf{Example \theexample.}}{\rm\par}

\begin{document}

\title{Correlation measures of a quantum state and information characteristics of a quantum channel}

\author{M.E. Shirokov\footnote{email:msh@mi.ras.ru}\\Steklov Mathematical Institute, Moscow, Russia}
\date{}
\maketitle

\begin{abstract}
 We discuss the interconnections between basic correlation measures of a bipartite quantum state and basic
information characteristics of a quantum channel, focusing on the benefits of these interconnections
for solving specific problems concerning the characteristics of both types.

We describe  properties of the (unoptimized and optimized) quantum discord in infinite-dimensional
bipartite systems. In particular, using the generalized Koashi-Winter relation, a simple condition is obtained that
guarantees that a state with zero quantum discord is quantum-classical. Two possible definitions of the quantum discord for states with infinite one-way classical correlation are proposed and analysed.

The generalized versions of Koashi-Winter and  Xi-Lu-Wang-Li relations are used to obtain   advanced continuity bounds for the  Holevo information at the outputs of a channel and its complementary channel (as functions of a channel for a given ensemble of input states), for the Holevo capacity and the unregularized private capacity of a quantum channel depending  either on the input dimension or on the input energy bound.

We also discuss the properties of quantum channels which are "doppelgangers" of the monotonicity of the quantum discord
and the entropy reduction of a local measurement under quantum channels acting on an unmeasured subsystem.
\end{abstract}

\maketitle
\tableofcontents

\section{Introduction}

There are two important areas of research in quantum information theory that have appeared since the very emergence of this scientific direction:
\begin{itemize}
  \item the study  of correlations of states of composite quantum systems;
  \item the study  of different information capacities of quantum channels.
\end{itemize}

These two directions are closely related, since correlations of quantum states are special resources that underlie many
quantum algorithms, in particular, protocols for transmitting information over a quantum channel. For example, it was observed many years ago that the
quantum entanglement -- purely quantum type of correlation -- can be used to increase the rate of classical information transmission over a quantum channel
in the different ways: by using entangled measurements for decoding  a classical message encoded in quantum states, by using entangled states for encoding a classical message and by using an entangled state shared by both parties of a communication system \cite{H-ISA,Hastings,BSST,New-R}.

Various characteristics are used to quantify the correlations of composite quantum states.  Important roles in the description of  correlations of bipartite quantum states are played by the quantum mutual information, the one-way classical correlation and the quantum discord \cite{L-mi,H&V,O&Z}. Roughly speaking, the quantum mutual information
describes the total correlation of a bipartite state, while the one-way classical correlation and the quantum discord
describe its classical and quantum components.

In the study of the abilities of quantum channels for transmission of classical information, the concepts
of Holevo capacity and mutual information of a quantum channel play a key role \cite{H-SCI,N&Ch,Wilde}. The relations between
these  characteristics and the correlation and entanglement measures are well known. For example, the constrained Holevo
capacity of a partial trace channel can be represented as the difference between the output entropy and the Entanglement of Formation,
the mutual information of a channel at a given input state is expressed via  the quantum mutual information of the corresponding bipartite state. By using this representation  the well known Koashi-Winter relation (cf.\cite{K&W}) can be rewritten as a relation between the constrained Holevo capacity and the one-way classical correlation. This gives an interpretation of the quantum discord as a difference between the mutual information of a partial trace channel and the constrained Holevo capacity of this channel. Another useful interconnection of this type can be obtained by using the Xi-Lu-Wang-Li relation \cite{Xi+} (see details in Sections 4 and 6.1).

The aim of this article is to discuss the interconnections between basic correlation measures and basic
information characteristics of  quantum channels with pointing the main attention on benefits of these interconnections
for solving the following tasks:
\begin{itemize}
  \item  computable estimates and continuity bounds for correlation measures;
  \item  local continuity analysis of correlation  measures;
  \item  computable estimates and continuity bounds for information characteristics of quantum channels;
  \item  local continuity analysis of information characteristics of quantum channels;
  \item description of the null set of the quantum discord in infinite-dimensional
bipartite systems.
\end{itemize}

The efficiency of the "interconnection technique" for solving the first two of the above tasks has been already demonstrated in \cite[Section 4.3]{QC}, where
it is used for deriving continuity bounds and local continuity conditions for the regularization of the one-way classical correlation
in both finite-dimensional and infinite-dimensional bipartite quantum systems. In this article  we  focus on solving the last three tasks.

The initial motivation of this research was to obtain advanced continuity bounds for the information capacities of quantum channels depending  either on the \emph{input dimension} (if it is finite) or on the \emph{input energy bound} (if the input dimension is infinite). Faithful continuity bounds of this type were obtained in \cite{CID}, but these continuity bounds are not too accurate and are determined by rather complex expressions (in the infinite-dimensional case). The interconnection technique allows us to obtain more accurate and simple  continuity bounds for the  Holevo information at the outputs of a channel and its complementary channel, for the  Holevo capacity and the unregularized private capacity of a quantum channel in both finite-dimensional and  infinite-dimensional cases.

The  article is organized as follows.

In Section 2, we describe the basic notation, definitions and
some auxilarily results used in the article.

In Section 3, we obtain several results on the properties of the entropy reduction of local measurements in a composite quantum system. These results are essentially used in the main part of the article.

In Section 4, we describe the generalized versions of Koashi-Winter and  Xi-Lu-Wang-Li relations.

In Section 5, properties of the  quantum discord in infinite-dimensional bipartite systems are considered. By proving the lower semicontinuity of the (unoptimized and optimized) quantum discord on the natural domain, we obtain a simple condition of local continuity (convergence) of the (unoptimized and optimized)  quantum discord and one-way classical correlation.
We apply the generalized Koashi-Winter relation to prove that a state $\omega$ with zero quantum discord is quantum-classical
provided that $\,\min\{S(\omega),S(\omega_B)\}<+\infty$, where $B$ is a measured subsystem. We also propose definitions
of the (unoptimized and optimized) quantum discord for states with infinite one-way classical correlation and discuss their properties.

Section 6 contains the main results of the article. By reformulating the generalized versions of Koashi-Winter and  Xi-Lu-Wang-Li relations in terms of a quantum channel and its complementary channel  in Section 6.1, we obtain
in Section 6.2.1 continuity bounds for the  Holevo information at the outputs of a channel and its complementary channel (as functions of a channel for a given ensemble of input states) in both finite-dimensional and  infinite-dimensional cases. Then, in Sections 6.2.2 and 6.2.3, we use these results to derive  continuity bounds for the  Holevo capacity and the unregularized private capacity of a quantum channel depending on the input dimension/energy. In Section 6.3,
we discuss the properties of quantum channels which are "doppelgangers" of the monotonicity of the quantum discord
and the entropy reduction of a local measurement under quantum channels acting on an unmeasured subsystem. In Section 6.4, we apply the interconnection technique to obtain bounds on the Holevo capacity and the entropic disturbance of a quantum channel via the
quantum discord. In Section 6.5, we use the generalized  Koashi-Winter relation and the results of Section 6.3 for the qualitative continuity  analysis of the constrained Holevo capacity and the output Holevo information as functions of pairs (channel, input state) and (channel, input ensemble) correspondingly.

In Section 7, we briefly summarize the overall results and mention the open questions appeared in this research.

\section{Preliminaries and basic notation}

Let $\mathcal{H}$ be a separable Hilbert space,
$\mathfrak{B}(\mathcal{H})$ the algebra of all bounded operators on $\mathcal{H}$ with the operator norm $\|\cdot\|$ and $\mathfrak{T}( \mathcal{H})$ the
Banach space of all trace-class
operators on $\mathcal{H}$  with the trace norm $\|\!\cdot\!\|_1$. Let
$\mathfrak{S}(\mathcal{H})$ be  the set of quantum states (positive operators
in $\mathfrak{T}(\mathcal{H})$ with unit trace) \cite{H-SCI,N&Ch,Wilde}.

Write $I_{\mathcal{H}}$ for the unit operator on a Hilbert space
$\mathcal{H}$ and  $\id_{\mathcal{\H}}$ for the identity
transformation of the Banach space $\mathfrak{T}(\mathcal{H})$.

The \emph{support} $\mathrm{supp}\rho$ of a positive operator $\rho\in\mathfrak{T}(\mathcal{H})$ is the closed subspace spanned by the eigenvectors of $\rho$ corresponding to its positive eigenvalues. The dimension of $\mathrm{supp}\rho$ is called the \emph{rank} of $\rho$ and is denoted by $\mathrm{rank}\rho$.

The \emph{von Neumann entropy} of a quantum state
$\rho \in \mathfrak{S}(\H)$ is  defined by the formula
$S(\rho)=\operatorname{Tr}\eta(\rho)$, where  $\eta(x)=-x\ln x$ if $x>0$
and $\eta(0)=0$. It is a concave lower semicontinuous function on the set~$\mathfrak{S}(\H)$ taking values in~$[0,+\infty]$ \cite{H-SCI,L-2,W}.
The von Neumann entropy satisfies the inequality
\begin{equation}\label{S-LAA-2}
S(p\rho+(1-p)\sigma)\leq pS(\rho)+(1-p)S(\sigma)+h_2(p)
\end{equation}
valid for any states  $\rho$ and $\sigma$ in $\S(\H)$ and $p\in(0,1)$, where $\,h_2(p)=\eta(p)+\eta(1-p)\,$ is the binary entropy \cite{N&Ch,Wilde}.

The \emph{quantum relative entropy} for two states $\rho$ and
$\sigma$ in $\mathfrak{S}(\mathcal{H})$ is defined as
\begin{equation*}
D(\rho\,\|\shs\sigma)=\sum_i\langle
i|\,\rho\ln\rho-\rho\ln\sigma\,|i\rangle,
\end{equation*}
where $\{|i\rangle\}$ is the orthonormal basis of
eigenvectors of the state $\rho$ and it is assumed that
$\,D(\rho\,\|\sigma)=+\infty\,$ if $\,\mathrm{supp}\rho\shs$ is not
contained in $\shs\mathrm{supp}\shs\sigma$ \cite{L-2,H-SCI,Wilde}.

The \emph{quantum mutual information} (QMI) of a state $\,\omega\,$ of a bipartite quantum system $AB$ is defined as
\begin{equation*}
I(A\!:\!B)_{\omega}=D(\omega\shs\Vert\shs\omega_{A}\otimes
\omega_{\shs B})=S(\omega_{A})+S(\omega_{\shs B})-S(\omega),
\end{equation*}
where $\omega_{A}=\Tr_B\omega$, $\omega_{B}=\Tr_A\omega$ and the second formula is valid if $\,S(\omega)\,$ is finite \cite{L-mi}.
Basic properties of the relative entropy show that $\,\omega\mapsto
I(A\!:\!B)_{\omega}\,$ is a lower semicontinuous function on the set
$\S(\H_{AB})$ taking values in $[0,+\infty]$  ($\H_{AB}=\H_{A}\otimes\H_{B}$).

The \emph{quantum conditional mutual} information (QCMI) of a state $\omega$ of a finite-dimensional tripartite quantum system $ABC$  is defined by the formula
\begin{equation}\label{CMI-e1}
I(A\!:\!B|C)_{\omega}=S(\omega_{AC})+S(\omega_{BC})-S(\omega_{ABC})-S(\omega_{C}).
\end{equation}
This quantity can be also expressed via the QMI as follows
\begin{equation*}
  I(A\!:\!B|C)_{\omega}=I(AC\!:\!B)_{\omega}-I(B\!:\!C)_{\omega}=I(A\!:\!BC)_{\omega}-I(A\!:\!C)_{\omega}.
\end{equation*}
If $\omega$ is a state of an infinite-dimensional tripartite quantum system $ABC$ then the right hand sides of (\ref{CMI-e1}) and
of the above two representations  may contain the uncertainty $"\infty-\infty"$. In this case one can define the QCMI by one of the
following expressions
\begin{equation}\label{QCMI-e1}
I(A\!:\!B|C)_{\omega}=\sup_{P_A}\left[\shs
I(A\!:\!BC)_{Q\omega Q}-I(A\!:\!C)_{Q\omega
Q}\shs\right]\!,\;\,Q=P_A\otimes I_{BC},
\end{equation}
\begin{equation}\label{QCMI-e2}
I(A\!:\!B|C)_{\omega}=\sup_{P_B}\left[\shs
I(AC\!:\!B)_{Q\omega Q}-I(B\!:\!C)_{Q\omega
Q}\shs\right]\!,\;\,Q=P_B\otimes I_{AC},
\end{equation}
where the suprema are taken over the sets of all finite rank projectors in
$\B(\H_A)$ and in $\B(\H_B)$ correspondingly and it is assumed that $I(X\!:\!Y)_{\sigma}=[\Tr
\sigma]I(X\!:\!Y)_{\sigma/\Tr\sigma}$ for any nonzero $\sigma$ in $\T_+(\H_{XY})$.

Expressions (\ref{QCMI-e1}) and (\ref{QCMI-e2}) are equivalent and coincide with the above formulae for any state $\omega$ at which these formulae are well defined. The QCMI defined by these  expressions is a nonnegative lower semicontinuous function on $\S(\H_{ABC})$ possessing all the basic properties of QMCI valid in the finite-dimensional case \cite[Theorem 2]{CMI}.

A finite or
countable collection $\{\rho_{k}\}$ of quantum states
with a  probability distribution $\{p_{k}\}$ is called (discrete) \emph{ensemble} and denoted by $\{p_k,\rho_k\}$. The state $\bar{\rho}=\sum_{k} p_k\rho_k$ is called  the \emph{average state} of  $\{p_k,\rho_k\}$.  The \emph{Holevo information} of an ensemble
$\{p_k,\rho_k\}$ is defined as
\begin{equation*}
\chi(\{p_k,\rho_k\})= \sum_{k} p_k D(\rho_k\|\bar{\rho})=S(\bar{\rho})-\sum_{k} p_kS(\rho_k),
\end{equation*}
where the second formula is valid if $S(\bar{\rho})$ is finite. This quantity is an upper bound on the classical information obtained from  quantum measurements
over the ensemble \cite{H-73}.

A state $\omega$ of a bipartite system $AB$ is called \emph{quantum-classical} (briefly, \emph{q-c state}) if it has the form
\begin{equation}\label{qc-def}
 \omega=\sum_{k} p_k\, \rho_k\otimes |k\rangle\langle k|,
\end{equation}
where $\{p_k,\rho_k\}$ is an ensembles of states in $\S(\H_A)$ and $\{|k\rangle\}$ is a fixed orthonormal basis in $\H_B$. It is essential that
$I(A\!:\!B)_{\omega}=\chi(\{p_k,\rho_k\})$ for any such state $\omega$ \cite{H-SCI,Wilde}.

A \emph{quantum operation} $\Phi$ from a system $A$ to a system $B$ is a linear completely positive  trace non-increasing map
from $\T(\H_A)$ to $\T(\H_B)$ \cite{H-SCI,Wilde}. A trace preserving operation is called \emph{quantum channel}.
By the Stinespring theorem  any quantum operation (correspondingly, channel) $\Phi$ from $A$ to $B$ can be represented as
\begin{equation*}
\Phi(\rho)=\Tr_E V_{\Phi}\rho V^*_{\Phi},\quad \rho\in\T(\H_A),
\end{equation*}
where $V_{\Phi}$ is a contraction (correspondingly, isometry) from the space $\H_A$ into the tensor product of the  space $\H_B$ and some
Hilbert space $\H_E$ called \emph{environment} \cite{H-SCI,Wilde}.

The quantum operation
\begin{equation}\label{comp-ch}
\widehat{\Phi}(\rho)=\Tr_B V_{\Phi}\rho V^*_{\Phi},\quad \rho\in\T(\H_A),
\end{equation}
from $A$ to $E$ is called \emph{complementary} to the operation $\Phi$ \cite{H-SCI,H-c-ch}. A complementary operation to an operation $\Phi$ is uniquely defined up to the isometrical equivalence \cite{H-c-ch}.

We will use the notion of \emph{strong convergence} of quantum channels and operations. A sequence $\{\Phi_n\}$ of quantum operations from $A$ to $B$ strongly converges to a quantum operation $\Phi_0$ if $\Phi_n(\rho)$ tends to $\Phi_0(\rho)$ as $\,n\to+\infty\,$ for any $\rho\in\S(\H_A)$ \cite{AQC}.

The \emph{constrained Holevo capacity} of a channel $\Phi:A\to B$ at a state $\rho$
in $\S(\H_A)$ is defined as
\begin{equation}\label{CHI-QC-def}
\bar{C}(\Phi,\rho)=\sup_{\sum_k p_k\rho_k=\rho}\chi(\{p_k,\Phi(\rho_k)\}),
\end{equation}
where the supremum is over all ensembles  $\{p_k,\rho_k\}$ of states in $\S(\H_A)$ with the average state $\rho$.
This quantity is related to the classical (unassisted) capacity of a channel \cite{H-SCI,H-Sh-2}.

The \emph{mutual information} $I(\Phi,\rho)$ of a  quantum channel $\Phi:A\to B$ at a state $\rho$ in $\S(\H_A)$
is defined as
\begin{equation}\label{MI-Ch-def}
I(\Phi,\rho)=I(B\!:\!R)_{\Phi\otimes\id_R(\hat{\rho})}=S(\rho)+S(\Phi(\rho))-S(\widehat{\Phi}(\rho)),
\end{equation}
where $\hat{\rho}$ is pure state in $\S(\H_{AR})$ such that $\Tr_R\hat{\rho}=\rho$ and
the second formula is valid if all the entropies involved are finite \cite{H-SCI,Wilde}.
This quantity is related to the classical entanglement-assisted capacity of a channel \cite{BSST,H-c-w-c}.

If $\Psi$ is a channel from $B$ to $C$ then $I(\Psi\circ\Phi,\rho)\leq I(\Psi,\Phi(\rho))$ \cite{H-SCI,Wilde}. We
will use\smallskip

\begin{lemma}\label{D-lemma} \emph{For arbitrary quantum channels $\,\Phi:A\to B$ and $\,\Psi:B\to C$ the nonnegative function
$\,\rho\mapsto I(\Psi,\Phi(\rho))-I(\Psi\circ\Phi,\rho)\,$ is lower semicontinuous on the set
$\,\{\rho\in\S(\H_A)\,|\,I(\Psi\circ\Phi,\rho)<+\infty\}$.}
\end{lemma}\smallskip

\emph{Proof.}  Let $\hat{\rho}\in\S(\H_{AR})$ be a given purification of an arbitrary state $\rho\in\S(\H_A)$ and
$\Phi(\rho)=\Tr_E V_\Phi\rho V^*_\Phi$ be the Stinespring representation of a quantum channel $\Phi$.\break Then $\sigma\doteq (V_\Phi\otimes I_R)\hat{\rho}\shs(V_\Phi^*\otimes I_R)$ is
a purification of the state $\Phi(\rho)$ and hence\break $I(\Psi,\Phi(\rho))=I(C\!:\!ER)_{\Psi\otimes\id_{ER}(\sigma)}$.
At the same time, $I(\Psi\circ\Phi,\rho)=I(C\!:\!R)_{\Psi\otimes\id_{ER}(\sigma)}$ as $\,\Tr_E \Psi\otimes\id_{ER}(\sigma)=(\Psi\circ\Phi)\otimes\id_{R}(\hat{\rho})$.
Thus, $I(\Psi,\Phi(\rho))-I(\Psi\circ\Phi,\rho)$ is the loss of the quantum mutual information $I(C\!:\!ER)_{\Psi\otimes\id_{ER}(\sigma)}$ under the partial trace over $E$.
Since for any sequence $\{\rho_n\}\subset\S(\H_A)$ converging to a state $\rho_0$ there is a sequence  $\{\hat{\rho}_n\}$
of purifications in $\S(\H_{AR})$ converging to a purification $\hat{\rho}_0$ of the state $\rho_0$, the claim of the lemma follows from Theorem 2 in \cite{LSE}. $\Box$\smallskip

If $H$ is a positive (semi-definite)  operator on a Hilbert space $\mathcal{H}$
(we will always assume that positive operators are self-adjoint) and $\rho$ is any positive operator in $\T(\H)$ then   the
quantity $\Tr H\rho$ is defined by the rule
\begin{equation*}
\Tr H\rho=
\left\{\begin{array}{l}
        \sup_n \Tr P_n H\rho\;\; \textrm{if}\;\;  \supp\rho\subseteq {\rm cl}(\mathcal{D}(H))\\
        +\infty\;\;\textrm{otherwise}
        \end{array}\right.,
\end{equation*}
where $P_n$ is the spectral projector of $H$ corresponding to the interval $[0,n]$ and ${\rm cl}(\mathcal{D}(H))$ is the closure of the domain $\mathcal{D}(H)$ of $H$. If $H$ is the Hamiltonian (the 
energy observable) of a quantum system described by the space $\H$, then
$\Tr H\rho$ is the mean energy of the 
state $\rho$ \cite{H-SCI}.

If the operator $H$ satisfies  the \emph{Gibbs condition}
\begin{equation}\label{H-cond}
  \Tr\, e^{-\beta H}<+\infty\quad\textrm{for all}\;\,\beta>0
\end{equation}
and $\{\rho_n\}$ is a sequence of states in $\S(\H)$ converging to a state $\rho_0$
such that $\,\sup_n \Tr H\rho_n<+\infty\,$ then (cf.~\cite{W})
$$
\exists\lim_{n\to+\infty}S(\rho_n)=S(\rho_0)<+\infty.
$$
We will use the function
\begin{equation}\label{F-def}
F_{H}(E)\doteq\sup_{\Tr H\rho\leq E}S(\rho)=S(\gamma_H(E))=\beta(E)E+\ln \Tr e^{-\beta(E)H},
\end{equation}
where
\begin{equation}\label{Gibbs}
\gamma_H(E)\doteq e^{-\beta(E) H}/\Tr e^{-\beta(E) H}
\end{equation}
is the \emph{Gibbs state} corresponding to the "energy" $E$,  the parameter $\beta(E)$ is determined by the equation
$\Tr H e^{-\beta H}=E\Tr e^{-\beta H}$ \cite{W}. The function $F_{H}(E)$ is  strictly increasing and concave  on $[E_0,+\infty)$,
where $E_0$ is the minimal eigenvalue of $H$ (the Gibbs condition (\ref{H-cond}) implies that the operator $H$ has a discrete spectrum of finite multiplicity) \cite{W-CB}. It is easy to see that
$F_{H}(E_0)=\ln m(E_0)$, where $m(E_0)$ is the multiplicity of $E_0$.\smallskip

We will use the following important result proved in \cite{D-A}.\smallskip

\begin{lemma}\label{D-A}  \emph{If a sequence $\{\rho_n\}$ of states converges to a state $\rho_0$ w.r.t. the weak operator topology then
the sequence $\{\rho_n\}$  converges to the state $\rho_0$ w.r.t. the trace norm.}
\end{lemma}

\section{Entropy reduction of local measurements}

Let $A$ and $B$ be quantum systems of any dimension and $\M=\{M_i\}$ be a discrete Positive Operator Valued Measure (POVM) on $\H_B$. To achieve the main aims of this article we will
need several results concerning the quantity
\begin{equation}\label{ER-E1}
ER(\omega, I_A\otimes \M)=S(\omega)-\sum_i p_i S(\omega_i),\quad \omega\in\S(\H_{AB}),
\end{equation}
where $p_i=\Tr M_i\omega_B$ and $\omega_i=p_i^{-1}(I_A\otimes\sqrt{M_i})\shs\omega\shs(I_A\otimes \sqrt{M_i})$
(if $p_i=0$ then we assume that $p_i S(\omega_i)=0$). This quantity is the \emph{entropy reduction} of the
measurement described by the POVM $I_A\otimes \M \doteq\{I_A\otimes M_i\}$ \cite{L-ER,Ozawa,Gr}, it can be extended to states with
infinite entropy by the expression
\begin{equation}\label{ER-E2}
ER(\omega, I_A\otimes \M)=I(\Psi_{I_A\otimes \M}, \omega),
\end{equation}
where $I(\Psi_{I_A\otimes \M}, \omega)$ is the mutual information of the quantum channel
$\Psi_{I_A\otimes \M}(\vartheta)=\sum_i[\Tr M_i\vartheta_B] |i\rangle\langle i|$ at a state $\omega$ (the channel $\Psi_{I_A\otimes \M}$ acts from the system $AB$ to such a system $E$ that $\dim\H_E=\mathrm{card}(\M)$, $\{|i\rangle\}$ is a basis in $\H_E$) \cite{QMG,Sh-ER}.\footnote{Strictly speaking, the quantity
(\ref{ER-E1}) is the entropy reduction of the efficient instrument\break $\{I_A\otimes\sqrt{M_i}\shs(\cdot)\shs I_A\otimes \sqrt{M_i}\}$, but the expression
(\ref{ER-E2}) shows that it is completely determined by the corresponding POVM $\,I_A\otimes \M \doteq\{I_A\otimes M_i\}$. The quantity
(\ref{ER-E1}) coincides with the information gain of any instrument (efficient or non-efficient) described by the POVM $I_A\otimes \M$ \cite{QMG,IGSI+}.}\smallskip

\begin{property}\label{ER-prop} \emph{Let $\M=\{M_i\}$ be a discrete POVM on $\H_B$ and $\,I_A\otimes \M \doteq\{I_A\otimes M_i\}$.}\smallskip

A) \emph{The function $\omega\mapsto ER(\omega, I_A\otimes \M)$ is nonnegative concave
and lower semicontinuous on $\S(\H_{AB})$. The inequalities
\begin{equation}\label{ER-UB}
ER(\omega, I_A\otimes \M)\leq\min\{S(\omega),S(\omega_B)\}
\end{equation}
and
\begin{equation}\label{ER-LAA-2}
\!ER(p\rho+(1-p)\sigma, I_A\otimes \M)\leq p ER(\rho, I_A\otimes \M)+(1-p)ER(\sigma, I_A\otimes \M)+h_2(p)
\end{equation}
hold (with possible values $+\infty$ in both sides) for all  $\omega, \rho, \sigma\in\S(\H_{AB})$ and $p\in[0,1]$, where $h_2(p)$ is the binary entropy.
}\smallskip

B) \emph{The function $\;\omega\mapsto ER(\Phi\otimes\id_B(\omega), I_A\otimes \M)-ER(\omega, I_A\otimes \M)$, where $\Phi:A\to A$ is any quantum channel,
is nonnegative and lower semicontinuous
on the set}
\begin{equation}\label{ER-set}
\left\{\shs\omega\in\S(\H_{AB})\,|\,ER(\omega,I_A\otimes \M)<+\infty\shs\right\}.
\end{equation}

C) \emph{If  $\{\omega_n\}\subset\S(\H_{AB})$ is a sequence converging to a state $\omega_0$ such that
\begin{equation}\label{S-conv}
  \lim_{n\to+\infty }S([\omega_n]_X)=S([\omega_0]_X)<+\infty,
\end{equation}
where $X$ is either $B$ or $AB$,
then}
\begin{equation}\label{ER-conv}
  \lim_{n\to+\infty }ER(\omega_n,I_A\otimes \M)=ER(\omega_0, I_A\otimes \M)<+\infty.
\end{equation}
\end{property}

\begin{remark}\label{at-rem}
Both upper bounds in (\ref{ER-UB}) are optimal in the sense that they are attained at some state $\omega$
in $\S(\H_{AB})$ for a particular POVM $\M=\{M_i\}$ on $\H_B$. Indeed, if $\omega=\rho\otimes\sigma$ (where $\rho\in\S(\H_A)$, $\sigma\in\S(\H_B)$) and $\M$ is any POVM
consisting of one rank operators then $ER(\omega,I_A\otimes \M)=S(\sigma)=S(\omega_B)$. If
$\omega=\sum_{i} p_i\, \rho_i\otimes |i\rangle\langle i|$ and $\M=\{|i\rangle\langle i|\}$, where $\{p_i\}$ is a probability distribution, $\{\rho_i\}$ is a
collection of pure states in $\S(\H_A)$ and $\{|i\rangle\langle i|\}$ is an orthonormal basis in $\H_B$, then
$ER(\omega,I_A\otimes \M)=H(\{p_i\})=S(\omega)$, where $H(\{p_i\})$ is the Shannon entropy of $\{p_i\}$.\smallskip
\end{remark}

\emph{Proof.} A) By representation (\ref{ER-E2}) the concavity
and lower semicontinuity  of the mutual information of a quantum channel (cf.~\cite{H-SCI}) imply the concavity
and lower semicontinuity of the function $f(\omega)=ER(\omega, I_A\otimes \M)$.
The inequality $ER(\omega, I_A\otimes \M)\leq S(\omega)$ follows from (\ref{ER-E1}),
the inequality $ER(\omega, I_A\otimes \M)\leq S(\omega_B)$ follows from the
inequality $ER(\omega, I_A\otimes \M)\leq ER(\omega_B, \M)$ which is a corollary of claim B proved below.

To show that the function $f(\omega)=ER(\omega, I_A\otimes \M)$ satisfies inequality (\ref{ER-LAA-2})
assume first that $\rho$ and $\sigma$ are states with finite entropy. In this case the required inequality follows, by the expression (\ref{ER-E1}), from the inequality (\ref{S-LAA-2})
and concavity of the function $\omega \mapsto \sum_i \tilde{S}((I_A\otimes\sqrt{M_i})\shs\omega(I_A\otimes \sqrt{M_i}))$, where $\tilde{S}$ denotes the
homogeneous extension of the entropy to the positive cone $\T_+(\H_{AB})$.

If $\rho$ and $\sigma$ are arbitrary states then consider the sequences of states
$\rho_n\doteq[\Tr P_n^{\rho}\rho]^{-1}P_n^{\rho}\rho$ and  $\sigma_n\doteq[\Tr P_n^{\sigma}\sigma]^{-1} P_n^{\sigma}\sigma$,
where $P_n^{\rho}$ and $P_n^{\sigma}$ are the spectral projectors of
the states $\rho$ and $\sigma$ corresponding to their $n$ maximal eigenvalues (taking the multiplicity into account).
By using the concavity and lower semicontinuity of the function $f(\omega)=ER(\omega, I_A\otimes \M)$ it is easy to show that
\begin{equation}\label{ER-conv-tmp}
  \lim_{n\to+\infty }f(\omega_n)=f(\omega_0)\leq+\infty,\quad \omega=\rho,\sigma, p\rho+(1-p)\sigma.
\end{equation}
Since inequality (\ref{ER-LAA-2}) holds for the states $\rho_n$ and $\sigma_n$ for all $n$ by the above observation,
the limit relations in (\ref{ER-conv-tmp}) imply the validity of this inequality for the states $\rho$ and $\sigma$.\smallskip

B) This claim follows from Lemma \ref{D-lemma} in Section 2, since $\Psi_{I_A\otimes \M}\circ(\Phi\otimes\id_B)=\Psi_{I_A\otimes \M}$.\smallskip

C) If (\ref{S-conv}) holds with $X=AB$ then (\ref{ER-conv}) follows from Proposition 1 in  \cite{Sh-ER}.
If (\ref{S-conv}) holds with $X=B$ then Proposition 1 in \cite{Sh-ER} implies
that
\begin{equation}\label{ER-conv+}
  \lim_{n\to+\infty }ER([\omega_n]_B,\M)=ER([\omega_0]_B, \M)<+\infty.
\end{equation}
By taking the quantum channel $\Phi(\rho)=[\Tr\rho\shs]\shs\tau$, where $\tau$ is a pure state in $\S(\H_A)$, we conclude from claim B
that the function $\omega\mapsto ER(\omega_B,\M)-ER(\omega, I_A\otimes \M)$
is nonnegative and lower semicontinuous on the set in (\ref{ER-set}). Thus, since the function $\omega\mapsto ER(\omega, I_A\otimes \M)$ is lower semicontinuous on $\S(\H_{AB})$
by claim A, relation (\ref{ER-conv+}) implies (\ref{ER-conv}).  $\Box$\smallskip

By the proof of Proposition \ref{ER-prop}C the following observation used below is valid\smallskip

\begin{corollary}\label{ER-cor} \emph{The function  $\,\omega\mapsto ER(\omega_B,\M)-ER(\omega,I_A\otimes \M)$ is
nonnegative and lower semicontinuous
on the set in (\ref{ER-set}).}
\end{corollary}\smallskip

It follows from Proposition \ref{ER-prop}A that the function $\omega\mapsto ER(\omega,I_A\otimes \M)$ belongs to the class $L_{1}^1(1,1)$ in the settings $A_1=AB$
and to the class $L^1_{2}(1,1)$ in the settings $A_1=B$, $A_2=A$  in terms of Section 3.1.3 in \cite{QC}. So, uniform continuity bounds for this function with the dimension/energy constraints can be obtained by using the general results presented in \cite[Section 3]{QC}.\smallskip

In particular, Theorem 3 in \cite[Section 3.1.3]{QC} implies the following\smallskip

\begin{corollary}\label{ER-CB} \emph{Let $\M=\{M_i\}$ be a discrete POVM on $\H_B$ and $I_A\otimes \M \doteq\{I_A\otimes M_i\}$. Let $\rho$ and $\sigma$ be arbitrary states in $\S(\H_{AB})$ and $\H_*$  the minimal subspace containing the supports of  $\rho$ and $\sigma$. If $\,d=\min\left\{\dim\H_B, \dim\H_*\right\}<+\infty\,$ then
\begin{equation}\label{ER-CB-f}
|ER(\rho,I_A\otimes \M)-ER(\sigma,I_A\otimes \M)|\leq \varepsilon \ln d+g(\varepsilon),
\end{equation}
where $\,\varepsilon=\textstyle\frac{1}{2}\|\rho-\sigma\|_1\,$ and}
\begin{equation}\label{g-def}
g(x)\doteq(x+1)\shs h_2\!\left(\frac{x}{x+1}\right)=(x+1)\ln(x+1)-x\ln x,\;\; x>0,\quad g(0)=0.
\end{equation}
\end{corollary}

\begin{remark}\label{ER-CB-r}
Continuity bound (\ref{ER-CB-f}) with $d=\dim\H_B$ is asymptotically tight for large $d$ (see Def.~1 in \cite[Section 3.2.1]{QC}) for any
POVM $\M$ consisting of one-rank operators. Indeed, if $\rho$ and $\sigma$ are, respectively, the chaotic state and any pure state in $\S(\H_{AB})$ then
$ER(\rho,I_A\otimes \M)=S(\rho_B)=\ln d$,  $ER(\sigma,I_A\otimes \M)=0$ and $\frac{1}{2}\|\rho-\sigma\|_1=1-1/(d\dim\H_A)$.\smallskip

Continuity bound (\ref{ER-CB-f}) with $d=\dim\H_*$ is also asymptotically tight for large $d$ for some POVM $\M$. Let $\dim\H_A=\dim\H_B=d$, $\H_*$ be the subspace
generated by the set $\{|i_A\otimes i_B\rangle\}_{i=1}^d$ of unit vectors in $\H_{AB}$, where $\{|i_X\rangle\}_{i=1}^d$ is an orthonormal basis in $\H_X$. If  $\M=\{|i_B\rangle\langle i_B|\}_{i=1}^d$, $\rho=(1/d)\sum_{i=1}^d|i_A\rangle\langle i_A|\otimes |i_B\rangle\langle i_B|$ and  $\sigma=|1_A\rangle\langle 1_A|\otimes |1_B\rangle\langle 1_B|$ then $ER(\rho,I_A\otimes \M)=\ln d$, $ER(\sigma,I_A\otimes \M)=0\,$ and $\frac{1}{2}\|\rho-\sigma\|_1=1-1/d$.
\end{remark}

\medskip

To obtain an infinite-dimensional version of Corollary  \ref{ER-CB} assume that $H$ is a positive operator on the space $\H_X$, where $X$ is either $B$ or $AB$, satisfying the condition
\begin{equation}\label{H-cond+}
  \lim_{\beta\rightarrow0^+}\left[\Tr\, e^{-\beta H}\right]^{\beta}=1,
\end{equation}
which is slightly stronger than the Gibbs condition (\ref{H-cond}). Condition (\ref{H-cond+}) is equivalent to the property
\begin{equation}\label{H-cond+a}
  F_{H}(E)=o\shs(\sqrt{E})\quad\textrm{as}\quad E\rightarrow+\infty
\end{equation}
of the function $F_H$ defined in (\ref{F-def}) \cite[Section 2.2]{QC}.

Theorem 6 in \cite[Section 3.2.3]{QC} implies, by the remark before Corollary  \ref{ER-CB}, the following\smallskip

\begin{corollary}\label{ER-CB+} \emph{Let $\,\M=\{M_i\}$ be a discrete POVM on $\H_B$ and  $I_A\otimes \M \doteq\{I_A\otimes M_i\}$. Let $H$ be a positive operator on the space $\H_{X}$, where $X$ is either $B$ or $AB$,  that satisfies condition (\ref{H-cond+}) and $F_{H}$ be the function defined in (\ref{F-def}). Then
\begin{equation}\label{ER-CB+f}
    |ER(\rho,I_A\otimes \M)-ER(\sigma,I_A\otimes \M)|\leq \delta F_{H}\!\!\left[\frac{2E}{\delta^2}\right]+g(\delta)
\end{equation}
for any states $\rho$ and $\sigma$ in $\S(\H_{AB})$ such that $\,\Tr H\rho_{X},\,\Tr H\sigma_{X}\leq E\,$ and
$$
\textit{either }\quad F(\rho,\sigma)\geq 1-\delta^2\geq0\quad \textit{ or }\quad(2-\varepsilon)\varepsilon\leq\delta^2\leq1,
$$
where $\,\varepsilon=\textstyle\frac{1}{2}\|\rho-\sigma\|_1$, $F(\rho,\sigma)=\|\sqrt{\rho}\sqrt{\sigma}\|_1^2\,$ (the fidelity of $\rho$ and $\sigma$) and $\,g(x)$ is the function defined in (\ref{g-def}).}
\end{corollary}
\smallskip

The r.h.s. of (\ref{ER-CB+f}) tends to zero as $\,\delta\to0^+$ due to the equivalence of  (\ref{H-cond+}) and (\ref{H-cond+a}).
\smallskip

An important example of an operator $H$ satisfying the condition (\ref{H-cond+}) is
the grounded Hamiltonian
\begin{equation}\label{H-osc}
H=\sum_{i=1}^{\ell}\hbar \omega_i a_i^*a_i
\end{equation}
of an $\ell$-mode quantum oscillator, where $a_i$ and $a^*_i$ are the annihilation and creation operators of the $i$-th mode \cite{H-SCI}.

In this case the function $F_H$ is bounded from above by the function
\begin{equation}\label{F-osc}
G_{\ell,\omega}(E)\doteq \ell\ln \frac{E+2E_0}{\ell E_*}+\ell,\quad E_0= \frac{1}{2}\sum_{i=1}^{\ell}\hbar \omega_i, \;\; E_*=\left[\prod_{i=1}^{\ell}\hbar\omega_i\right]^{1/\ell}.
\end{equation}
It is essential that this upper bound is $\varepsilon$-sharp for large $E$ \cite{CID}.

Thus, if the system $X$ in Corollary \ref{ER-CB+} is an  $\ell$-mode quantum oscillator then one can
use the function $G_{\ell,\omega}$ instead of $F_H$ in (\ref{ER-CB+f}).
In this case
the continuity bound (\ref{ER-CB+f}) with $X=B$ and $\,\delta=\sqrt{1-F(\rho,\sigma)}\,$ is asymptotically tight for large $E$ for any
POVM $\M$ consisting of one-rank operators. This can be shown by modifying the arguments from Remark \ref{ER-CB-r}  or can be deduced from the asymptotical tightness of continuity bound (\ref{CHI-CPHI-EC+}) in Section 6.2.1 for ensembles of pure states
(which is proved by using (\ref{ER-CB+f})).

The continuity bound (\ref{ER-CB+f}) expressed via $\|\rho-\sigma\|_1$ is simple but not too accurate. More accurate (asymptotically tight) continuity bounds for the function
$\,\omega\mapsto ER(\omega,I_A\otimes \M)\,$ depending on the trace norm distance can be obtained  by applying Theorem 7 in \cite[Section 3.2.4]{QC}.
Assume for simplicity that
\begin{equation}\label{H-0}
\inf\limits_{\|\varphi\|=1}\langle\varphi|H|\varphi\rangle=0.
\end{equation}
Let $G$ be a continuous function on $\mathbb{R}_+$ such that
\begin{equation}\label{G-c1}
G(E)\geq F_{H}(E)\quad \forall E>0,\quad G(E)=o\shs(\sqrt{E})\quad\textrm{as}\quad E\rightarrow+\infty
\end{equation}
and
\begin{equation}\label{G-c2}
G(E_1)< G(E_2),\quad G(E_1)/\sqrt{E_1}\geq G(E_2)/\sqrt{E_2}
\end{equation}
for any $E_2>E_1>0$ (general conditions for existence of such a function are discussed in \cite[Section 3.2.4]{QC}).

If $H$ is the grounded Hamiltonian (\ref{H-osc}) of an $\ell$-mode quantum oscillator then the function
$G_{\ell,\omega}$ defined in (\ref{F-osc}) satisfies conditions (\ref{G-c1}) and (\ref{G-c2}) \cite{CID,QC}.

Let $d_0$ be the minimal natural number such that  $\,\ln d_0>G(0)\,$. For given  $\,E,\varepsilon,t>0$ and $C,D\geq 0$ we put
\begin{equation}\label{CB-exp}
\!\mathbb{CB}_{\shs t}(E,\varepsilon\,|\,C,D)=C\varepsilon(1+4t)\!\left(G\!\left[\!\frac{E}{(\varepsilon t)^2}\!\right]+1/d_0+\ln2\right)+D(2g(\varepsilon t)+g(\varepsilon(1+2t)))
\end{equation}
and $\,T=(1/\varepsilon)\min\{1, \sqrt{E/G^{-1}(\ln d_0)}\}$.

The classification of the function
$\,\omega\mapsto ER(\omega,I_A\otimes \M)\,$ mentioned before Corollary \ref{ER-CB} allows us to apply Theorem 7 in \cite[Section 3.2]{QC} to obtain the following
\smallskip\pagebreak
\begin{corollary}\label{ER-CB++} \emph{Let $\,\M=\{M_i\}$ be a discrete POVM on $\H_B$ and $\,I_A\otimes \M \doteq\{I_A\otimes M_i\}$. Let $H$ be a positive operator on the space $\H_{X}$, where $X$ is either $B$ or $AB$,  that satisfies conditions (\ref{H-cond+}) and (\ref{H-0}). If  $G$ is an arbitrary continuous function on $\mathbb{R}_+$ satisfying conditions  (\ref{G-c1}) and (\ref{G-c2}) then
\begin{equation}\label{ER-CB++f}
 |ER(\rho,I_A\otimes \M)-ER(\sigma,I_A\otimes \M)|\leq \min_{\shs t\in(0,T]} \mathbb{CB}_t(E,\varepsilon\,|\,1,1)
\end{equation}
for any states $\rho$ and $\sigma$ in $\S(\H_{AB})$ such that $\,\Tr H\rho_{X},\,\Tr H\sigma_{X}\leq E\,$ and $\,\textstyle\frac{1}{2}\|\rho-\sigma\|_1\leq\varepsilon$.
}\end{corollary}
\smallskip

The r.h.s. of (\ref{ER-CB++f}) tends to zero as $\,\varepsilon\to0^+$ due to the second property of the function $G$ in (\ref{G-c1}).
\smallskip

\begin{remark}\label{ER-CB++r} Let $X$ be an $\ell$-mode quantum oscillator, $\,H$ its grounded Hamiltonian defined in (\ref{H-osc})
and $G=G_{\ell,\omega}$ be the function defined in (\ref{F-osc}). Then continuity bound (\ref{ER-CB++f}) is asymptotically tight
for large $E$ in the case $X=B$ provided that $\M$ is a POVM consisting of one-rank operators (see Def.~1 in \cite[Section 3.2.1]{QC}). Since $G_{\ell,\omega}(E)=F_H(E)(1+o(1))$
as $E\to+\infty$, to prove this claim it suffices, by the last claim of Theorem 7 in \cite[Section 3.2]{QC}, to note that
$$
ER(\rho\otimes\gamma_H(E),I_A\otimes \M)=S(\gamma_H(E))=F_H(E),
$$
where $\rho$ is any state in $\S(\H_{A})$ and $\gamma_H(E)$ is the Gibbs state (\ref{Gibbs}) of $B$ corresponding to the "energy" $E$.
\end{remark}

\section{The generalized versions of the Koashi-Winter and Xi-Lu-Wang-Li relations}

The \emph{one-way classical correlation} is proposed by Henderson and Vedral in \cite{H&V} as a measure of classical correlations of a state  of a bipartite system $AB$. For an arbitrary state $\omega$ in $\S(\H_{AB})$ it is defined as
\begin{equation}\label{CB-def}
C_B(\omega)=\sup_{\M\in \mathfrak{M}_B}\chi(\{p_i,\omega^i_{A}\}),
\end{equation}
where the supremum  is taken over the set $\mathfrak{M}_B$ of all discrete POVM
$\M=\{M_i\}$ on the space $\H_B$, $p_i=\Tr M_i\omega_B$
is the probability of  the $i$-th outcome, $\omega^i_{A}=p_i^{-1}\Tr_B(I_A\otimes M_i)\omega$ is the posteriori state of  system $A$ corresponding to  the $i$-th outcome
(if $p_i=0$ then it is assumed that the ensemble $\{p_i,\omega^i_{A}\}$ has no state in the $i$-th position).  It is easy to show that the supremum in (\ref{CB-def}) can be taken only over the set $\mathfrak{M}^0_B$ of all  POVM $\M=\{M_i\}$ consisting of one-rank operators \cite{K&W}.

The function $C_B$ on $\S(\H_{AB})$ is nonnegative,  non-increasing under local channels and lower semicontinuous (it is continuous if and only if $\,\min\{\dim\H_A,\dim\H_B\}<+\infty$ \cite[Section 4.3]{QC}). It is equal to the von Neumann entropy of $\omega_A$  at any pure state $\omega$ and  coincides with the QMI $I(A\!:\!B)_{\omega}$ at any  q-c state $\omega$ (i.e. a state $\omega$ having the form (\ref{qc-def})) \cite{H&V,MI-B,Xi+}.\smallskip

The \emph{quantum discord}  is the difference between the QMI and the one-way classical correlation:
\begin{equation}\label{q-d}
D_B(\omega)\doteq I(A\!:\!B)_{\omega}-C_B(\omega)=\inf_{\M\in \mathfrak{M}_B}\left(I(A\!:\!B)_{\omega}-\chi(\{p_i,\omega^i_{A}\})\right).
\end{equation}
It is treated as a quantity describing the quantum component of correlation of a state $\omega$ of a bipartite system $AB$. This definition slightly differs from the original definition proposed by Ollivier and Zurek in \cite{O&Z}, where  the quantum discord is defined by the expression similar to (\ref{q-d}) in which $C_B(\omega)$ is defined by formula (\ref{CB-def}) with the supremum over all von Neumann measurements only. The advantage of the definition (\ref{q-d}) used below is that it provides, due to Naimark's theorem, the invariance of
the quantum discord w.r.t. to the embedding of $\H_{AB}\doteq\H_A\otimes\H_B$ into $\H_{AB'}\doteq\H_A\otimes\H_{B'}$, where $\H_{B'}$ is any space containing $\H_B$ (the Ollivier-Zurek definition does not possess this property) \cite{W-pc}.

The function  $D_B$ is well defined (as a nonnegative number or $+\infty$) by formula (\ref{q-d}) for any state $\omega$
with finite $C_B(\omega)$.
The quantum discord  is invariant under local unitary trasformations
and  non-increasing under local channels acting on $A$, it is equal to the von Neumann entropy of $\omega_A$  at any pure state $\omega$ and to zero at any q-c state. \cite{Xi,Piani+,Piani,Xi+,Datta}. If $A$ and $B$ are finite-dimensional systems then any
state  with zero discord is a q-c state \cite{Datta}. In Section 5.2 we will show that the same is true
for a state  $\omega$ of an infinite-dimensional system  $AB$ provided that $\,\min\{S(\omega),S(\omega_B)\}<+\infty$.\smallskip

For a POVM $\M=\{M_i\}$ in $\mathfrak{M}_B$ introduce the q-c channel
\begin{equation}\label{Psi-M}
\Psi_{\M}(\rho)=\sum_{i}[\Tr M_i\rho\shs]|i\rangle\langle i|
\end{equation}
from $B$ to $E$ determined by any fixed basis $\{|i\rangle\}$ in a  Hilbert space $\H_E$ (such that $\dim\H_E=\mathrm{card}(\M)$). Then we can rewrite definitions (\ref{CB-def}) and (\ref{q-d}) as follows
\begin{equation*}
C_B(\omega)=\sup_{\M\in\mathfrak{M}_B}I(A\!:\!E)_{\id_A\otimes\Psi_{\M}(\omega)},\quad D_B(\omega)=\inf_{\M\in\mathfrak{M}_B}\left[I(A\!:\!B)_{\omega}-I(A\!:\!E)_{\id_A\otimes\Psi_{\M}(\omega)}\right].
\end{equation*}

For a given $\M=\{M_i\}\in \mathfrak{M}_B$  introduce the
 \emph{unoptimized one-way classical correlation} and \emph{unoptimized quantum discord} defined for any state $\omega\in\S(\H_{AB})$ as
\begin{equation}\label{UCBDB}
C_B^{\shs\M}(\omega)=I(A\!:\!E)_{\id_A\otimes\Psi_{\M}(\omega)}\quad \textrm{and} \quad  D_B^{\shs\M}(\omega)=I(A\!:\!B)_{\omega}-I(A\!:\!E)_{\id_A\otimes\Psi_{\M}(\omega)}.
\end{equation}
The  unoptimized quantum discord $D_B^{\shs\M}(\omega)$ is  well defined (as a nonnegative number or $+\infty$) on the subset of $\S(\H_{AB})$ consisting of states $\omega$ with finite $C_B^{\shs\M}(\omega)$. We will see in Section 5.3 that the function $\omega \mapsto D_B^{\shs\M}(\omega)$ has a lower semicontinuous extension to the whole set $\S(\H_{AB})$. The unoptimized quantum discord has operational interpretations described in \cite{NQD,NQD+}.
\smallskip

The \emph{constrained Holevo capacity of  partial trace} at a state $\omega$ of a bipartite system $AC$ is defined as
\begin{equation*}
\chi_A(\omega)=\sup_{\sum_k p_k\omega_k=\omega}\sum_{k}p_k D([\omega_k]_A\|\shs\omega_A)=\sup_{\hat{\omega}}I(A\!:\!E)_{\hat{\omega}},
\end{equation*}
where the first supremum is over all discrete ensembles $\{p_k, \omega_k\}$ of  states in $\S(\H_{AC})$ with the average state $\omega$ and
the second supremum is over all extensions of $\omega$ to a q-c state $\hat{\omega}$ in $\S(\H_{ACE})$ having the form
$\hat{\omega}=\sum_k p_k \omega_k\otimes |k\rangle\langle k|$, where $\{|k\rangle\}$ is an orthonormal basis in a separable Hilbert space $\H_{E}$.

The quantity $\chi_A(\omega)$ is the constrained Holevo capacity
$\bar{C}(\Theta_{\!A},\omega)$ of the partial trace channel $\Theta_{\!A}:\vartheta\mapsto \vartheta_A$ from $AC$ to $A$ at a state $\omega$ defined in (\ref{CHI-QC-def}).
It is easy to see that
\begin{equation*}
\chi_A(\omega)\leq \min\{S(\omega),S(\omega_A)\}\quad \forall\omega\in\S(\H_{AC}).
\end{equation*}

Koashi and Winter proved in \cite{K&W} that
\begin{equation}\label{KWS}
C_B(\omega_{AB})+E_{F}(\omega_{AC})=S(\omega_{A})
\end{equation}
for any pure state $\omega$ of a finite-dimensional tripartite system $ABC$, where
$E_{F}$ is the Entanglement of Formation of a state in $\S(\H_{AC})$ \cite{B&Co}. They also showed that
$"\leq"$ holds in  (\ref{KWS}) for any mixed state $\omega$ in $\S(\H_{ABC})$.

By updating the arguments from the proof of Theorem 1 in \cite{K&W} one can obtain
the generalized Koashi-Winter relation
\begin{equation}\label{KWS-gen}
C_B(\omega_{AB})=\chi_A(\omega_{AC})
\end{equation}
valid for any pure state $\omega$ of an infinite-dimensional system $ABC$. If $S(\omega_A)<+\infty$ then (\ref{KWS-gen}) is reduced to the standard Koashi-Winter relation (\ref{KWS}) due to the representations
\begin{equation*}
\chi_A(\omega)=S(\omega_A)-E_{F,d}(\omega)=S(\omega_A)-E_{F,c}(\omega),
\end{equation*}
where $E_{F,d}$ and $E_{F,c}$ are the discrete and continuous versions of
the Entanglement of Formation (which coincide due to the condition $S(\omega_A)<+\infty$ \cite[Section 4.4]{QC}).
The same arguments from  \cite{K&W} show that
\begin{equation}\label{KWS-gen+}
C_B(\omega_{AB})\leq\chi_A(\omega_{AC})
\end{equation}
for any mixed state $\omega$ in $\S(\H_{ABC})$. The advantage of relations  (\ref{KWS-gen}) and (\ref{KWS-gen+}) is their \emph{independence of the condition} $S(\omega_A)<+\infty$.\smallskip

The proofs of relations (\ref{KWS-gen}) and (\ref{KWS-gen+}) are based on the equality
\begin{equation}\label{KWS-gen+un}
C^{\M}_B(\omega_{AB})=\chi(\Theta_{\!A}(\mu_{\omega,\M}))\doteq\chi(\{p_i,\Theta_{\!A}(\rho_i)\})
\end{equation}
valid for any POVM $\M=\{M_i\}\in\mathfrak{M}_B$ and any state $\omega\in\S(\H_{ABC})$, where
$\Theta_{\!A}:\vartheta\mapsto \vartheta_A$ is a channel from $AC$ to $A$ and
$\mu_{\omega,\M}=\{p_i, \rho_i\}$, $p_i=\Tr M_i\omega_{B}$,  $\rho_i=p_i^{-1}\Tr_B (M_i\otimes I_{AC})\omega$. The equality (\ref{KWS-gen+un}) is a direct corollary of the definition of $C^{\M}_B$.
\smallskip

By using the Koashi-Winter relation (\ref{KWS}) Xi, Lu, Wang and Li proved in  \cite[Section III]{Xi+} the equality
\begin{equation}\label{Xi++}
C_B(\omega_{AB})+D_B(\omega_{BC})=S(\omega_{B})
\end{equation}
valid for any pure state $\omega$ of a finite-dimensional tripartite system $ABC$. This equality remains valid
for any pure state $\omega$ of an infinite-dimensional system $ABC$ provided that $S(\omega_B)<+\infty$.

We will need a version of the Xi-Lu-Wang-Li relation (\ref{Xi++}) which holds \emph{independently of the condition}
$S(\omega_B)<+\infty$. To obtain such a version note that
\begin{equation}\label{Xi++un}
C^{\M}_B(\omega_{AB})=ER(\omega_{BC},\M\otimes I_C)\leq+\infty
\end{equation}
for any pure state $\omega\in \S(\H_{ABC})$ and any POVM $\M\in\mathfrak{M}_B$, where
$ER(\omega_{BC},\M\otimes I_C)$ is the entropy reduction of the local measurement described by the POVM $\,\M\otimes I_C=\{M_i\otimes I_C\}$ (see Section 3) and $\mathfrak{M}_B$
is the set of all POVM on $\H_B$.

If $S(\omega_{A})=S(\omega_{BC})<+\infty$ then the equality (\ref{Xi++un}) follows from the definitions (\ref{ER-E1}) and (\ref{UCBDB}), since
the purity of the states $\,\omega_i\doteq p_i^{-1}(\sqrt{M_i}\otimes I_{AC})\shs\omega(\sqrt{M_i}\otimes I_{AC})$ implies
$$
S(p_i^{-1}\Tr_B (I_{A}\otimes M_i)\omega_{AB})=S(\Tr_{BC}\omega_i)=S(\Tr_{A}\omega_i)=S(p_i^{-1}(\sqrt{M_i}\otimes I_{C})\shs\omega_{BC}(\sqrt{M_i}\otimes I_{C}))
$$
provided that $p_i=\Tr M_i\omega_{B}\neq0$.

If $S(\omega_A)=+\infty$ then take any sequence $\{P_n\}$ of finite rank projectors in $\B(\H_A)$
strongly converging to the unit operator $I_A$. Consider the sequence of pure states $\omega_n=c_n^{-1}(P_n\otimes I_{BC})\omega (P_n\otimes I_{BC})$, $c_n=\Tr P_n\omega_A$,
converging to the pure state $\omega$. Since $S([\omega_n]_A)<+\infty$, the above observation shows that (\ref{Xi++un}) holds with $\omega=\omega_n$ for all $n$.
Thus, to prove (\ref{Xi++un}) it suffices to show that
\begin{equation}\label{tmp-1}
\lim_{n\to+\infty}C^{\M}_B([\omega_n]_{AB})=C^{\M}_B(\omega_{AB})\leq+\infty
\end{equation}
and
\begin{equation}\label{tmp-2}
\lim_{n\to+\infty}ER([\omega_n]_{BC},\M\otimes I_C)=ER(\omega_{BC},\M\otimes I_C)\leq+\infty.
\end{equation}

The limit relation (\ref{tmp-1}) follows from the lower semicontinuity of the function $C^{\M}_B$ (which is a corollary of the same property of QMI) and
the inequality $c_nC^{\M}_B([\omega_n]_{AB})\leq C^{\M}_B(\omega_{AB})\leq+\infty$ valid for any $n$, where $c_n=\Tr P_n\omega_A$.
This inequality can be derived from  the monotonicity of the QMI under local quantum operations (see, f.i. \cite[Lemma 9]{CMI}).
The limit relation (\ref{tmp-2}) can be easily proved by using the concavity and  lower semicontinuity of the function $\vartheta\mapsto ER(\vartheta,\M\otimes I_C)$ on $\S(\H_{BC})$ (Proposition \ref{ER-prop} in Section 3),
since $c_n[\omega_n]_{BC}\leq\omega_{BC}$ for all $n$, where $c_n=\Tr P_n\omega_A$.

Thus, the equality (\ref{Xi++un}) is proved. By taking the supremum over all POVM $\M$ in $\mathfrak{M}_B$ we obtain
\begin{equation}\label{Xi+++}
C_B(\omega_{AB})=\sup_{\M\in\mathfrak{M}_B}ER(\omega_{BC},\M\otimes I_C)=\sup_{\M\in\mathfrak{M}^0_B}ER(\omega_{BC},\M\otimes I_C)\leq+\infty
\end{equation}
for an arbitrary pure state $\omega\in \S(\H_{ABC})$, where  $\mathfrak{M}^0_B$
is the set of all POVM on $\H_B$ consisting of one-rank operators. The second equality in (\ref{Xi+++}) holds, since the supremum in the definition of  $C_B(\omega_{AB})$ can be taken over the set $\mathfrak{M}^0_B$ \cite{K&W}.\smallskip

If $S(\omega_B)<+\infty$ then $ER(\omega_{BC},\M\otimes I_C)=S(\omega_B)-D^{\M}_B(\omega_{BC})$ for any POVM $\M$ in $\mathfrak{M}^0_B$ (see Proposition \ref{ext-DB}E in Section 5)
and hence (\ref{Xi+++}) turns into the Xi-Lu-Wang-Li relation (\ref{Xi++}). So, we will call the equality (\ref{Xi+++})
the \emph{generalized Xi-Lu-Wang-Li relation}.

The generalized Koashi-Winter relation (\ref{KWS-gen}) and its regularized version are used in \cite{QC} to obtain continuity bounds and local continuity conditions
for the function $C_B$ and its regularization. In Sections 5 and 6 we will consider another applications of this relation, the generalized Xi-Lu-Wang-Li relation (\ref{Xi+++})
and their "unoptimized" versions (\ref{KWS-gen+un}) and (\ref{Xi++un}).

\section{Quantum discord in infinite-dimensional systems}

\subsection{Lower semicontinuity of the (unoptimized and optimized) quantum discord and its corollary}

If one of  systems $A$ and $B$ is finite-dimensional then the quantum discord $D_B$ (defined in (\ref{q-d})) is a uniformly continuous function on the space $\S(\H_{AB})$. This follows from the continuity bounds for the quantum discord presented in \cite{Xi} and
in  \cite[Section 4.3.1]{QC}.
The same is true for the unoptimized quantum discord
$D_B^{\M}$ for any given POVM $\M$ on $\H_B$. This can be shown easily by using Theorem 3 in \cite[Section 3.1.3]{QC} and the well known properties of the QMI.

If both systems $A$ and $B$ are infinite-dimensional then the quantum discord $D_B$ is well defined by formula (\ref{q-d}) only on the set
of states with finite one-way classical correlation $C_B$. It is not continuous on this set and may take infinite values.
The same singular properties hold for the unoptimized quantum discord
$D_B^{\M}$ for any given POVM $\M$ on $\H_B$, which is well defined by the second formula in (\ref{UCBDB}) only on the set
of states with finite unoptimized one-way classical correlation $C^{\M}_B$ (it is easy to see that $C^{\M}_B$ is finite at any state in $\S(\H_{AB})$ if and only if $\M$ has a finite number of outcomes).

By definition, the optimized (resp.  unoptimized) quantum discord $D_B$ (resp. $D^{\M}_B$) is a difference between the lower semicontinuous functions $I(A\!:\!B)$ and $C_B$ (resp. $C^{\M}_B$) on $\S(\H_{AB})$. The following proposition shows that the function $D_B$ (resp. $D^{\M}_B$) is itself
lower semicontinuous on its natural domain.\smallskip

\begin{property}\label{DB-LS} A) \emph{For an arbitrary POVM $\,\M$ on the space $\H_B$ the function $D_B^{\shs\M}$ (defined in (\ref{UCBDB})) is lower semicontinuous on the set}
$$\left\{\omega\in\S(\H_{AB})\,|\,C^{\shs\M}_B(\omega)<+\infty\right\}.$$

B) \emph{The function $D_B$ (defined in (\ref{q-d})) is lower semicontinuous on the set}
$$\left\{\omega\in\S(\H_{AB})\,|\,C_B(\omega)<+\infty\right\}.$$
\end{property}\smallskip

\emph{Proof.} Assume that
$V_{\M}:\H_B\rightarrow\H_{EF}$ is the Stinespring isometry of the channel $\Psi_{\M}:B\to E$ defined in (\ref{Psi-M}). Then we have
\begin{equation}\label{u-r}
D_B^{\shs\M}(\omega)=I(A\!:\!B)_{\omega}-I(A\!:\!E)_{\id_A\otimes\Psi_{\M}(\omega)}=I(A\!:\!EF)_{\tilde{\omega}}-I(A\!:\!E)_{\tilde{\omega}}=I(A\!:\!F|E)_{\tilde{\omega}}
\end{equation}
for any state $\omega\in\S(\H_{AB})$, where $\tilde{\omega}=(I_A\otimes V_{\M})\shs\omega\shs(I_A\otimes V^*_{\M})$.
This representation was used by Piani in  \cite{Piani} to prove the monotonicity of (unoptimized and optimized) quantum discord under local channels acting on a unmeasured subsystem, i.e.
the validity of the inequalities
\begin{equation}\label{un-DB-m}
  D_B^{\M}(\Phi\otimes \id_B(\omega))\leq D^{\M}_B(\omega)
\end{equation}
and
\begin{equation}\label{DB-m}
  D_B(\Phi\otimes \id_B(\omega))\leq D_B(\omega)
\end{equation}
for any state $\omega$ in $\S(\H_{AB})$ with finite QMI and any channel $\Phi:A\to A$.

Claim A of the proposition follows directly from the representation (\ref{u-r}) and the lower semicontinuity of the quantum conditional mutual information established in \cite[Theorem 2]{CMI}.

Claim B can be proved by showing that
\begin{equation}\label{DB-rep}
 D_B(\omega)=\sup_{\Phi\in\F_A}D_B(\Phi\otimes \id_B(\omega))
\end{equation}
for any state $\omega$ with finite $C_B(\omega)$, where $\F_A$ is the set of all channels $\Phi:A\to A$ with a finite-dimensional output, since
the function $\omega\mapsto D_B(\Phi\otimes \id_B(\omega))$ is (uniformly) continuous on $\S(\H_{AB})$ for any $\Phi$ in $\F_A$  by the remark at the beginning of this section.

The inequality $"\leq"$ in (\ref{DB-rep}) follows from the monotonicity property (\ref{DB-m}).
To prove the converse inequality it suffices to show that
\begin{equation}\label{VerNewE}
\lim_{n\to+\infty}I(A\!:\!B)_{\Phi_n\otimes \id_B(\omega)}=I(A\!:\!B)_{\omega}
\quad
\textup{and}
\quad
\lim_{n\to+\infty}C_B(\Phi_n\otimes \id_B(\omega))=C_B(\omega),
\end{equation}
where $\{\Phi_n\}$ is any sequence in $\F_A$ strongly converging to the identity channel $\id_A$ (it means that $\Phi_n(\rho)$ tends to $\rho$ as $n\to+\infty$ for any $\rho\in\S(\H_A)$).  These relations
follow from the lower semicontinuity of the functions  $\omega\mapsto I(A\!:\!B)_{\omega}$ and $\omega\mapsto C_B(\omega)$, since
$I(A\!:\!B)_{\Phi_n\otimes \id_B(\omega)}\leq I(A\!:\!B)_{\omega}$ and $C_B(\Phi_n\otimes \id_B(\omega))\leq C_B(\omega)$
for all $n$ by the monotonicity of the QMI and the one-way classical correlation under local channels. $\Box$

The lower semicontinuity of the (unoptimized and optimized) one-way classical correlation and Proposition \ref{DB-LS} imply the following
result.\smallskip

\begin{corollary}\label{DB-LS-c}
\emph{If $\,\{\omega_n\}$ is a sequence in  $\,\mathfrak{S}(\mathcal{H}_{AB})$ converging to a state $\omega_0$ such that
\begin{equation}\label{I-cont}
 \lim_{n\to+\infty }\!I(A\!:\!B)_{\omega_n}=I(A\!:\!B)_{\omega_0}<+\infty
\end{equation}
then}
\begin{equation*}
  \lim_{n\to+\infty }\!C_B(\omega_n)=C_B(\omega_0)<+\infty\quad\textit{and}\quad\lim_{n\to+\infty }\!D_B(\omega_n)=D_B(\omega_0)<+\infty.
\end{equation*}

\emph{If $\,\M$ is any POVM on $\H_B$ then (\ref{I-cont}) implies that}
\begin{equation*}
  \lim_{n\to+\infty }\!C^{\M}_B(\omega_n)=C^{\M}_B(\omega_0)<+\infty\quad\textit{and}\quad\lim_{n\to+\infty }\!D^{\M}_B(\omega_n)=D^{\M}_B(\omega_0)<+\infty.
\end{equation*}
\end{corollary}\smallskip

The main claim of Corollary \ref{DB-LS-c} has a clear physical interpretation: it states that \emph{local continuity of the total correlation
implies local continuity of its classical and quantum components.}\smallskip

Since there exist simple sufficient conditions for the validity of (\ref{I-cont}) presented in Section 5.1 in \cite{LSE},
Corollary \ref{DB-LS-c} gives a  practical way to prove the local continuity (convergence) of the one-way classical correlation
and the quantum discord.

\subsection{On states with zero quantum discord}

If $\omega$ is a state of a finite-dimensional bipartite system $AB$ then
$D_B(\omega)=0$ if and only if $\omega$ is a quantum-classical (q-c) state, i.e. a state having the form (\ref{qc-def}) \cite{O&Z,Datta}.
The nontrivial implication
\begin{equation}\label{n-imp}
  \left\{\shs D_B(\omega)=0\shs\right\}\quad \Rightarrow\quad \left\{\shs\omega \textrm{ is a q-c state\shs}\right\}
\end{equation}
is proved by noting that the infimum in  (\ref{q-d}) is always attained at some POVM $\M_{\omega}$ in $\mathfrak{M}_B^0$ and by
proving that the equality $D^{\M_{\omega}}_B(\omega)=0$ implies that
$\omega$ is a q-c state. The last step can be done by using the  characterisation
of a tripartite state at which the QCMI is equal to zero presented in  \cite{SSA-EQ} (see the proof of Theorem 2 in \cite{Datta}).

If $\omega$ is a q-c state of an infinite-dimensional bipartite system $AB$ then
it follows from the definition (\ref{q-d}) that $D_B(\omega)=0$, but it is not clear how to prove the
converse implication (\ref{n-imp}). One of the obstacles preventing us to prove
this implication is the question of attainability of the infimum in (\ref{q-d}).

It turns out that implication (\ref{n-imp}) can be established when either $\,S(\omega)<+\infty\,$ or $\,S(\omega_B)<+\infty\,$
by using the  generalized Koashi-Winter relation
(\ref{KWS-gen}) and the characterization of the input states of a quantum channel for which
the constrained Holevo capacity coincides with the mutual information obtained in \cite{TIN}.\smallskip

\begin{property}\label{null-QD}
\emph{The implication (\ref{n-imp}) holds for a state $\omega$ of an infinite-dimensional bipartite system $AB$
provided that either $\,S(\omega)<+\infty\,$ or $\,S(\omega_B)<+\infty$.}
\end{property}\smallskip

\emph{Proof.}  Assume first that $S(\omega_B)<+\infty$. Let $\hat{\omega}$ be a pure state in $\S(\H_{ABC})$ such that $\hat{\omega}_{AB}=\omega$.
If $D_B(\omega)=0$ then the generalized Koashi-Winter relation
(\ref{KWS-gen}) implies that $\chi_A(\hat{\omega}_{AC})=I(A\!:\!B)_{\hat{\omega}}$. This equality
means that the constrained Holevo capacity $\bar{C}(\Theta_{\!A},\hat{\omega}_{AC})$ of the partial trace channel $\Theta_{\!A}:\rho\mapsto \rho_A$, $\rho\in\S(\H_{AC})$,
at the state $\hat{\omega}_{AC}$ is equal to the mutual information $I(\Theta_{\!A},\hat{\omega}_{AC})$ of the channel $\Theta_{\!A}$
at the state $\hat{\omega}_{AC}$. So, since $S(\hat{\omega}_{AC})=S(\omega_B)<+\infty$, Theorem 3 in \cite{TIN} implies that
$$
\hat{\omega}_{AC}=\sum_k|\varphi_k\rangle\langle\varphi_k|,
$$
where $\{\varphi_k\}$ is an orthogonal system of vectors in $\H_{AC}$ such that $\Theta_{\!A}(|\varphi_k\rangle\langle\varphi_j|)=0$ for all $k\neq j$.
Let $|\varphi'_k\rangle=U|\varphi_k\rangle$, where $U$ is any unitary operator from $\H_{AC}$ onto $\H_B$. Consider the pure state
$\vartheta=\sum_{k,j}[\|\varphi_k\|\|\varphi_j\|]^{-1}|\varphi_k\otimes\varphi'_k\rangle\langle\varphi_j\otimes\varphi'_j|$ in $\S(\H_{ABC})$. Then $\vartheta_{AC}=\hat{\omega}_{AC}$
and the condition $\Theta_{\!A}(|\varphi_k\rangle\langle\varphi_j|)=0$ implies that $\vartheta_{AB}$ is a q-c state in $\S(\H_{AB})$. Since $\hat{\omega}$ and $\vartheta$
are two purifications of the state $\hat{\omega}_{AC}$, there is a partial isometry $W$ in $\B(\H_B)$ such that
$$
(I_{AC}\otimes W)\shs\hat{\omega}\shs(I_{AC}\otimes W^*)=\vartheta\quad \textrm{and} \quad (I_{AC}\otimes W^*)\shs\vartheta\shs(I_{AC}\otimes W)=\hat{\omega}.
$$
It follows that $\,\omega=\hat{\omega}_{AB}=(I_{A}\otimes W^*)\shs\vartheta_{AB}\shs (I_{A}\otimes W)\,$ is a q-c state.

Assume now that $S(\omega)<+\infty$ and $D_B(\omega)=0$. Let $\{P_n\}$ be an increasing sequence of finite-rank projectors in $\B(\H_A)$ strongly
converging to the unit operator $I_A$. Consider the sequence of states $\omega_n=c^{-1}_n (P_n\otimes I_B)\shs\omega\shs(P_n\otimes I_B)$, where $c_n=\Tr P_n\omega_A$.
By Lemma \ref{DB-GMP} below we have $c_nD_B(\omega_n)\leq D_B(\omega)=0$ and hence $D_B(\omega_n)=0$ for all $n$ large enough.
By Lemma 4 in \cite{L-2} the assumption $\,S(\omega)<+\infty\,$ implies that $\,S(\omega_n)<+\infty$ for all $n$. So, since $S([\omega_n]_A)<+\infty$, it follows from the triangle inequality for the entropy  that $S([\omega_n]_B)<+\infty$ for all $n$.  Thus, the above part of the proof
implies that $\,\omega_n$ is a q-c state in $\S(\H_{AB})$ for each $n$. Since the set of all q-c states is a closed subset of $\S(\H_{AB})$ by Lemma \ref{Q-C-closed} in the Appendix,
the state $\omega$ is a q-c state as a limit of the sequence $\,\{\omega_n\}$ of q-c states. $\Box$\smallskip

The generalization of monotonicy property (\ref{DB-m}) presented in the following lemma is proved  by using representation (\ref{u-r}) and the mototonicity of QCMI under local trace-non-preserving operations (see, f.i., \cite[Lemma 9]{CMI}).\smallskip
\begin{lemma}\label{DB-GMP}  \emph{Let $\omega$ be a state in $\S(\H_{AB})$ such that $C_B(\omega)<+\infty$ and $\,\Phi:A\to A$ a quantum operation   such that $\Tr\shs\Phi(\omega_A)\neq0$. Then}
\begin{equation*}
(\Tr\shs\Phi(\omega_A))D_B\!\left(\frac{\Phi\otimes \id_B(\omega)}{\Tr\shs\Phi(\omega_A)}\right)\leq D_B(\omega)\leq+\infty.
\end{equation*}
\end{lemma}
\smallskip

The question of how to prove the implication (\ref{n-imp}) in the case $\,S(\omega)=S(\omega_B)=+\infty\,$ remains open. But we may
put forward\smallskip

\textbf{Conjecture 1.} \textit{The implication (\ref{n-imp}) holds for any state $\omega$ in $\S(\H_{AB})$.}\smallskip

If this conjecture is true then implication (\ref{n-imp}) also holds for any state $\omega$ in $\S(\H_{AB})$ with $D_B$ replaced by the extensions $\widehat{D}_B$ and $\widetilde{D}_B$ of quantum discord to the set $\S(\H_{AB})$ considered in Section 5.3
(see Proposition \ref{ext-DB+}B).  

\subsection{On definitions of quantum discord for states with infinite one-way classical correlation}

For any POVM $\M$ on  $\H_B$ the unoptimized quantum discord $D_B^{\shs\M}(\omega)$ is well defined (as a nonnegative number or $+\infty$)  by the second formula in (\ref{UCBDB}) for any state $\omega$ in $\S(\H_{AB})$ with finite $C_B^{\shs\M}(\omega)$.  Correspondingly, the (optimazed) quantum discord $D_B(\omega)$ is well defined  in (\ref{q-d}) for any state $\omega\in\S(\H_{AB})$ with finite $C_B(\omega)$. At the same time, we can speak about quantum discord of states with infinite $C_B$. For example, it is reasonable to assume that q-c states  with infinite QMI have zero discord.

\subsubsection{The case of $D_B^{\shs\M}$}

To obtain appropriate extension of $D_B^{\shs\M}$ to the whole space $\S(\H_{AB})$ one can use representation (\ref{u-r}), i.e
one can define the extended unoptimized quantum discord as
\begin{equation}\label{e-UQD}
\widehat{D}_B^{\shs\M}(\omega)\doteq I(A\!:\!F|E)_{\tilde{\omega}},
\end{equation}
where $\tilde{\omega}=(I_A\otimes V_{\M})\shs\omega\shs (I_A\otimes V^*_{\M})$, $V_{\M}:\H_B\rightarrow\H_{EF}$ is the Stinespring isometry of the channel $\Psi_{\M}:B\to E$ defined in (\ref{Psi-M}) and it is assumed that $I(A\!:\!F|E)$ is the extended QCMI defined by the equivalent expressions (\ref{QCMI-e1}) and (\ref{QCMI-e2}).

The properties of the extended QCMI allow us to prove the following\smallskip

\begin{property}\label{ext-DB} \emph{Let $\M$ be an arbitrary discrete POVM on  $\H_B$.}\smallskip

A) \emph{The function $\widehat{D}_B^{\shs\M}$ (defined in (\ref{e-UQD})) is a unique nonnegative lower semicontinuous extension of $D_B^{\shs\M}$ to the set $\S(\H_{AB})$
possessing monotonicity property (\ref{un-DB-m}). It can be expressed as
\begin{equation}\label{e-UQD+}
 \widehat{D}_B^{\shs\M}(\omega)=\sup_{\Phi\in\F_A}D_B^{\shs\M}(\Phi\otimes \id_B(\omega))=\lim_{n\to+\infty}D_B^{\shs\M}(\Phi_n\otimes \id_B(\omega)),
\end{equation}
where $\F_A$ is the set of all channels $\Phi:A\to A$ with a finite-dimensional output and $\{\Phi_n\}$ is any sequence in $\F_A$ strongly converging to the identity channel $\,\id_A$.}
\smallskip

B) \emph{If $\,\M=\{|k\rangle\langle k|\}$, where $\{|k\rangle\}$ is an orthonormal basis in $\H_B$, then $\widehat{D}_B^{\shs\M}(\omega)=0$ for any q-c state $\omega$ determined by the basis $\{|k\rangle\}$ (i.e. having the form (\ref{qc-def})).}\smallskip

C) \emph{The function $\widehat{D}_B^{\shs\M}$ is invariant under local unitary transformations of $A$ and
\begin{equation*}
  (\Tr\shs\Phi(\omega_A))\widehat{D}_B^{\shs\M}\!\left(\frac{\Phi\otimes \id_B(\omega)}{\Tr\shs\Phi(\omega_A)}\right)\leq \widehat{D}_B^{\shs\M}(\omega)
\end{equation*}
for any quantum operation $\Phi:A\to A$ and any state $\omega$ in $\S(\H_{AB})$ s.t. $\Tr\shs\Phi(\omega_A)\neq0$.}\smallskip

D) \emph{The inequalities
\begin{equation}\label{DB-LAA-1}
\widehat{D}_B^{\shs\M}(p\rho+(1-p)\sigma) \geq p \widehat{D}_B^{\shs\M}(\rho)+(1-p)\widehat{D}_B^{\shs\M}(\sigma)-h_2(p)
\end{equation}
and
\begin{equation*}
\widehat{D}_B^{\shs\M}(p\rho+(1-p)\sigma) \leq p \widehat{D}_B^{\shs\M}(\rho)+(1-p)\widehat{D}_B^{\shs\M}(\sigma)+h_2(p)
\end{equation*}
hold for any states $\rho$ and $\sigma$ in  $\S(\H_{AB})$ and $p\in[0,1]$ with possible values $+\infty$ in both sides.}\smallskip

E) \emph{If $\,\M\in\mathfrak{M}_B^0\,$ (i.e. $\M$ consists of one-rank operators) then the equality
\begin{equation}\label{Heq}
\widehat{D}_B^{\shs\M}(\omega)+ER(\omega, I_A\otimes\M)=S(\omega_B)
\end{equation}
holds for any state $\omega$ in $\S(\H_{AB})$ with possible values $\,+\infty\,$ in both sides.}\footnote{$ER(\omega, I_A\otimes\M)$ is the entropy reduction of the POVM $I_A\otimes\M$ at a state $\omega$ defined in Section 3).}
\end{property}\smallskip

\emph{Proof.} A) It follows from (\ref{u-r}) that $\widehat{D}_B^{\shs\M}(\omega)=D_B^{\shs\M}(\omega)$ for any state $\omega$ with finite QMI.
Since the functions $\omega\mapsto I(A\!:\!B)_{\omega}\,$ and $C^{\M}_B(\omega)$ are lower semicontinuous on $\S(\H_{AB})$ and do not increase under local channels, the expression (\ref{e-UQD+}) proved below allows us to show that
$\widehat{D}_B^{\shs\M}(\omega)=D_B^{\shs\M}(\omega)=+\infty\,$ if $\,I(A\!:\!B)_{\omega}=+\infty\,$ and $\,C^{\M}_B(\omega)<+\infty$.\smallskip

The nonnegativity, lower semicontinuity and monotonicity property (\ref{un-DB-m}) of the function $\,\widehat{D}_B^{\shs\M}\,$ follow from the
corresponding properties of the extended QCMI described in Theorem 2 in  \cite{CMI}. If $\widetilde{D}_B^{\shs\M}$ is an other extension of $D_B^{\shs\M}$ possessing these properties then take any  state $\omega$ in $\S(\H_{AB})$ and consider the sequence of states $\omega_n\doteq\Pi_n\otimes\id_B(\omega)$  defined by means of any given sequence $\{\Pi_n\}$ of channels in $\mathfrak{F}_A$ strongly converging to the identity channel  $\id_A$. The lower semicontinuity and the monotonicity property of $\widehat{D}_B^{\shs\M}$
and $\widetilde{D}_B^{\shs\M}$ imply that
\begin{equation}\label{tmp-r}
\widehat{D}_B^{\shs\M}(\omega)=\lim_{n\to+\infty}D_B^{\shs\M}(\omega_n)=\widetilde{D}_B^{\shs\M}(\omega).
\end{equation}
Both representations in (\ref{e-UQD+}) are easily proved by using the lower semicontinuity and the monotonicity property of $\widehat{D}_B^{\shs\M}$.\smallskip

B) If $\omega$ is a q-c state determined by the basis $\{|k\rangle\}$ then all the states $\omega_n$ defined in the proof of claim A are q-c states determined by the basis $\{|k\rangle\}$ with finite QMI and hence
$\widehat{D}_B^{\shs\M}(\omega_n)=D_B^{\shs\M}(\omega_n)=0$. It follows from (\ref{tmp-r}) that $\widehat{D}_B^{\shs\M}(\omega)=0$.\smallskip

Claims C and D directly follow from the corresponding properties of the extended QCMI \cite{CMI},\cite[Section 5.2]{QC}.\smallskip

E) Assume first that $S(\omega_B)<+\infty$. If also $S(\omega)<+\infty\,$ then $S(\omega_A)<+\infty\,$ and the assumption $\M=\{M_i\}\in\mathfrak{M}^0_B$ implies that $S([\omega_i]_A)=S(\omega_i)$ for each $i$ such that $\,p_i=\Tr M_i\omega_B\neq0$, where  $\,\omega_i=p_i^{-1}(I_A\otimes \sqrt{M_i})\shs\omega\shs (I_A\otimes \sqrt{M_i})$. So, we have
\begin{equation*}
\begin{array}{rl}
D_B^{\shs\mathbb{M}}(\omega)\!\!& \displaystyle=\;S(\omega_A)+S(\omega_B)-S(\omega)-\left[S(\omega_A)-\sum_i p_i S([\omega_i]_A)\right]\\&\displaystyle=\;S(\omega_B)-\left[S(\omega)-\sum_i p_i S(\omega_i)\right]=S(\omega_B)-ER(\omega, I_A\otimes\M).
\end{array}
\end{equation*}
If $S(\omega)=+\infty$  then consider the sequence of states $\,\omega_n\doteq\Pi_n\otimes\id_B(\omega)\,$   defined by means of any given sequence $\{\Pi_n\}$ of channels in $\mathfrak{F}_A$ strongly converging to the identity channel $\id_A$. Since $S(\omega_n)\leq S([\omega_n]_A)+S(\omega_B)<+\infty$,
equality (\ref{Heq}) holds with $\omega=\omega_n$ for all $n$ by the above observation. Thus, to prove
the validity of (\ref{Heq}) for the state $\omega$ it suffices to show that
\begin{equation}\label{D-l-r}
\lim_{n\to+\infty}D_B^{\shs\M}(\omega_n)=\widehat{D}_B^{\shs\M}(\omega)\quad \textrm{and} \quad \lim_{n\to+\infty}ER(\omega_n, I_A\otimes\M)=ER(\omega, I_A\otimes\M).
\end{equation}
The first limit relation in (\ref{D-l-r}) follows from (\ref{e-UQD+}).
The second relation in (\ref{D-l-r}) follows from Proposition \ref{ER-prop}C in Section 3, since $[\omega_n]_B=\omega_B$ for all $n$ and $S(\omega_B)<+\infty$.

Assume now that $S(\omega_B)=+\infty$. Show first that
\begin{equation}\label{imp-ipl}
  \{S(\omega)<+\infty\}\quad \Rightarrow\quad \{\widehat{D}_B^{\shs\M}(\omega)=+\infty\}.
\end{equation}
Let $\{P_n\}$ be a sequence of projectors in $\B(\H_A)$ strongly converging
to the unit operator $I_A$. Consider the sequence of states $\,\omega_n=c_n^{-1}(P_n\otimes I_B)\shs\omega\shs (P_n\otimes I_B)$, where
$c_n=\Tr P_n\omega_A$. It follows from claim C that
\begin{equation}\label{DB-m-t}
c_n\widehat{D}_B^{\shs\M}(\omega_n)\leq \widehat{D}_B^{\shs\M}(\omega).
\end{equation}
Since $\,S([\omega_n]_A)<+\infty\,$ and $\,c_nS(\omega_n)\leq S(\omega)<+\infty\,$ by Lemma 4 in \cite{L-2}, we have $S([\omega_n]_B)<+\infty$  for any $n$.
By the above part of the proof equality (\ref{Heq})
holds with $\omega=\omega_n$  for all $n$. So, since $ER(\omega_n, I_A\otimes\M)\leq S(\omega_n)\leq c_n^{-1}S(\omega)$ for all $n$
and $S([\omega_n]_B)$ tends to $S(\omega_B)=+\infty\,$ as $\,n\to+\infty$, we conclude from (\ref{DB-m-t}) that
$\widehat{D}_B^{\shs\M}(\omega)=+\infty$.

To complete the proof of E we have to show that the finiteness of $\widehat{D}_B^{\shs\M}(\omega)$ implies
$ER(\omega, I_A\otimes\M)=+\infty$  (provided that $S(\omega_B)=+\infty$). Let $\omega=\sum_{i=1}^{+\infty}\lambda_i|\varphi_i\rangle\langle\varphi_i|$ be the spectral decomposition of $\omega$
and $\omega_n=c_n^{-1}\sum_{i=1}^{n}\lambda_i|\varphi_i\rangle\langle\varphi_i|$, where $c_n=\sum_{i=1}^{n}\lambda_i$. The concavity of the entropy reduction and inequality (\ref{DB-LAA-1}) imply that
\begin{equation}\label{2-ineq}
c_n ER(\omega_n, I_A\otimes\M)\leq ER(\omega, I_A\otimes\M)\quad \textrm{and} \quad
c_n\widehat{D}_B^{\shs\M}(\omega_n)\leq\widehat{D}_B^{\shs\M}(\omega)+h_2(c_n)<+\infty.
\end{equation}
Thus, it follows from (\ref{imp-ipl}) that $S([\omega_n]_B)<+\infty$ for all $n$ because $S(\omega_n)<+\infty$.  By the above part of the proof equality (\ref{Heq})
holds with $\omega=\omega_n$ for all $n$. So, since $S([\omega_n]_B)$ tends to $S(\omega_B)=+\infty\,$ as $\,n\to+\infty$, we conclude from the
inequalities in (\ref{2-ineq}) that $ER(\omega, I_A\otimes\M)=+\infty$.  $\Box$\smallskip

Proposition \ref{ext-DB}E implies the following observation used below\smallskip

\begin{corollary}\label{ext-DB-c} \emph{If $\,\M\in\mathfrak{M}_B^0\,$ then the inequality
\begin{equation*}%
\widehat{D}_B^{\shs\M}(\omega)\geq S(\omega_B)-S(\omega)
\end{equation*}
holds (with possible values $\,+\infty\,$ in both sides) for any state $\omega$ in $\S(\H_{AB})$ with finite entropy.}
\end{corollary}

\subsubsection{The case of $D_B$}

Now consider the question of definition of the (optimized) quantum discord for any state in $\S(\H_{AB})$.
The first way to define the extended quantum discord is the following
\begin{equation}\label{D-B-E-1}
  \widehat{D}_B(\omega)=\inf_{\M\in\mathfrak{M}_B}\widehat{D}_B^{\shs\M}(\omega),\quad \omega\in\S(\H_{AB}),
\end{equation}
where $\widehat{D}_B^{\shs\M}$ is the extension of  unoptimized quantum discord $D_B^{\shs\M}$ defined in (\ref{e-UQD}).
Since $\widehat{D}_B^{\shs\M}(\omega)=D_B^{\shs\M}(\omega)$ provided that $C_B^{\shs\M}(\omega)<+\infty$, we have
$\widehat{D}_B(\omega)=D_B(\omega)$ for any state $\omega$ with finite $C_B(\omega)$.

Another way inspired by  expression (\ref{e-UQD+}) is to define  the extended quantum discord
as
\begin{equation}\label{D-B-E-2}
  \widetilde{D}_B(\omega)=\sup_{\Phi\in\F_A}D_B(\Phi\otimes \id_B(\omega)),\quad \omega\in\S(\H_{AB}),
\end{equation}
where $\F_A$ is the set of all channels $\Phi:A\to A$ with a finite-dimensional output.
By using the monotonicity property (\ref{DB-m}) of quantum discord and the limit relations in (\ref{VerNewE}) it is easy to show that $\widetilde{D}_B(\omega)=D_B(\omega)$ for any state $\omega$ with finite $C_B(\omega)$.

Thus, both functions  $\widehat{D}_B$ and $\widetilde{D}_B$ are extensions of $D_B$ (defined in (\ref{q-d}) for all states with finite $C_B$) to the whole space $\S(\H_{AB})$.  By representation (\ref{e-UQD+}) the minimax
inequality implies that
\begin{equation}\label{ext-ineq}
  \widetilde{D}_B(\omega)\leq \widehat{D}_B(\omega)\quad \forall \omega\in\S(\H_{AB}).
\end{equation}

Properties of these two extensions are described in the following\smallskip

\begin{property}\label{ext-DB+} A) \emph{The function $\widetilde{D}_B$ (defined in (\ref{D-B-E-2})) is a unique nonnegative lower semicontinuous extension of $D_B$ to the set $\S(\H_{AB})$
possessing monotonicity property (\ref{DB-m}). It can be expressed as
\begin{equation}\label{e-UQD++}
 \widetilde{D}_B(\omega)=\lim_{n\to+\infty}D_B(\Phi_n\otimes \id_B(\omega)),
\end{equation}
where $\{\Phi_n\}$ is any sequence of channels in  $\F_A$ strongly converging to the channel $\,\id_A$.}\smallskip

B) \emph{The functions $\widehat{D}_B$ and $\widetilde{D}_B$ are equal to zero on the set $\S_{\rm qc}(\H_{AB})$ of all q-c states  (i.e. states having the form (\ref{qc-def}) with any basis $\{|k\rangle\}$) in $\S(\H_{AB})$. If
Conjecture 1 in Section 5.2 is true, then}
\begin{equation}\label{B-exp}
\widetilde{D}_B^{-1}(0)=\widehat{D}_B^{-1}(0)=\S_{\rm qc}(\H_{AB}).
\end{equation}

C) \emph{The functions $\widehat{D}_B$ and $\widetilde{D}_B$ are invariant under local unitary transformations and
\begin{equation}\label{DB-m++}
  (\Tr\shs\Phi(\omega_A))D^*_B\!\left(\frac{\Phi\otimes \id_B(\omega)}{\Tr\shs\Phi(\omega_A)}\right)\leq D^*_B(\omega),\quad D^*_B=\widehat{D}_B,\widetilde{D}_B
\end{equation}
for any quantum operation $\Phi:A\to A$ and any state $\omega$ in $\S(\H_{AB})$ s.t. $\Tr\shs\Phi(\omega_A)\neq0$.}
\smallskip

D) \emph{The inequalities
\begin{equation}\label{DB-LAA-1+}
D^{*}_B(p\rho+(1-p)\sigma) \geq p D^{*}_B(\rho)+(1-p)D^{*}_B(\sigma)-h_2(p),\quad D^*_B=\widehat{D}_B,\widetilde{D}_B
\end{equation}
hold for any states $\rho$ and $\sigma$ in  $\S(\H_{AB})$ and $p\in[0,1]$ (with possible values $\,+\infty\,$ in one or both sides).}\smallskip


E) \emph{For any state $\omega$ in $\S(\H_{AB})$ such that $\,S(\omega)<+\infty\,$ the following inequalities hold
\begin{equation}\label{new-D-ineq}
D^*_B(\omega)\geq S(\omega_B)-S(\omega),\quad \quad D^*_B=\widehat{D}_B,\widetilde{D}_B
\end{equation}
(with possible values $\,+\infty\,$ in one or both sides). In particular, $\widehat{D}_B(\omega)=\widetilde{D}_B(\omega)=+\infty\;$ for any state $\omega$ in $\S(\H_{AB})$ such that $\,S(\omega)<+\infty\,$ and $\,C_B(\omega)=+\infty$.}
\end{property}\smallskip

\emph{Proof.} A) The lower semicontinuity of the function $\widetilde{D}_B$ on $\S(\H_{AB})$ follows from its definition, since the function $\omega\mapsto D_B(\Phi\otimes \id_B(\omega))$
is (uniformly) continuous on $\S(\H_{AB})$ for any channel $\Phi$ in  $\F_A$ by the remark at the beginning of Section 5.1. To prove that $\widetilde{D}_B$ is a unique extension of $D_B$ with the stated properties one can use the same arguments as in the proof of the corresponding assertion of Proposition \ref{ext-DB}.

Representation (\ref{e-UQD++}) follows from the lower semicontinuity of $\widetilde{D}_B$ and its definition.\smallskip

B) If $\omega$ is a q-c state  then $\widehat{D}_B(\omega)=0$ by Proposition \ref{ext-DB}B. It follows from (\ref{ext-ineq}) that $\widetilde{D}_B(\omega)=0$ as well.\smallskip

Assume that
Conjecture 1 in Section 5.2 is true. By inequality (\ref{ext-ineq}) to prove (\ref{B-exp}) it suffices to show that $\widetilde{D}_B(\omega)=0$ implies that $\omega$ is a q-c state.
Let $\{\Phi_n\}$ be any sequence of channels in  $\F_A$ strongly converging to the channel $\,\id_A$.
Then the monotonicity property  (\ref{DB-m}) implies that $D_B(\Phi_n\otimes \id_B(\omega))=0$ for all $n$. So, by the assumed validity of Conjecture 1,
$\Phi_n\otimes \id_B(\omega)$ is a q-c state for each $n$. Since  $\,\Phi_n\otimes \id_B(\omega)\,$ tends to $\,\omega\,$ as $\,n\to+\infty$, Lemma \ref{Q-C-closed} in the Appendix shows that $\omega$ is a q-c state.\smallskip

C) Inequality (\ref{DB-m++}) for the function $\widehat{D}_B$ follows from the corresponding inequality for the function $\widehat{D}^{\shs\M}_B$
in Proposition \ref{ext-DB}.

Let $\omega_0$ be an arbitrary state in $\S(\H_{AB})$ such that $\Tr\shs\Phi([\omega_0]_A)\neq0$ and $\{\Psi_n\}$ any sequence channels in  $\F_A$ strongly converging to the identity channel $\id_A$.
Since the inequality (\ref{DB-m++}) with  $\omega=\omega_n\doteq\Psi_n\otimes \id_B(\omega_0)$ and $D^*_B=D_B$ holds for all $n$ by Lemma \ref{DB-GMP} in Section 5.2, representation (\ref{e-UQD++}) implies that
$$
\widetilde{D}_B(\omega_0)=\lim_{n\to+\infty}D_B(\omega_n)\geq\liminf_{n\to+\infty}c_nD_B(c_n^{-1}\Phi\otimes \id_B(\omega_n))\geq c_0\widetilde{D}_B(c_0^{-1}\Phi\otimes \id_B(\omega_0)),
$$
where $c_n=\Tr\shs\Phi([\omega_n]_A)$, $n\geq0$. The last inequality follows from the lower semicontinuity of $\widetilde{D}_B$ on $\S(\H_{AB})$.\smallskip

D) Inequality (\ref{DB-LAA-1+}) for the function $\widehat{D}_B$
follows, by definition (\ref{D-B-E-1}), from the corresponding inequality for the function $\widehat{D}_B^{\shs\M}$
(inequality (\ref{DB-LAA-1}) in  Proposition \ref{ext-DB}).

Inequality (\ref{DB-LAA-1+}) for the function $\widetilde{D}_B$ is derived
by using expression (\ref{e-UQD++}) from the corresponding inequality for the function $D_B$.\smallskip

E) Inequality (\ref{new-D-ineq}) for $D_B^*=\widehat{D}_B$ directly follows from Corollary \ref{ext-DB-c} in Section 5.3.1. 

To prove (\ref{new-D-ineq}) for $D_B^*=\widetilde{D}_B$ take a sequence $\{P_n\}$  of finite rank projectors in $\B(\H_A)$
strongly converging to the unit operator $I_A$. Consider the sequence of quantum operations $\Phi_n(\rho)=P_n\rho P_n$.  By Lemma 4 in \cite{L-2} we have
$c_nS(\omega_n)\leq S(\omega)<+\infty$, where $c_n=\Tr P_n\omega_A$ and $\omega_n=c_n^{-1}\Phi_n\otimes\id_B(\omega)$. So, Corollary \ref{ext-DB-c} in Section 5.3.1 implies that
\begin{equation}\label{new-D-ineq+}
c_n D_B(\omega_n)\geq c_nS([\omega_n]_B)-S(\omega)\quad \forall n.
\end{equation}
Since $c_n[\omega_n]_B\leq\omega_B$ for all $n$, by using the concavity and lower semicontinuity of the entropy it is easy to show that $S([\omega_n]_B)$ tends to
$S(\omega_B)\leq+\infty$.  Since Proposition \ref{ext-DB+}C
shows that $c_n D_B(\omega_n)\leq \widetilde{D}_B(\omega)$ for all $n$, the lower semicontinuity of $\widetilde{D}_B$ implies that
$D_B(\omega_n)=\widetilde{D}_B(\omega_n)$ tends to $\widetilde{D}_B(\omega)\leq+\infty$. So, inequality (\ref{new-D-ineq}) for $D_B^*=\widetilde{D}_B$ 
is proved by passing to the limit as $\,n\to+\infty\,$ in (\ref{new-D-ineq+}).

The last claim of E is due to the fact that the  condition $\,C_B(\omega)=+\infty$ implies $\,S(\omega_B)=+\infty$. $\Box$\smallskip

Thus, at the moment we have two extensions $\widehat{D}_B$ and $\widetilde{D}_B$ of the quantum discord
to states with infinite value of $C_B$ defined, respectively, by expressions (\ref{D-B-E-1}) and (\ref{D-B-E-2}).
Proposition \ref{ext-DB+} shows that $\widehat{D}_B(\omega)=\widetilde{D}_B(\omega)$ provided that
\begin{itemize}
\item  $\omega$ is a state such that either $\,C_B(\omega)<+\infty$ or $\,S(\omega)<+\infty$;
\item $\omega$ is a q-c state (in particular, a q-c state with infinite QMI).
\end{itemize}

Thus, it is reasonable to put forward the following \smallskip

\textbf{Conjecture 2.} \textit{The functions $\widehat{D}_B$ and $\widetilde{D}_B$ coincide on $\S(\H_{AB})$.}
\smallskip

The main problem that prevents to prove this conjecture
consists in the necessity to take the infimum in the definition of quantum discord over POVM with unbounded number of outcomes.
Proposition \ref{ext-DB+}A implies that Conjecture 2 can be proved by showing that $\widehat{D}_B$ is a lower semicontinuous function on $\S(\H_{AB})$.

\section{Applications to quantum channels characteristics}

\subsection{The Koashi-Winter and Xi-Lu-Wang-Li relations in terms of a channel}

Let $\Phi$ be a quantum channel from $A$ to $B$ with the Stinespring representation $\Phi(\rho)=\Tr_E V_{\Phi}\rho V_{\Phi}^*$ determined
by a given isometry $V_{\Phi}:\H_A\rightarrow \H_{BE}$ (see details in Section 2). For an input state $\rho\in\S(\H_A)$ introduce the state $\omega_{\Phi}^{\rho}=(V_{\Phi}\otimes I_R)\shs \hat{\rho}\shs (V_{\Phi}^*\otimes I_R)$ in $\S(\H_{BER})$, where $\hat{\rho}$ is a
given pure state in $\S(\H_{AR})$ such that $\Tr_R\hat{\rho}=\rho$. Then the mutual information $I(\Phi,\rho)$ and the constrained Holevo capacity $\bar{C}(\Phi,\rho)$ of the channel $\Phi$ at the state $\rho$ (defined, respectively, in (\ref{MI-Ch-def}) and (\ref{CHI-QC-def})) can be expressed as
\begin{equation}\label{MI-rep}
I(\Phi,\rho)=I(B\!:\!R)_{\omega_{\Phi}^{\rho}}=I(B\!:\!R)_{\Phi\otimes\id_R(\hat{\rho})}
\end{equation}
and
\begin{equation}\label{CHI-rep}
\bar{C}(\Phi,\rho)=\chi_B([\omega_{\Phi}^{\rho}]_{BE})=C_R([\omega_{\Phi}^{\rho}]_{BR})=C_R(\Phi\otimes\id_R(\hat{\rho})),
\end{equation}
where $C_R$  is the one-way classical correlation in $BR$ with  measured system $R$ and the second equality is due to the generalized Koashi-Winter relation (\ref{KWS-gen}) valid  as $\omega_{\Phi}^{\rho}$ is a pure state in $\S(\H_{BRE})$. It follows that
\begin{equation}\label{D-rep}
D_R(\Phi\otimes\id_R(\hat{\rho}))=D_R([\omega_{\Phi}^{\rho}]_{BR})=I(\Phi,\rho)-\bar{C}(\Phi,\rho),
\end{equation}
i.e. the quantum discord $D_R(\Phi\otimes\id_R(\hat{\rho}))$ characterizes the gain
of entanglement assistance in transmission of classical information over the channel $\Phi$ \cite{BSST,H-SCI,Wilde}.

The constrained Holevo capacity $\bar{C}(\widehat{\Phi},\rho)$ of any channel $\widehat{\Phi}$ complementary to the channel $\Phi$ (see  Section 2)
can be expressed as
\begin{equation}\label{ER-rep}
\begin{array}{rl}
 \bar{C}(\widehat{\Phi},\rho)=\chi_E([\omega_{\Phi}^{\rho}]_{BE}) = C_R([\omega_{\Phi}^{\rho}]_{ER})\!\! &=\;\sup_{\M\in\mathfrak{M}_R}ER([\omega_{\Phi}^{\rho}]_{BR},I_{B}\otimes\M)\\\\&=\;\sup_{\M\in\mathfrak{M}_R}ER(\Phi\otimes\id_R(\hat{\rho}),I_{B}\otimes\M)
\end{array}
\end{equation}
for any input state $\rho\in\S(\H_A)$ and its purification $\hat{\rho}\in\S(\H_{AR})$, where $\mathfrak{M}_R$ is the set of all discrete POVM on $\H_R$ (since all the channels complementary to the channel $\Phi$ are isometrically equivalent (cf.\cite{H-c-ch}) to each other, the quantity $\bar{C}(\widehat{\Phi},\rho)$ is completely determined by $\Phi$ and $\rho$).
The second equality in (\ref{ER-rep}) is due to the generalized Koashi-Winter relation (\ref{KWS-gen}), while the third one
follows from the generalized Xi-Lu-Wang-Li relation (\ref{Xi+++}).  The supremum in (\ref{ER-rep}) can be taken only over the set $\mathfrak{M}^0_R$ of all POVM on $\H_R$ consisting of one-rank operators (this is shown in Section 4).

For a given ensemble $\mu=\{p_i,\rho_i\}$ of states in $\S(\H_A)$ with the average state $\rho$ denote by $\M_\mu=\{M^{\mu}_i\}$ a POVM on $\H_R$ such that
$p_i\rho_i=\Tr_R (I_A\otimes M^{\mu}_i)\shs \hat{\rho}$ for all $i$, where $\hat{\rho}\in\S(\H_{AR})$ is a given purification of $\rho$ (such POVM always exists by the  Schrodinger-Gisin-Hughston-Jozsa-Wootters theorem \cite{Schr,Gisin,HJW}).
If $\mu=\{p_i,\rho_i\}$ is an ensemble of pure states then the POVM  $\M_\mu=\{M^{\mu}_i\}$ can be chosen in $\mathfrak{M}_R^0$. The "unoptimazed"  relations  (\ref{KWS-gen+un}) and (\ref{Xi++un}) imply that

\begin{equation}\label{MI-CHI-rep+}
C^{\shs\M_{\mu}}_R(\Phi\otimes\id_R(\hat{\rho}))=C^{\shs\M_{\mu}}_R([\omega_{\Phi}^{\rho}]_{BR})=\chi(\Phi(\mu))\doteq\chi(\{p_i,\Phi(\rho_i)\}),
\end{equation}
\begin{equation}\label{D-rep+}
D^{\shs\M_{\mu}}_R(\Phi\otimes\id_R(\hat{\rho}))=D^{\shs\M_{\mu}}_R([\omega_{\Phi}^{\rho}]_{BR})=I(\Phi,\rho)-\chi(\Phi(\mu))\qquad\quad\,
\end{equation}
and
\begin{equation}\label{ER-rep+}
\begin{array}{rl}
ER(\Phi\otimes\id_R(\hat{\rho}), I_B\otimes\M_\mu)\!\!&=\;ER([\omega_{\Phi}^{\rho}]_{BR}, I_B\otimes\M_\mu)\\\\&=\;C^{\shs\M_{\mu}}_R([\omega_{\Phi}^{\rho}]_{ER})=\chi(\widehat{\Phi}(\mu))\doteq\chi(\{p_i,\widehat{\Phi}(\rho_i)\}).
\end{array}
\end{equation}
The output Holevo
information $\chi(\widehat{\Phi}(\mu))$ of a complementary channel is interpreted as a bound on the amount of classical information "obtained" by the environment (or "eavesdropper" in terms of secret communications) \cite{H-SCI,Wilde}.  The quantity $\chi(\widehat{\Phi}(\mu))$ is completely determined by $\Phi$ and $\mu$ due to  the isometrical equivalence
of all the channels complementary to the channel $\Phi$ \cite{H-c-ch}.\smallskip

Below we will consider different applications of the relations (\ref{MI-rep})-(\ref{ER-rep}) and (\ref{MI-CHI-rep+})-(\ref{ER-rep+}).

\subsection{New continuity bounds for characteristics of a quantum channel depending on input dimension/energy}

\subsubsection{Continuity bounds for the output Holevo information of a quantum channel and a complementary channel}


Representations (\ref{MI-CHI-rep+}) and (\ref{ER-rep+}) along with the results of Sections 4.3.1-2 in \cite{QC} and Corollaries \ref{ER-CB}-\ref{ER-CB++} in Section 3
allow us to obtain uniform continuity
bounds for the functions  $\Phi\mapsto\chi(\Phi(\mu))$ and $\Phi\mapsto\chi(\widehat{\Phi}(\mu))$  valid for any (discrete or continuous) ensemble $\mu$
that depend either on the input dimension of $\Phi$ (if it is finite) or on the energy bound on the mean energy of $\mu$.
The new continuity bounds for the function  $\Phi\mapsto\chi(\Phi(\mu))$
essentially refine the corresponding results obtained in \cite{CID}.

A \textit{generalized (continuous) ensemble} of states in $\S(\H)$ is defined as a Borel probability measure on the set $\S(\H)$ \cite{H-SCI,H-Sh-2}.
The set $\mathcal{P}(\mathcal{H})$ of all generalized  ensembles of states in $\S(\H)$ contains the  subset $\mathcal{P}_0(\mathcal{H})$ of discrete ensembles  corresponding to discrete measures. We will assume that the set $\mathcal{P}(\mathcal{H})$ is equipped with the weak convergence topology \cite{Bill,H-Sh-2}. So, a sequence of ensembles $\mu_n$ in $\P(\H)$  converges to an ensemble $\mu_0$ in $\P(\H)$ if $\;\lim_{n\to+\infty}\int f(\rho)\mu_n(d\rho)=\int f(\rho)\mu_0(d\rho)\;$ for any continuous bounded function $f$ on $\S(\H)$. It is easy to see that $\mathcal{P}_0(\mathcal{H})$ is a dense subset of $\mathcal{P}(\mathcal{H})$.

The average state $\bar{\rho}(\mu)$ of a generalized
ensemble $\mu \in \mathcal{P}(\mathcal{H})$ is defined as the barycenter of
$\mu $, i.e.  $\bar{\rho}(\mu)=\int_{\mathfrak{S}(\mathcal{H})}\rho \mu (d\rho)$ (where $\int_{\mathfrak{S}(\H)}$ denotes the Bochner integral).

For an ensemble $\mu \in \mathcal{P}(\mathcal{H}_{A})$ its image $\mathrm{\Phi}(\mu) $
under a quantum channel $\mathrm{\Phi}:A\rightarrow B\,$ is defined as the
ensemble in $\mathcal{P}(\mathcal{H}_{B})$ corresponding to the measure $\mu
\circ \mathrm{\Phi} ^{-1}$ on $\mathfrak{S}(\mathcal{H}_{B})$, i.e.
$\,\mathrm{\Phi} (\mu )[\mathfrak{S}_{B}]=\mu[\mathrm{\Phi} ^{-1}(\mathfrak{S}_{B})]\,$ for any Borel subset
$\mathfrak{S}_{B}$ of $\mathfrak{S}(\mathcal{H}_{B})$, where $\mathrm{\Phi} ^{-1}(\mathfrak{S}_{B})$ is the pre-image of $\mathfrak{S}_{B}$ under
the map $\mathrm{\Phi} $. If $\mu =\{p _{k},\rho _{k}\}$ then  $\mathrm{\Phi} (\mu)=\{p _{k},\mathrm{\Phi}(\rho_{k})\}$.

For a given channel $\,\mathrm{\Phi}:A\rightarrow B\,$ the output Holevo information of a
generalized ensemble $\mu$ in $\mathcal{P}(\H_A)$ is defined as
\begin{equation}\label{H-Q}
\!\chi(\Phi(\mu))=\int_{\mathfrak{S}(\mathcal{H}_A)} D(\mathrm{\Phi}(\rho)\shs \|\shs \mathrm{\Phi}(\bar{\rho}(\mu)))\mu (d\rho )=S(\mathrm{\Phi}(\bar{\rho}(\mu
)))-\int_{\mathfrak{S}(\mathcal{H}_A)} S(\mathrm{\Phi}(\rho))\mu (d\rho ),  
\end{equation}
where the second formula is valid under the condition $S(\mathrm{\Phi}(\bar{\rho}(\mu)))<+\infty$ \cite{H-Sh-2}.

We will use the following simple\smallskip

\begin{lemma}\label{sl} \emph{Let  $\mu$ be a generalized ensemble in $\P(\H_A)$. There is a sequence $\{\mu_n\}$ of
finite ensembles in $\P(\H_A)$ weakly converging to $\mu$ such that $\,\bar{\rho}(\mu_n)=\bar{\rho}(\mu)\,$ for all $n$ and
$\;\lim\limits_{n\to+\infty}\chi(\Phi(\mu_n))=\chi(\Phi(\mu))\leq+\infty\;$
for any quantum channel  $\,\Phi:A\to B$.}
\end{lemma}\smallskip

\emph{Proof.} By applying the construction from the proof of Lemma 1 in \cite{H-Sh-2} to the ensemble $\mu$ one can obtain a sequence $\{\mu_n\}$ of
finite ensembles in $\P(\H_A)$ weakly converging to $\mu$ such that $\,\bar{\rho}(\mu_n)=\bar{\rho}(\mu)\,$ and
$\,\chi(\Phi(\mu_n))\leq\chi(\Phi(\mu))\,$ for all $n$ and any quantum channel  $\,\Phi:A\to B$ (the last inequality is due to the joint convexity of the relative entropy). So,  the lower semicontinuity of the function $\nu\mapsto\chi(\Phi(\nu))$ on $\P(\H_A)$  (cf.\cite[Proposition 1]{H-Sh-2}) implies that $\,\chi(\Phi(\mu_n))\,$ tends to  $\,\chi(\Phi(\mu))\,$ as $\,n\to+\infty$. $\Box$
\smallskip

To formulate our first result recall that the diamond norm of a Hermitian preserving linear map $\Theta:\T(\H_A)\to\T(\H_B)$  is defined as
\begin{equation}\label{DN-def}
\|\Theta\|_{\diamond}\doteq \sup_{\rho\in\S(\H_{AR})}\|\Theta\otimes \id_R(\rho)\|_1,
\end{equation}
where $R$ is a system isomorphic to $A$ \cite{Kit}. It is also called the norm of complete boundedness \cite{Paul}. The metric induced by the diamond norm is a basic measure of divergence between finite-dimensional
quantum channels \cite[Section 9]{Wilde}).\smallskip

\begin{property}\label{CHI-PHI} \emph{Let $\,d=\dim\H_A<+\infty\,$ and $\mu$ be an arbitrary generalized ensemble of states in $\S(\H_A)$.  Then
\begin{equation}\label{CHI-PHI+}
 |\chi(\Phi(\mu))-\chi(\Psi(\mu))|\leq \varepsilon \ln d+g(\varepsilon)
\end{equation}
and
\begin{equation}\label{CHI-PHI++}
 |\chi(\widehat{\Phi}(\mu))-\chi(\widehat{\Psi}(\mu))|\leq \varepsilon \ln d+g(\varepsilon)
\end{equation}
for any channels $\Phi$ and $\Psi$ from $A$ to $B$ such that $\frac{1}{2}\|\Phi-\Psi\|_{\diamond}\leq\varepsilon$, where  $g$ is the function defined in (\ref{g-def}).}\smallskip

\emph{Continuity bounds (\ref{CHI-PHI+}) and (\ref{CHI-PHI++}) are asymptotically tight for large $d$ (Def.1 in \cite[Section 3.2.1]{QC}). Moreover, for any $d\geq2$ and $\varepsilon\in[0,1]$ there exist channels $\Phi$ and $\Psi$ from a $d$-dimensional system $A$ to some system $B$ and an input ensemble $\mu$  such that}
\begin{equation}\label{double-eq}
\textstyle\frac{1}{2}\|\Phi-\Psi\|_{\diamond}=\varepsilon\quad and\quad |\chi(\Phi(\mu))-\chi(\Psi(\mu))|=|\chi(\widehat{\Phi}(\mu))-\chi(\widehat{\Psi}(\mu))|= \varepsilon \ln d.
\end{equation}
\end{property}

\textbf{Note:} We may assume in Proposition \ref{CHI-PHI} that $\,d=\rank\bar{\rho}(\mu)$.\smallskip

\begin{remark}\label{CHI-PHI++r} Proposition 6 in \cite{CID} implies that
\begin{equation}\label{OLD-CHI-PHI}
 |\chi(\Phi(\mu))-\chi(\Psi(\mu))|\leq \varepsilon \ln d+\varepsilon \ln 2+g(\varepsilon)
\end{equation}
for any channels $\Phi$ and $\Psi$ from $A$ to $B$ such that $\beta(\Phi,\Psi)\leq\varepsilon$ and any ensemble $\mu$ of states in $\S(\H_A)$, where
$d=\dim\H_A$ and $\beta(\Phi,\Psi)$ is the Bures distance between $\Phi$ and $\Psi$
defined as
\begin{equation}\label{BD-def}
\beta(\Phi,\Psi)\doteq \sup_{\rho\in\S(\H_{AR})}\beta(\Phi\otimes\id_R(\rho),\Psi\otimes\id_R(\rho))
\end{equation}
(here $R$ is a quantum system isomorphic to $A$ and
$\beta$ in the r.h.s. denotes the Bures distance
between states in $\S(\H_{BR})$ \cite{H-SCI,Wilde}).
Since $\frac{1}{2}\|\Phi-\Psi\|_{\diamond}\leq \beta(\Phi,\Psi)$,
continuity bound (\ref{CHI-PHI+}) is  \emph{strictly sharper} than (\ref{OLD-CHI-PHI}).\smallskip

The continuity bound (\ref{CHI-PHI++}) has no analogues constructed earlier (as far as I know).
\end{remark}\smallskip

\emph{Proof of Proposition \ref{CHI-PHI}.} It follows from the proof of Proposition 14 in \cite[Section 4.3.1]{QC} that the continuity bound for the function $C_B$
depending on $d_B$ in the case $\rho_B=\sigma_B$  (presented in this proposition) remains valid for the function
$C^{\shs\M}_B$ provided that $\M$ is a POVM in $\mathfrak{M}_B$ with a finite number of outcomes. It implies, in out notation, that
$$
|C^{\shs\M}_R(\Phi\otimes \id_R (\hat{\rho}))-C^{\shs\M}_R(\Psi\otimes \id_R (\hat{\rho}))|\leq \varepsilon \ln\dim\H_R+g(\varepsilon),\;\; \varepsilon=\textstyle\frac{1}{2}\|\Phi\otimes \id_R (\hat{\rho})-\Psi\otimes \id_R (\hat{\rho})\|_1,
$$
for any POVM $\M$ in $\mathfrak{M}_R$ with a finite number of outcomes and any pure state $\hat{\rho}$ in $\S(\H_{AR})$, $R\cong A$. So, if $\mu$ is a finite ensemble then by taking $\M=\M_{\mu}$ in the above inequality and by using
(\ref{MI-CHI-rep+}) along with (\ref{DN-def}) we obtain (\ref{CHI-PHI+}).

Inequality (\ref{CHI-PHI++}) for any discrete (in particular, finite)  ensemble $\mu$ is proved by using representation (\ref{ER-rep+}) and  the continuity bound for the entropy reduction of a local measurement from Corollary \ref{ER-CB} in Section 3.

If $\mu$ is an arbitrary generalized ensemble then Lemma \ref{sl} implies the existence of a sequence $\{\mu_n\}$ of finite ensembles weakly converging to
$\mu$ such that
$\;
\lim_{n\rightarrow+\infty}\chi(\Theta(\mu_n))=\chi(\Theta(\mu)),\;$  $\Theta=\Phi,\Psi,\widehat{\Phi},\widehat{\Psi},\,$
and $\bar{\rho}(\mu_n)=\bar{\rho}(\mu)$ for all $n$.
By the above part of the proof,  inequalities  (\ref{CHI-PHI+}) and (\ref{CHI-PHI++}) hold with $\mu=\mu_n$ for all $n$. Thus, the above limit relations  imply the validity of (\ref{CHI-PHI+}) and (\ref{CHI-PHI++}) for the ensemble $\mu$.

To prove the last claim consider the family of quantum  erasure channels $\Omega_p$ from a $d$-dimensional quantum system $A$ to its
$(d+1)$-dimensional extension $B$  defined as
\begin{equation}\label{era-ch}
\Omega_p(\rho)=(1-p)\rho+p\shs[\Tr\rho] |\tau_0\rangle\langle\tau_0|,\quad \rho\in\S(\H_A),
\end{equation}
where $\tau_0$ is a unit vector in $\H_B$ orthogonal to $\H_A\subset\H_B$, $p\in[0,1]$.
It is easy to see that $\,\widehat{\Omega}_p=\Omega_{1-p}$, $\,\frac{1}{2}\|\Omega_{p}-\Omega_{q}\|_{\diamond}=|p-q|\,$ and $\,\chi(\Omega_p(\mu))=(1-p)\chi(\mu)\,$ for any ensemble $\mu$ of input states. It follows that
(\ref{double-eq}) holds for any ensemble $\mu$ of pure states with the chaotic average state and the channels $\Phi=\Omega_p$ and $\Psi=\Omega_q$, where $p$ and $q$ are any numbers in $[0,1]$ such that $|p-q|=\varepsilon$.
$\Box$ \smallskip

The \emph{privacy} of a channel $\Phi:A\to B$ at an ensemble $\mu$ of states in $\S(\H_A)$ is defined as (cf.\cite{H-SCI})
\begin{equation}\label{pi-def}
\pi(\Phi,\mu)=\chi(\Phi(\mu))-\chi(\widehat{\Phi}(\mu)).
\end{equation}
This quantity is related to private capacity of a channel (see Section 6.2.3 below).\smallskip

Proposition \ref{CHI-PHI} gives a continuity bound for the function $\Phi\mapsto\pi(\Phi,\mu)$.\smallskip

\begin{corollary}\label{PI-PHI} \emph{Let $\,d=\dim\H_A<+\infty\,$ and $\mu$ be an arbitrary generalized ensemble of states in $\S(\H_A)$.  Then
\begin{equation}\label{PI-PHI+}
 |\pi(\Phi,\mu)-\pi(\Psi,\mu)|\leq 2\varepsilon \ln d+2g(\varepsilon)
\end{equation}
for any channels $\,\Phi$ and $\,\Psi$ from $A$ to $B$ such that $\,\frac{1}{2}\|\Phi-\Psi\|_{\diamond}\leq\varepsilon$, where  $g$ is the function defined in (\ref{g-def}).}\smallskip

\emph{Continuity bound (\ref{PI-PHI+}) is asymptotically tight for large $d$ (Def.1 in  \cite[Section 3.2.1]{QC}). Moreover, for any $d\geq2$ and $\varepsilon\in[0,1]$ there exist channels $\,\Phi$ and $\,\Psi$ from a $d$-dimensional system $A$ to some system $B$ and an input ensemble $\mu$  such that}
\begin{equation*}
\textstyle\frac{1}{2}\|\Phi-\Psi\|_{\diamond}=\varepsilon\quad and\quad |\pi(\Phi,\mu)-\pi(\Psi,\mu)|= 2\varepsilon \ln d.
\end{equation*}
\end{corollary}

\textbf{Note:} We may assume in Corollary \ref{PI-PHI} that $\,d=\rank\bar{\rho}(\mu)$.
\smallskip

\emph{Proof.}  It suffices to prove the last claim. This can be done easily by using the family of erasure channels $\Omega_p$ defined in (\ref{era-ch}) and the observations after (\ref{era-ch}). $\Box$\smallskip

To obtain an infinite-dimensional version of Proposition \ref{CHI-PHI}  assume that $H$ is a positive operator on the space $\H_A$ satisfying the condition
(\ref{H-cond+}). We will treat $H$ as the Hamiltonian (the energy observable) of an input system $A$ of a quantum channel.

A physically relevant measure of divergence between infinite-dimensional
quantum channels from $A$ to $B$ is induced by the energy-constrained diamond norm on the space of all Hermitian preserving linear maps from $\T(\H_A)$  to $\T(\H_B)$
defined as
\begin{equation}\label{EC-DN-def}
\|\Theta\|^H_{\diamond,E}\doteq \sup_{\rho\in\S(\H_{AR}):\Tr H\rho_A\leq E}\|\Theta\otimes \id_R(\rho)\|_1,
\end{equation}
where $R$ is an infinite-dimensional quantum system \cite{Sh-ECN,W-ECN} (this norm differs from the eponymous norm used in \cite{Lupo,Pir}).

A topologically equivalent measure of divergence between infinite-dimensional
quantum channels is induced by the energy-constrained Bures distance
\begin{equation}\label{EC-BD-def}
\beta^H_{E}(\Phi,\Psi)\doteq \sup_{\rho\in\S(\H_{AR}):\Tr H\rho_A\leq E}\beta(\Phi\otimes\id_R(\rho),\Psi\otimes\id_R(\rho))
\end{equation}
between quantum channels $\Phi$ and $\Psi$ from $A$  to $B$, where $\beta$ in the r.h.s. denotes the Bures distance
between states in  $\S(\H_{BR})$ \cite{CID}.

If $H$ is a densely defined operator with a discrete spectrum of finite multiplicity and  $E$ is greater than the minimal eigenvalue $E_0$ of $H$ then the norm $\|\Theta\|^H_{\diamond,E}$
and the distance $\beta^H_{E}(\Phi,\Psi)$ generate the strong convergence topology on the set of all quantum channels from $A$ to any given system $B$ \cite{Sh-ECN,CID}.

The following proposition gives continuity bounds for the functions $\Phi\mapsto\chi(\Phi(\mu))$ and $\Phi\mapsto\chi(\widehat{\Phi}(\mu))$ under the constraint on the mean energy of $\mu$
expressed in terms of the energy-constrained Bures distance and the energy-constrained diamond norm.\smallskip

\begin{property}\label{CHI-PHI-EC} \emph{Let $H$ be a positive operator on the space $\H_A$ satisfying  condition
(\ref{H-cond+}) and $F_{H}$ be the function defined in (\ref{F-def}). Let $\,E>E_0$ and $\varepsilon\in (0,1]$ be arbitrary. Let $\mu$ be a generalized ensemble of states in $\S(\H_A)$ such that $\Tr H\bar{\rho}(\mu)\leq E$.}\smallskip

A) \emph{For any quantum channels $\Phi$ and $\Psi$ from $A$ to $B$ such that either $\beta^H_E(\Phi,\Psi)\leq\varepsilon$ or $\,\sqrt{\|\Phi-\Psi\|^H_{\diamond, E}}\leq\varepsilon\,$ the following inequalities hold
\begin{equation}\label{CHI-PHI-EC+}
    |\chi(\Phi(\mu))-\chi(\Psi(\mu))|\leq \varepsilon F_{H}\!\!\left[\frac{2E}{\varepsilon^2}\right]+g(\varepsilon)
\end{equation}
and}
\begin{equation}\label{CHI-CPHI-EC+}
    |\chi(\widehat{\Phi}(\mu))-\chi(\widehat{\Psi}(\mu))|\leq \varepsilon F_{H}\!\!\left[\frac{2E}{\varepsilon^2}\right]+g(\varepsilon).
\end{equation}

B) \emph{If, in addition, the operator $H$ satisfies  condition (\ref{H-0}) and
$G$ is any continuous function on $\mathbb{R}_+$ satisfying conditions (\ref{G-c1}) and (\ref{G-c2}) then
\begin{equation}\label{CHI-PHI-EC++}
 |\chi(\Phi(\mu))-\chi(\Psi(\mu))|\leq \min_{\shs t\in(0,T]} \mathbb{CB}_t(E,\varepsilon\,|\,1,2)
\end{equation}
and
\begin{equation}\label{CHI-CPHI-EC++}
 |\chi(\widehat{\Phi}(\mu))-\chi(\widehat{\Psi}(\mu))|\leq \min_{\shs t\in(0,T]} \mathbb{CB}_t(E,\varepsilon\,|\,1,1)
\end{equation}
for any quantum channels $\Phi$ and $\Psi$ from $A$ to $B$ such that $\frac{1}{2}\|\Phi-\Psi\|^H_{\diamond, E}\leq\varepsilon$, where $\mathbb{CB}_{\shs t}(E,\varepsilon\,|\,C,D)$ is defined in (\ref{CB-exp}) and $\,T$ is defined by the formula after (\ref{CB-exp}).}\smallskip

C) \emph{If $A$ is an $\ell$-mode quantum oscillator,  $H$ is its grounded Hamiltonian defined in (\ref{H-osc})
and the function $G_{\ell,\omega}$ defined in (\ref{F-osc}) is used in the role of $G$ then
the continuity bounds (\ref{CHI-PHI-EC+})-(\ref{CHI-CPHI-EC++}) are asymptotically tight for large $E$ (Def.1 in \cite[Section 3.2.1]{QC}).}
\end{property}
\smallskip

The right hand sides of (\ref{CHI-PHI-EC+})-(\ref{CHI-CPHI-EC++}) tend to zero as $\varepsilon\to0^+$. This follows from the equivalence of (\ref{H-cond+})
and (\ref{H-cond+a}) and the second property of the function $G$ in (\ref{G-c1}).\smallskip

\begin{remark}\label{xyz} The continuity bounds (\ref{CHI-PHI-EC+}) and (\ref{CHI-PHI-EC++}) \emph{essentially refine}
the continuity bound for the function $\Phi\mapsto\chi(\Phi(\mu))$ given by Proposition 8 in \cite{CID}. The continuity bounds  (\ref{CHI-CPHI-EC+}) and (\ref{CHI-CPHI-EC++}) have no analogues constructed earlier (as far as I know).
\end{remark}\smallskip

\emph{Proof.} Assume first that $\mu$ is a discrete ensemble of states in $\S(\H_A)$ and take a pure state  $\hat{\rho}\in\S(\H_{AR})$ such that $\hat{\rho}_A=\bar{\rho}(\mu)$ and $\Tr H_R\hat{\rho}_R\leq E$, where $R$ is an infinite-dimensional system and $H_R$ is an operator on $\H_R$ unitary equivalent to the operator $H$ (this can be done due to the condition $\Tr H\bar{\rho}(\mu)\leq E$).
Let $\M_{\mu}$ be the  POVM on $\H_R$ corresponding to the ensemble $\mu$ (it is defined at the end of Section 6.1).

Inequalities (\ref{CHI-CPHI-EC+}) and (\ref{CHI-CPHI-EC++}) are proved by using relation (\ref{ER-rep+}) and the continuity bounds for the
function  $\omega\mapsto ER(\omega, I_B\otimes\M)$ on $\S(\H_{BR})$ given by Corollaries \ref{ER-CB+} and \ref{ER-CB++} in Section 3 (with $X=R$) along with the definitions (\ref{EC-DN-def}) and (\ref{EC-BD-def}) and the simple relations
$$
\sqrt{1-F(\varrho,\varsigma)}\leq \sqrt{2-2\sqrt{F(\varrho,\varsigma)}}=\beta(\varrho,\varsigma)\leq\sqrt{\|\varrho-\varsigma\|_1}
$$
between the fidelity $F(\varrho,\varsigma)\doteq\|\sqrt{\varrho}\sqrt{\varsigma}\|^2_1$, the Bures distance and the trace norm distance between quantum states $\varrho$ and $\varsigma$ \cite{H-SCI,Wilde}.

Inequality (\ref{CHI-PHI-EC+}) is derived from (\ref{CHI-CPHI-EC+})  by  using the following observations:
\begin{enumerate}[I)]
  \item $\Phi=\widehat{\widehat{\Phi}}$ and $\Psi=\widehat{\widehat{\Psi}}$;
  \item for any quantum channels $\Phi$ and $\Psi$ from $A$ to $B$ there exist complementary channels $\widehat{\Phi}$ and $\widehat{\Psi}$ from $A$ to some system $E$
  such that $\beta^H_E(\widehat{\Phi},\widehat{\Psi})\leq\beta^H_E(\Phi,\Psi)$ \cite[Corollary 2]{CID}.
\end{enumerate}

The definition (\ref{UCBDB}) of the unoptimized one-way classical correlation and the properties of the quantum mutual
information imply that for any POVM $\M$ on $\H_R$ the function $\omega\mapsto C^{\M}_R(\omega)$ on $\S(\H_{BR})$ belongs to the class $L_2^1(1,2)$ in the settings $A_1=R$, $A_2=B$ within the notation introduced in \cite[Section 3.1.2]{QC} (see the proof of Lemma 1 in \cite[Seciton 4.3.1]{QC}).

Thus, inequality (\ref{CHI-PHI-EC++}) is proved by applying Theorem 7 in \cite[Section 3.2.4]{QC} to the function $\omega\mapsto C^{\M}_R(\omega)$  on $\S(\H_{BR})$ (with $A_1=R,A_2=B$ and $H=H_R$) and by using relation (\ref{MI-CHI-rep+}) along with definition (\ref{EC-DN-def}).

If $\mu$ is an arbitrary generalized ensemble then the validity of (\ref{CHI-PHI-EC+})-(\ref{CHI-CPHI-EC++})
is proved by the same approximation arguments (based on Lemma \ref{sl}) as in the proof of Proposition \ref{CHI-PHI}.

To prove claim C we use the family of erasure channels $\Omega_p$ defined by (\ref{era-ch})  assuming that $A$ is an $\ell$-mode quantum oscillator. By the observation after (\ref{era-ch}) we have
\begin{equation}\label{double-eq+}
\begin{array}{rl}
 |\chi(\Omega_p(\mu))-\chi(\Omega_q(\mu))|\!\!&=\;|\chi(\widehat{\Omega}_p(\mu))-\chi(\widehat{\Omega}_q(\mu))|\\\\&=\;|p-q|\shs \chi(\mu)=|p-q|\shs S(\gamma_H(E))=|p-q|\shs F_H(E)
\end{array}
\end{equation}
for any ensemble $\mu$
of pure states in $\S(\H_A)$ such that $\bar{\rho}(\mu)=\gamma_H(E)$ -- the Gibbs state (\ref{Gibbs}) corresponding to the "energy" $E$.

The asymptotical tightness of
continuity bounds (\ref{CHI-PHI-EC+}) and (\ref{CHI-CPHI-EC+}) follows from (\ref{double-eq+}) with $\,p=1/2\,$ and $\,q=1/2\pm x$, since $\,F_H(E)=O(\ln E)\,$ as $\,E\to+\infty\,$ and
$$
\beta^H_E(\Omega_{1/2-x},\Omega_{1/2})\leq\beta(\Omega_{1/2-x},\Omega_{1/2})\leq x+o(x)
$$
for small $x$ (this is shown in the proof of Theorem 1 in  \cite{CID}).\smallskip

If $G=G_{\ell,\omega}$ then it is easy to see that
$$\liminf_{E\to+\infty}\inf_{\varepsilon\in(0,1]} \min_{t\in[0,T]}\frac{\mathbb{CB}_t(E,\varepsilon\,|\,C,D)}{G(E)\varepsilon}=C.$$

Thus, the asymptotical tightness of
continuity bounds (\ref{CHI-PHI-EC++}) and (\ref{CHI-CPHI-EC++}) follows from (\ref{double-eq+}), since  $\,\frac{1}{2}\|\Omega_p-\Omega_q\|^H_{\diamond, E}\leq\frac{1}{2}\|\Omega_p-\Omega_q\|_{\diamond}=|p-q|\,$ and $\,G_{\ell,\omega}(E)=F_H(E)+o(1)\,$ as $\,E\to+\infty$.
$\Box$\smallskip

Proposition \ref{CHI-PHI-EC} gives continuity bounds for the function $\Phi\mapsto\pi(\Phi,\mu)$ (defined in (\ref{pi-def})) under the constraint on the mean energy of $\mu$
expressed in terms of the energy-constrained Bures distance and the energy-constrained diamond norm.\smallskip

\begin{corollary}\label{PI-CPHI-EC} \emph{Let $H$ be a positive operator on the space $\H_A$ satisfying  the condition
(\ref{H-cond+}) and $F_{H}$ be the function defined in (\ref{F-def}). Let $\,E>E_0$ and $\varepsilon\in (0,1]$ be arbitrary. Let $\mu$ be a generalized ensemble of  states in $\S(\H_A)$ such that $\Tr H\bar{\rho}(\mu)\leq E$.}\smallskip

A) \emph{For any quantum channels $\Phi$ and $\Psi$ from $A$ to $B$ such that either $\beta^H_E(\Phi,\Psi)\leq\varepsilon$ or $\sqrt{\|\Phi-\Psi\|^H_{\diamond, E}}\leq\varepsilon$ the following inequality holds}
\begin{equation}\label{PI-CPHI-EC+}
    |\pi(\Phi,\mu)-\pi(\Psi,\mu)|\leq 2\varepsilon F_{H}\!\!\left[\frac{2E}{\varepsilon^2}\right]+2g(\varepsilon).
\end{equation}

B) \emph{If, in addition, the operator $H$ satisfies  condition (\ref{H-0}) and
$G$ is any continuous function on $\mathbb{R}_+$ satisfying conditions (\ref{G-c1}) and (\ref{G-c2}) then
\begin{equation}\label{PI-CPHI-EC++}
 |\pi(\Phi,\mu)-\pi(\Psi,\mu)|\leq \min_{t\in[0,T]}\mathbb{CB}_t(E,\varepsilon\,|\,2,3)
\end{equation}
for any quantum channels $\Phi$ and $\Psi$ from $A$ to $B$ such that $\frac{1}{2}\|\Phi-\Psi\|^H_{\diamond, E}\leq\varepsilon$, where $\mathbb{CB}_{\shs t}(E,\varepsilon\,|\,2,3)$ is defined in (\ref{CB-exp}) and $\,T$ is defined by the formula after (\ref{CB-exp}).}
\smallskip

C) \emph{If $A$ is an $\ell$-mode quantum oscillator,  $H$ is its grounded Hamiltonian defined in (\ref{H-osc})
and the function $G_{\ell,\omega}$ defined in (\ref{F-osc}) is used in the role of $G$ then
the continuity bounds (\ref{PI-CPHI-EC+})
and (\ref{PI-CPHI-EC++})  are asymptotically tight for large $E$ (Def.1 in \cite[Sec.3.2.1]{QC}).}
\end{corollary}\smallskip

The right hand sides of (\ref{PI-CPHI-EC+}) and (\ref{PI-CPHI-EC++}) tend to zero as $\varepsilon\to0^+$. This follows from the equivalence of (\ref{H-cond+})
and (\ref{H-cond+a}) and the second property of the function $G$ in (\ref{G-c1}).
\smallskip

\emph{Proof.} Claims A and B follow from Proposition \ref{CHI-PHI-EC}. It suffices to note that
$$
\mathbb{CB}_{\shs t}(E,\varepsilon\,|\,1,2)+\mathbb{CB}_{\shs t}(E,\varepsilon\,|\,1,1)=\mathbb{CB}_{\shs t}(E,\varepsilon\,|\,2,3).
$$

Claim C is proved by using the family of erasure channels $\Omega_p$ defined in (\ref{era-ch}), where $A$ is an $\ell$-mode quantum oscillator. By the observation after (\ref{era-ch}) we have
\begin{equation}\label{double-eq++}
|\pi(\Omega_p,\mu)-\pi(\Omega_q,\mu)|=2|p-q|\shs\chi(\mu)=2|p-q|\shs S(\gamma_H(E))=2|p-q|\shs F_H(E)
\end{equation}
for any ensemble $\mu$
of pure states in $\S(\H_A)$ such that $\bar{\rho}(\mu)=\gamma_H(E)$ -- the Gibbs state (\ref{Gibbs})  corresponding to the "energy" $E$.

The asymptotical  tightness of continuity bounds (\ref{PI-CPHI-EC+})
and (\ref{PI-CPHI-EC++}) follows from (\ref{double-eq++}) and the arguments used in the proof of claim C of Proposition \ref{CHI-PHI-EC}. $\Box$

\subsubsection{Continuity bounds for the Holevo capacity of a quantum channel}

Propositions \ref{CHI-PHI} and \ref{CHI-PHI-EC} allow us to obtain  continuity bounds for the unconstrained and energy-constrained Holevo capacity of a
quantum channel (cf.\cite{H-SCI,Wilde,H-c-w-c}) defined, respectively, as
$$
\bar{C}(\Phi)=\sup_{\mu\in\P(\H_A)}\chi(\Phi(\mu))\quad \textrm{and} \quad   \bar{C}(\Phi,H,E)=\sup_{\mu\in\P(\H_A):\Tr H\bar{\rho}(\mu)\leq E}\chi(\Phi(\mu)).
$$
These continuity bounds\emph{ depend either on the input dimension $\dim\H_A$ or on the input energy bound $E$} (in contrast to the continuity bounds obtained in \cite{L&S,Sh-ECN,W-ECN}). They essentially refine the continuity bounds of this type obtained previously \cite{CID}.

The following proposition gives continuity bounds for the function $\Phi\mapsto\bar{C}(\Phi)$ in terms of the diamond norm distance either between the channels themselves or between complementary channels.\smallskip

\begin{property}\label{CHI-CAP}  \emph{Let $\,d=\dim\H_A<+\infty\,$ and $g(x)$ be the function defined in (\ref{g-def}). Then
\begin{equation}\label{CHI-CAP+}
 |\bar{C}(\Phi)-\bar{C}(\Psi)|\leq \varepsilon \ln d+g(\varepsilon)
\end{equation}
for any channels $\,\Phi$ and $\,\Psi$ from $A$ to $B$ such that $\frac{1}{2}\|\Phi-\Psi\|_{\diamond}\leq\varepsilon$.} \smallskip

\emph{Inequality (\ref{CHI-CAP+}) holds for  channels $\Phi$ and $\Psi$ from $A$ to $B$ if there exist complementary
channels $\widehat{\Phi}$ and $\widehat{\Psi}$ from $A$ to some system $E$ such that $\frac{1}{2}\|\widehat{\Phi}-\widehat{\Psi}\|_{\diamond}\leq\varepsilon$ (the notion of a channel $\widehat{\Phi}$ complementary to a channel $\Phi$ is described in Section 2).}

\emph{The continuity bound (\ref{CHI-CAP+}) and the continuity bound  obtained from (\ref{CHI-CAP+}) by replacing the condition   $\frac{1}{2}\|\Phi-\Psi\|_{\diamond}\leq\varepsilon\,$ with  $\frac{1}{2}\|\widehat{\Phi}-\widehat{\Psi}\|_{\diamond}\leq\varepsilon\,$
are asymptotically tight for large $d$ (Def.1 in \cite[Section 3.2.1]{QC}). Moreover, for any $d\geq2$ and $\varepsilon\in[0,1]$ there exist channels $\Phi$ and $\,\Psi$ from a $d$-dimensional system $A$ to some system $B$  such that}
\begin{equation*}
\textstyle\frac{1}{2}\|\Phi-\Psi\|_{\diamond}=\frac{1}{2}\|\widehat{\Phi}-\widehat{\Psi}\|_{\diamond}=\varepsilon\quad and\quad |\bar{C}(\Phi)-\bar{C}(\Psi)|= \varepsilon \ln d.
\end{equation*}
\end{property}

\emph{Proof.} The main  claim follows directly from the definition of $\bar{C}(\Phi)$ and the continuity bound (\ref{CHI-PHI+}) in Proposition \ref{CHI-PHI}.

Since $\Phi=\widehat{\widehat{\Phi}}$ and $\Psi=\widehat{\widehat{\Psi}}$, the second claim  follows from the continuity bound (\ref{CHI-PHI++}) in Proposition \ref{CHI-PHI}.

The last claim can be proved by using the family of erasure channels $\Omega_p$ defined in (\ref{era-ch}) and the observations after (\ref{era-ch}), which show, in particular, that $\,\bar{C}(\Omega_p)=(1-p)\ln d$.  $\Box$\smallskip

\begin{remark}\label{CHI-CAP-r+} Theorem 1 in \cite{CID} states that
\begin{equation}\label{OLD-CHI-CAP+}
 |\bar{C}(\Phi)-\bar{C}(\Psi)|\leq \varepsilon \ln d+\varepsilon \ln 2+g(\varepsilon)
\end{equation}
for any channels $\Phi$ and $\Psi$ from $A$ to $B$ such that $\beta(\Phi,\Psi)\leq\varepsilon$, where
$d=\dim\H_A$ and $\beta(\Phi,\Psi)$ is the Bures distance between $\Phi$ and $\Psi$
defined in (\ref{BD-def}). Since $\frac{1}{2}\|\Phi-\Psi\|_{\diamond}\leq \beta(\Phi,\Psi)$,
the continuity bound (\ref{CHI-CAP+}) is  strictly sharper than (\ref{OLD-CHI-CAP+}).\smallskip

The main advantage of continuity bound (\ref{CHI-CAP+}) (compared to
(\ref{OLD-CHI-CAP+})) is the linear dependence on $\,\varepsilon=\frac{1}{2}\|\Phi-\Psi\|_{\diamond}$
of the first term in the r.h.s. of (\ref{CHI-CAP+}).
\end{remark}\smallskip

\begin{remark}\label{CHI-CAP-r} The second claim of Proposition \ref{CHI-CAP} allows us to essentially refine the estimates
for $|\bar{C}(\Phi)-\bar{C}(\Psi)|$ given by (\ref{CHI-CAP+}) in some cases. To show this assume that $\,\H_A=\H_B=\H_2\,$ is a two-dimensional space and
$$
\Phi(\rho)=[\Tr\rho]|0\rangle\langle0|,\quad \Psi(\rho)=\sum_{i=0}^1\langle i|\rho|i\rangle|i\rangle\langle i|,\quad \rho\in\S(\H_2),
$$
where $\{|0\rangle,|1\rangle\}$ is an orthonormal basis in $\H_2$. Then $V_{\Phi}: |\varphi\rangle\mapsto |0\otimes\varphi\rangle$
and $V_{\Psi}: |\varphi\rangle\mapsto \langle 0|\varphi\rangle|0\otimes 0\rangle+\langle 1|\varphi\rangle|1\otimes 1\rangle$
are the Stinespring isometries for the channels $\Phi$ and $\Psi$ acting from $\H_2$ into $\H_2\otimes\H_2$. Thus, the channels
$\id_{\H_2}$ and $\Psi$ are complementary to the channels $\Phi$ and $\Psi$, i.e. $\widehat{\Phi}=\id_{\H_2}$ and $\widehat{\Psi}=\Psi$  (in this case $\H_E=\H_2$). It is easy to see that
$$
\|\Phi-\Psi\|_{\diamond}=2\quad \textrm{and} \quad\|\widehat{\Phi}-\widehat{\Psi}\|_{\diamond}=1.
$$
Thus, the first and second claims of Proposition \ref{CHI-CAP} give, respectively, the estimates
$$
\ln2=|\bar{C}(\Phi)-\bar{C}(\Psi)|\leq\ln2+g(1)=3\ln2
$$
and
$$
\ln2=|\bar{C}(\Phi)-\bar{C}(\Psi)|\leq(1/2)\ln2+g(1/2)=\ln2+(3/2)\ln(3/2).
$$
We see that the second bound is essentially sharper than the first one. This example also
shows that the term $g(\varepsilon)$ in (\ref{CHI-CAP+}) with  $\varepsilon=\frac{1}{2}\|\widehat{\Phi}-\widehat{\Psi}\|_{\diamond}$  cannot be removed.  $\Box$
\end{remark}

The following proposition gives continuity bounds for the function $\Phi\mapsto\bar{C}(\Phi,H,E)$ in terms of the energy-constrained Bures distance and the energy-constrained diamond norm distance (either between  the channels themselves or between complementary channels).\smallskip

\begin{property}\label{CHI-CAP-EC} \emph{Let $H$ be a positive operator on the space $\H_A$ satisfying condition
(\ref{H-cond+}) and $F_{H}$ be the function defined in (\ref{F-def}). Let $\,E>E_0$ (the minimal eigenvalue of $H$) and $\varepsilon\in (0,1]$ be arbitrary.} \smallskip

A) \emph{For any quantum channels $\Phi$ and $\Psi$ from $A$ to $B$ such that  either $\beta^H_E(\Phi,\Psi)\leq\varepsilon$ or $\sqrt{\|\Phi-\Psi\|^H_{\diamond, E}}\leq\varepsilon$ the following inequality holds
\begin{equation}\label{CHI-CAP++}
    |\bar{C}(\Phi,H,E)-\bar{C}(\Psi,H,E)|\leq \varepsilon F_{H}\!\!\left[\frac{2E}{\varepsilon^2}\right]+g(\varepsilon).
\end{equation}
Inequality (\ref{CHI-CAP++}) holds for  quantum channels $\Phi$ and $\Psi$ from $A$ to $B$
if there exist complementary channels $\widehat{\Phi}$ and $\widehat{\Psi}$ from $A$ to some system $E$ such that  either $\,\beta^H_E(\widehat{\Phi},\widehat{\Psi})\leq\varepsilon\,$ or $\sqrt{\|\widehat{\Phi}-\widehat{\Psi}\|^H_{\diamond, E}}\leq\varepsilon$.}\smallskip

B) \emph{If, in addition, the operator $H$ satisfies  condition (\ref{H-0}) and
$G$ is any continuous function on $\mathbb{R}_+$ satisfying conditions (\ref{G-c1}) and (\ref{G-c2}) then
\begin{equation}\label{CHI-CAP+++}
 |\bar{C}(\Phi,H,E)-\bar{C}(\Psi,H,E)|\leq \min_{\shs t\in(0,T]} \mathbb{CB}_t(E,\varepsilon\,|\,1,2)
\end{equation}
for any quantum channels $\Phi$ and $\Psi$ from $A$ to $B$ such that $\frac{1}{2}\|\Phi-\Psi\|^H_{\diamond, E}\leq\varepsilon$, where $\mathbb{CB}_{\shs t}(E,\varepsilon\,|\,1,2)$ is defined in (\ref{CB-exp}) and $\,T$ is defined by the formula after (\ref{CB-exp}).}
\smallskip

\emph{Inequality (\ref{CHI-CAP+++}) holds for  quantum channels $\Phi$ and $\Psi$ from $A$ to $B$
if there exist complementary channels $\widehat{\Phi}$ and $\widehat{\Psi}$ from $A$ to some system $E$ such that $\frac{1}{2}\|\widehat{\Phi}-\widehat{\Psi}\|^H_{\diamond, E}\leq\varepsilon$. Moreover, in this case (\ref{CHI-CAP+++}) holds with $\mathbb{CB}_t(E,\varepsilon\,|\,1,2)$ replaced by $\mathbb{CB}_t(E,\varepsilon\,|\,1,1)$.}

\smallskip

C) \emph{If $A$ is an $\ell$-mode quantum oscillator, $H$ is its grounded Hamiltonian  defined in (\ref{H-osc})
and the function $G_{\ell,\omega}$ defined in (\ref{F-osc}) is used in the role of $G$ then the continuity bounds  (\ref{CHI-CAP++}) and (\ref{CHI-CAP+++}) are
asymptotically tight for large $E$ (Def.1 in \cite[Sec.3.2.1]{QC}).}
\end{property}
\smallskip

The right hand sides of (\ref{CHI-CAP++}) and (\ref{CHI-CAP+++}) tend to zero as $\varepsilon\to0^+$. This follows from the equivalence of (\ref{H-cond+})
and (\ref{H-cond+a}) and the second property of the function $G$ in (\ref{G-c1}).
\smallskip

\emph{Proof.} Claims A and B of the proposition are easily derived from  Proposition \ref{CHI-PHI-EC} in Section 6.2.1 by using the observations $\rm I$ and $\rm II$ from the proof of Proposition \ref{CHI-PHI-EC}.

Claim C is proved by exploiting the family of erasure channels $\Omega_p$ defined in (\ref{era-ch}), where $A$ is an $\ell$-mode quantum oscillator. The observation after (\ref{era-ch}) imply that
$\,|\bar{C}(\Omega_p,H,E)-\bar{C}(\Omega_q,H,E)|=|p-q|\shs F_H(E)$. Using this equality and the arguments from the proof of claim C of Proposition \ref{CHI-PHI-EC} it is easy to show the asymptotical  tightness of continuity bounds (\ref{CHI-CAP++}) and (\ref{CHI-CAP+++}). $\Box$\smallskip

\begin{remark}\label{sup-rem} It is easy to see that continuity bound (\ref{CHI-CAP++}) is substantially simpler and sharper than
the continuity bound for the function $\Phi\mapsto\bar{C}(\Phi,H,E)$ in terms of $\beta^H_E(\Phi,\Psi)$ presented in Theorem 2 in \cite{CID}.\smallskip

An essential advantage of continuity bound (\ref{CHI-CAP+++}) is the linear dependence of its main term on $\,\varepsilon=\frac{1}{2}\|\Phi-\Psi\|^H_{\diamond, E}$
in the case when  $\,G(E)=O(\ln E)\,$ as $\,E\to +\infty\,$ (in particular, if $H$ is the Hamiltonian of a multi-mode quantum oscillator
and $G=G_{\ell,\omega}$).\smallskip

Another advantage of the continuity bounds (\ref{CHI-CAP++}) and  (\ref{CHI-CAP+++}) is the possibility to replace the condition of closeness of $\Phi$
and $\Psi$ in terms of $\beta^H_E(\Phi,\Psi)$ (resp. $\|\Phi-\Psi\|^H_{\diamond, E}$) by the closeness condition in terms of  $\beta^H_E(\widehat{\Phi},\widehat{\Psi})$ (resp.   $\|\widehat{\Phi}-\widehat{\Psi}\|^H_{\diamond, E}$). One can construct
an analog of the example in Remark \ref{CHI-CAP-r} showing the benefit of such  replacements.\smallskip
\end{remark}

\subsubsection{Continuity bounds for the unregularized private capacity of a quantum channel}

The \emph{private capacity} of a quantum channel determines the ultimate rate of transmission of  classical information over this channel  with the
additional requirement that almost no information is sent to the environment (see details in \cite{H-SCI,Wilde}). By the Devetak theorem (cf.\cite{Dev}) the private capacity of a channel $\Phi$ between finite-dimensional systems $A$ and $B$  is given by
the regularized expression
\begin{equation*}
C_\mathrm{p}(\Phi)=\lim_{n\rightarrow +\infty }n^{-1}\bar{C}_\mathrm{p}(\Phi^{\otimes n}),
\end{equation*}
where
\begin{equation*}
\bar{C}_\mathrm{p}(\Phi)=\sup_{\mu\in\P(\H_A)}\pi(\Phi,\mu),\qquad \pi(\Phi,\mu)\doteq\chi(\Phi(\mu)-\chi(\widehat{\Phi}(\mu)),
\end{equation*}
($\P(\H_A)$ is the set of all ensembles of states in $\S(\H_A)$, $\,\widehat{\Phi}\,$ is a complementary channel to the channel $\Phi$ defined in (\ref{comp-ch})).

There are many channels $\Phi$ for which $C_\mathrm{p}(\Phi)=\bar{C}_\mathrm{p}(\Phi)$. This holds, in particular, for any degradable
channel $\Phi$, i.e. such a channel $\Phi$ that $\widehat{\Phi}=\Theta\circ\Phi$ for some channel $\Theta$ \cite{D-ch}. We will call $\bar{C}_\mathrm{p}(\Phi)$ the \emph{unregularized private capacity} of $\Phi$.

Propositions \ref{CHI-PHI} implies a continuity bound for the function $\Phi\mapsto \bar{C}_\mathrm{p}(\Phi)$ expressed
in terms of  the diamond norm defined in (\ref{DN-def}). This continuity bound refines the
continuity bound for this function obtained in \cite[Theorem 1]{CID}.\smallskip

\begin{property}\label{PC-CAP}  \emph{Let $\,d=\dim\H_A<+\infty\,$ and $g(x)$ be the function defined in (\ref{g-def}). Then
\begin{equation}\label{PC-CAP+}
 |\bar{C}_\mathrm{p}(\Phi)-\bar{C}_\mathrm{p}(\Psi)|\leq 2\varepsilon \ln d+2g(\varepsilon)
\end{equation}
for any channels $\Phi$ and $\Psi$ from $A$ to $B$ such that $\,\frac{1}{2}\|\Phi-\Psi\|_{\diamond}\leq\varepsilon$.} \smallskip

\emph{Inequality (\ref{PC-CAP+}) holds for channels $\Phi$ and $\Psi$ from $A$ to $B$ if there exist complementary
channels $\widehat{\Phi}$ and $\widehat{\Psi}$ from $A$ to some system $E$ such that $\,\frac{1}{2}\|\widehat{\Phi}-\widehat{\Psi}\|_{\diamond}\leq\varepsilon$.}\smallskip

\emph{The continuity bound (\ref{PC-CAP+}) and the continuity bound  obtained from (\ref{PC-CAP+}) by replacing the condition   $\frac{1}{2}\|\Phi-\Psi\|_{\diamond}\leq\varepsilon\,$ with  $\frac{1}{2}\|\widehat{\Phi}-\widehat{\Psi}\|_{\diamond}\leq\varepsilon\,$
are asymptotically tight for large $d$ (Def.1 in \cite[Section 3.2.1]{QC}). Moreover, for any $d\geq2$ and $\varepsilon\in[0,1]$ there exist channels $\Phi$ and $\Psi$ from  an $d$-dimensional system $A$ to some system $B$  such that}
\begin{equation*}
\textstyle\frac{1}{2}\|\Phi-\Psi\|_{\diamond}=\frac{1}{2}\|\widehat{\Phi}-\widehat{\Psi}\|_{\diamond}=\varepsilon\quad and\quad |\bar{C}_\mathrm{p}(\Phi)-\bar{C}_\mathrm{p}(\Psi)|= 2\varepsilon \ln d.
\end{equation*}

\end{property}

\emph{Proof.} The main claim of the proposition follows directly from Corollary \ref{PI-PHI} in Section 6.2.1.

Since $\Phi=\widehat{\widehat{\Phi}}$ and $\Psi=\widehat{\widehat{\Psi}}$, the second claim is also derived from  Corollary \ref{PI-PHI} by noting that $\,\pi(\Theta,\mu)=-\pi(\widehat{\Theta},\mu)$, $\Theta=\Phi,\Psi$.

The last claim can be proved by using the family of erasure channels $\Omega_p$ defined in (\ref{era-ch}) and the observations after (\ref{era-ch}), which show, in particular, that $\,\bar{C}_{\rm p}(\Omega_p)=(1-2p)\ln d$. $\Box$

\begin{remark}\label{PC-CAP-r+} \emph{Theorem 1 in \cite{CID} states that
\begin{equation}\label{OLD-PC-CAP+}
 |\bar{C}_\mathrm{p}(\Phi)-\bar{C}_\mathrm{p}(\Psi)|\leq 2\varepsilon \ln d+2\varepsilon \ln 2+2g(\varepsilon)
\end{equation}
for any channels $\Phi$ and $\Psi$ from $A$ to $B$ such that $\beta(\Phi,\Psi)\leq\varepsilon$, where
$d=\dim\H_A$ and $\beta(\Phi,\Psi)$ is the Bures distance between $\Phi$ and $\Psi$
defined in (\ref{BD-def}). Since $\frac{1}{2}\|\Phi-\Psi\|_{\diamond}\leq \beta(\Phi,\Psi)$,
the continuity bound (\ref{PC-CAP+}) is  strictly sharper than (\ref{OLD-PC-CAP+}).}

\emph{The main advantage of continuity bound (\ref{PC-CAP+}) (compared to
(\ref{OLD-PC-CAP+})) is the linear dependence on $\,\varepsilon=\frac{1}{2}\|\Phi-\Psi\|_{\diamond}$
of the first term in the r.h.s. of (\ref{PC-CAP+}).}
\end{remark}

\smallskip

The unregularized private capacity of a channel $\Phi:A\to B$ with the energy type constraint  determined by a positive operator $H$ on $\H_A$
is defined as (cf.\cite{Wilde+++})
$$
\bar{C}_\mathrm{p}(\Phi,H,E)=\sup_{\mu\in\P(\H_A):\Tr H\bar{\rho}(\mu)\leq E}\pi(\Phi,\mu),\qquad \pi(\Phi,\mu)\doteq\chi(\Phi(\mu)-\chi(\widehat{\Phi}(\mu)).
$$

The following proposition gives continuity bounds for the function $\,\Phi\mapsto\bar{C}_\mathrm{p}(\Phi,H,E)\,$ in terms of the energy-constrained Bures distance and the energy-constrained diamond norm  defined, respectively, in (\ref{EC-BD-def}) and (\ref{EC-DN-def}).\smallskip

\begin{property}\label{PC-EC} \emph{Let $H$ be a positive operator on the space $\H_A$ satisfying  condition
(\ref{H-cond+}) and $F_{H}$ be the function defined in (\ref{F-def}). Let $\,E>E_0$ (the minimal eigenvalue of $H$) and $\varepsilon\in (0,1]$ be arbitrary.}\smallskip

A) \emph{For any quantum channels $\Phi$ and $\Psi$ from $A$ to $B$ such that  either $\beta^H_E(\Phi,\Psi)\leq\varepsilon$ or $\sqrt{\|\Phi-\Psi\|^H_{\diamond, E}}\leq\varepsilon$ the following inequality holds
\begin{equation}\label{PC-EC+}
    |\bar{C}_\mathrm{p}(\Phi,H,E)-\bar{C}_\mathrm{p}(\Psi,H,E)|\leq 2\varepsilon F_{H}\!\!\left[\frac{2E}{\varepsilon^2}\right]+2g(\varepsilon).
\end{equation}
Inequality (\ref{PC-EC+}) holds for  quantum channels $\Phi$ and $\Psi$ from $A$ to $B$
if there exist complementary channels $\widehat{\Phi}$ and $\widehat{\Psi}$ from $A$ to some system $E$ such that  either $\,\beta^H_E(\widehat{\Phi},\widehat{\Psi})\leq\varepsilon\,$ or $\sqrt{\|\widehat{\Phi}-\widehat{\Psi}\|^H_{\diamond, E}}\leq\varepsilon$.}
\smallskip

B) \emph{If, in addition, the operator $H$ satisfies  condition (\ref{H-0}) and
$G$ is any continuous function on $\mathbb{R}_+$ satisfying conditions (\ref{G-c1}) and (\ref{G-c2}) then
\begin{equation}\label{PC-EC++}
 |\bar{C}_\mathrm{p}(\Phi,H,E)-\bar{C}_\mathrm{p}(\Psi,H,E)|\leq \min_{\shs t\in(0,T]} \mathbb{CB}_t(E,\varepsilon\,|\,2,3)
\end{equation}
for any channels $\Phi$ and $\Psi$ from $A$ to $B$ such that   $\,\frac{1}{2}\|\Phi-\Psi\|^H_{\diamond, E}\leq\varepsilon$, where
$\mathbb{CB}_{\shs t}(E,\varepsilon\,|\,2,3)$ is defined in (\ref{CB-exp}),  $\,T$ is defined by the formula after (\ref{CB-exp}).
Inequality (\ref{PC-EC++}) holds for  quantum channels $\Phi$ and $\Psi$ from $A$ to $B$
if there exist complementary channels $\widehat{\Phi}$ and $\widehat{\Psi}$ from $A$ to some system $E$ such that $\frac{1}{2}\|\widehat{\Phi}-\widehat{\Psi}\|^H_{\diamond, E}\leq\varepsilon$.}
\smallskip

C) \emph{If $A$ is an $\ell$-mode quantum oscillator, $H$ is its grounded Hamiltonian  defined in (\ref{H-osc})
and the function $G_{\ell,\omega}$ defined in (\ref{F-osc}) is used in the role of $G$ then the continuity bounds (\ref{PC-EC+}) and (\ref{PC-EC++}) are asymptotically tight for large $E$ (Def.1 in \cite[Sec.3.2.1]{QC}).}
\end{property}
\smallskip

The right hand sides of (\ref{PC-EC+}) and (\ref{PC-EC++}) tend to zero as $\varepsilon\to0^+$. This follows from the equivalence of (\ref{H-cond+})
and (\ref{H-cond+a}) and the second property of the function $G$ in (\ref{G-c1}).
\smallskip

\emph{Proof.} Claims A and B of the proposition follow  from  Corollary \ref{PI-CPHI-EC} in Section 6.2.1.

Claim C is proved by exploiting the family of erasure channels $\Omega_p$ defined in (\ref{era-ch}), where $A$ is an $\ell$-mode quantum oscillator. The observations after (\ref{era-ch}) imply that
$\,|\bar{C}_{\rm p}(\Omega_p,H,E)-\bar{C}_{\rm p}(\Omega_q,H,E)|=2|p-q|\shs F_H(E)$. Using this equality and the arguments from the proof of claim C of Proposition \ref{CHI-PHI-EC} it is easy to show the asymptotical  tightness of continuity bounds (\ref{PC-EC+})  and (\ref{PC-EC++}). $\Box$\smallskip

\begin{remark}\label{sup-rem++}  It is easy to see that continuity bound (\ref{PC-EC+}) is substantially simpler and sharper than
the continuity bound for the function $\Phi\mapsto\bar{C}_\mathrm{p}(\Phi,H,E)$ in terms of $\beta^H_E(\Phi,\Psi)$ presented in Theorem 2 in \cite{CID}.\smallskip

An essential advantage of continuity bound (\ref{PC-EC++}) is the linear dependence of its main term on $\,\varepsilon=\frac{1}{2}\|\Phi-\Psi\|^H_{\diamond, E}$
in the case when    $\,G(E)=O(\ln E)\,$ as $\,E\to +\infty\,$ (in particular, if $H$ is the Hamiltonian of a multi-mode quantum oscillator
and $G=G_{\ell,\omega}$).
\end{remark}

\subsection{Monotonicity of the quantum discord and the entropy reduction of a local measurement in terms of a channel}

By relation (\ref{MI-rep}) the monotonicity of the QMI of a bipartite state under local channels is equivalent to the monotonicity of the mutual information of a channel under a concatenation, which means that $\,I(\Psi\circ\Phi,\rho)\leq I(\Phi, \rho)\,$  for any state  $\rho$ in $\S(\H_A)$ and any channels $\Phi:A\to B$, $\Psi:B\to C$.

By relation (\ref{CHI-rep}) (resp. (\ref{MI-CHI-rep+})) the monotonicity of the optimized (resp. unoptimazed) one-way classical correlation of a bipartite state under local channels acting on an unmeasured subsystem is equivalent to the monotonicity of the constrained Holevo capacity (resp. the output Holevo information)  under a concatenation, which means that
$\,\bar{C}(\Psi\circ\Phi,\rho)\leq \bar{C}(\Phi, \rho)\,$  for any state  $\rho$ in $\S(\H_A)$
(resp. $\,\chi(\Psi\circ\Phi(\mu))\leq\chi(\Phi(\mu))\,$ for any ensemble $\mu$ of  states in $\S(\H_A)$) and any channels $\Phi:A\to B$, $\Psi:B\to C$.

In this subsection we  discuss the properties of quantum channels which are "doppelgangers" of the monotonicity of the (optimized and  unoptimazed) quantum
discord and the entropy reduction of a local measurement under quantum channels acting on an unmeasured subsystem.

By  relation (\ref{D-rep+}) the monotonicity property (\ref{un-DB-m}) of unoptimazed quantum discord is equivalent to the
monotonicity of the quantity  $\,\Upsilon(\Phi,\mu)\doteq I(\Phi,\bar{\rho}(\mu))-\chi(\Phi(\mu))\,$ under a concatenation, i.e. the validity of the inequality
\begin{equation}\label{D-m}
  \Upsilon(\Psi\circ\Phi,\mu)\leq\Upsilon(\Phi,\mu)
\end{equation}
for any channels $\Phi:A\to B$, $\Psi:B\to C$ and any discrete ensemble $\mu$ of states in $\S(\H_A)$  such that $\chi(\Phi(\mu))<+\infty\,$ (this condition implies that  $\chi(\Psi\circ\Phi(\mu))<+\infty$, so, both quantities in (\ref{D-m}) are well defined).\smallskip

Note first that (\ref{D-m}) holds for generalized (continuous) ensembles as well.\smallskip

\begin{property}\label{D-m+} \emph{Inequality (\ref{D-m}) holds for any channels $\Phi:A\to B$, $\Psi:B\to C$ and any generalized ensemble $\mu\in\P(\H_A)$ such that $\chi(\Phi(\mu))<+\infty$.}
\end{property}

\smallskip
\emph{Proof.}
If $\mu$ is a generalized ensemble in $\P(\H_A)$ then Lemma \ref{sl} in Section 6.2.1 implies the existence of a sequence $\{\mu_n\}$ of finite ensembles weakly converging to
$\mu$ such that
$\,\lim_{n\rightarrow+\infty}\chi(\Theta(\mu_n))=\chi(\Theta(\mu))$,$\; \Theta=\Phi,\Psi\circ\Phi\,$
and $\bar{\rho}(\mu_n)=\bar{\rho}(\mu)$ for all $n$. As noted before,  inequality (\ref{D-m}) holds with $\mu=\mu_n$ for all $n$. Thus, the above limit relations  imply the validity of (\ref{D-m}) for the ensemble $\mu$. $\square$

By Proposition \ref{D-m+} we have
\begin{equation}\label{D-m++}
0\leq\chi(\Phi(\mu))-\chi(\Psi\circ\Phi(\mu))\leq I(\Phi,\bar{\rho}(\mu))-I(\Psi\circ\Phi,\bar{\rho}(\mu))\leq+\infty
\end{equation}
for any channels $\Phi:A\to B$, $\Psi:B\to C$ and any  ensemble $\mu\in\P(\H_A)$ such that
$I(\Psi\circ\Phi,\bar{\rho}(\mu))<+\infty$ (this condition implies $\chi(\Psi\circ\Phi(\mu))<+\infty$, so, both quantities in (\ref{D-m++}) are well defined).\smallskip

\begin{example}\label{SG-e} Let $\{\Phi_t\}_{t\in\mathbb{R}_+}$ be a semigroup of quantum channels from a system $A$ to itself
and $\mu$ be an ensemble in $\P(\H_A)$ such that $S(\bar{\rho}(\mu))$ is finite. By monotonicity of the relative entropy both functions $F(t)\doteq I(\Phi_t,\bar{\rho}(\mu))$ and $G(t)\doteq\chi(\Phi_t(\mu))$ are non-increasing on $\mathbb{R}_+$. It is clear that $\,0\leq G(t)\leq F(t)<+\infty\,$ for all $t\geq 0$. It follows from (\ref{D-m++}) that
$$
0\leq G(t)-G(t+s)\leq F(t)-F(t+s)\quad \forall t,s\geq 0,
$$
i.e. \emph{the rate of decreasing of $G(t)$ does no exceed the rate of decreasing of $F(t)$}. This can be used to estimate
the rate of decreasing of the hardly computable function $G(t)$, since the function $F(t)$ can be explicitly calculated
for many quantum dynamical semigroups, f.i., for semigroups of one-mode Gaussian channels \cite[Ch.12]{H-SCI}.
\end{example}\smallskip

The monotonicity of $\Upsilon(\Phi,\rho)$  will be used in Section 6.5 for local continuity analysis of characteristics of quantum channels.\smallskip

By relation (\ref{D-rep}) the monotonicity property (\ref{DB-m}) of (optimazed) quantum discord is equivalent to the
monotonicity of the quantity $\,\Delta(\Phi,\rho)\doteq I(\Phi,\rho)-\bar{C}(\Phi,\rho)\,$ under a concatenation, i.e.  the validity of the inequality
\begin{equation}\label{U-m}
  \Delta(\Psi\circ\Phi,\rho)\leq\Delta(\Phi,\rho)
\end{equation}
for any channels $\Phi:A\to B$, $\Psi:B\to C$ and any input state $\rho\in\S(\H_A)$ such that $\bar{C}(\Phi,\rho)<+\infty$ (and hence $\bar{C}(\Psi\circ\Phi,\rho)<+\infty$). It is clear that (\ref{U-m}) can be obtained
from  (\ref{D-m}) by taking the infimum over all discrete ensembles $\mu$ with the average state $\rho$ (this corresponds to the infimum over all POVM in the definition of quantum discord).\smallskip

By  relation (\ref{ER-rep+}) the monotonicity property of the entropy reduction of a local measurement
under a quantum channel acting on an unmeasured subsystem presented in Proposition \ref{ER-prop}B in Section 3 is equivalent to the validity of the inequality
\begin{equation}\label{NMP}
\chi(\widehat{\Psi\circ\Phi}(\mu))\geq\chi(\widehat{\Phi}(\mu))
\end{equation}
for any channels $\Phi:A\to B$, $\Psi:B\to C$ and any discrete ensemble $\mu$ of  states in $\S(\H_A)$. Inequality (\ref{NMP}) can be directly derived from the monotonicity of the relative entropy by using the explicit expressions for the channels $\widehat{\Psi\circ\Phi}$ and $\widehat{\Phi}$ via the Stinespring isometries of $\Phi$ and $\Psi$.
\smallskip

Inequality (\ref{NMP}) implies, in particular, that the entropy exchange (cf.\cite{H-SCI}) $S(\Phi,\rho)$ of a channel $\Phi:A\to B$ at an input state $\rho$ (defined as $S(\widehat{\Phi}(\rho))$) does not decrease under a
concatenation with any channel $\Psi:B\to C$ not decreasing the von Neumann entropy, i.e. $S(\Psi\circ\Phi,\rho)\geq S(\Phi,\rho)$ provided that $S(\Psi(\sigma))\geq S(\sigma)$ for any $\sigma$ in $\S(\H_B)$.

\subsection{Bounds on the Holevo capacity and the entropic disturbance via the quantum discord}

Let $\,\Phi :A\rightarrow B\,$ be a quantum channel and  $\,\widehat{\Phi} :A\rightarrow E\,$ be a
channel complementary to the channel $\Phi$ (see Section 2).

If $\rho$ is a state in $\S(\H_A)$ with finite entropy $S(\rho)$ then it follows from (\ref{D-rep}) that
\begin{equation}\label{chi-d}
  S(\rho)-\bar{C}(\Phi,\rho)=I(\widehat{\Phi},\rho)-\bar{C}(\widehat{\Phi},\rho)=D_R(\widehat{\Phi}\otimes\id_R(\hat{\rho})),
\end{equation}
where $\hat{\rho}$ is a pure state in $\S(\H_{AR})$ such that $\hat{\rho}_A=\rho$ and $D_R(\cdot)$ denotes the quantum discord of
a state in $\S(\H_{ER})$ (with a measured system $R$). The first equality in (\ref{chi-d}) follows from the fact that the states
$\Phi(\rho)$ and $\widehat{\Phi}(\rho)$ have the same positive parts of the spectrum for any pure state $\rho$  in $\S(\H_A)$ \cite{TIN}.

If $\mu$ is a discrete ensemble in $\P(\H_A)$ with finite Holevo information $\chi(\mu)$ then representation (\ref{D-rep+}) implies the following  "unoptimized" version of relation (\ref{chi-d}):
\begin{equation}\label{ED-UB++}
\chi(\mu)-\chi(\Phi(\mu))\leq I(\widehat{\Phi},\bar{\rho}(\mu))-\chi(\widehat{\Phi}(\mu))= D^{\M_{\mu}}_R(\widehat{\Phi}\otimes\id_R(\hat{\rho}(\mu))),
\end{equation}
where $\hat{\rho}(\mu)$ is a pure state in $\S(\H_{AR})$ such that $[\hat{\rho}(\mu)]_A=\bar{\rho}(\mu)$ and $\M_{\mu}$ is the POVM on $\H_R$ corresponding to
the ensemble $\mu$ by the rule described at the end Section 6.1  (before (\ref{MI-CHI-rep+})). If
$\mu$ is an ensemble of pure states then $"="$ holds in (\ref{ED-UB++}) and the POVM $\M_{\mu}$ can be chosen consisting of one-rank operators.

The first inequality in (\ref{ED-UB++}) is easily proved if $\mu=\{p_i,\rho_i\}$ is a discrete ensemble such that $S(\bar{\rho}(\mu))$ and $S(\Phi(\bar{\rho}(\mu)))$ are finite. Indeed, in this case it is reduced to the inequality
$$
\sum_i p_i S(\Phi(\rho_i))\leq \sum_i p_i (S(\widehat{\Phi}(\rho_i))+S(\rho_i)),
$$
which follows from the triangle inequality for the von Neumann entropy \cite{H-SCI,Wilde}. Using this, the validity
of the first inequality in (\ref{ED-UB++}) for any (discrete  or continuous) ensemble $\mu$ with finite $\chi(\mu)$ can be proved by a simple approximation.

The quantity in the l.h.s. of (\ref{ED-UB++}) is  the decrease of the Holevo information of an ensemble $\mu$ under
action of a channel $\Phi$. It is  called the \emph{entropic disturbance} in \cite{ED-1,ED-2,CSR}.

Relations (\ref{chi-d}) and $(\ref{ED-UB++})$ can be used to obtain estimates for the constrained Holevo capacity
and for the entropic disturbance by using estimates for the (optimized and unoptimaized) quantum discord (that can be derived, in particular, from  continuity
bounds for these quantities).\smallskip

Below we will show how to use relation (\ref{chi-d}) to obtain a lower bound on the Holevo capacity $\bar{C}(\Phi)$ of
a finite-dimensional channel $\Phi$ in terms of its complementary channel.\smallskip

\begin{property}\label{CHI-LB} \emph{Let $\,\Phi :A\rightarrow B\,$ be a quantum channel and $\,\widehat{\Phi} :A\rightarrow E\,$ be a
channel complementary to the channel $\,\Phi$. Then
\begin{equation}\label{CHI-LB+}
\bar{C}(\Phi)\geq (1-\varepsilon_{\widehat{\Phi}})\ln d_A-g(\varepsilon_{\widehat{\Phi}}),\;\quad \varepsilon_{\widehat{\Phi}}=\inf_{\omega\in\S_{\rm m}(\H_{AR})}\inf_{\vartheta\in\C}\textstyle\frac{1}{2}\|\widehat{\Phi}\otimes\id_R(\omega)-\vartheta \|_1,
\end{equation}
where $d_A=\dim\H_A<+\infty$, $g(x)$ is the function defined in (\ref{g-def}), $R\cong A$, $\S_{\rm m}(\H_{AR})$ is the set of all maximally entangled pure states in $\S(\H_{AR})$ and $\,\C$ is the set of all
q-c states in $\S(\H_{ER})$ of the form}
$$
(1/d_A)\sum_{i=1}^{d_A} \rho_i\otimes |i\rangle\langle i|, \quad \{\rho_i\}_{i=1}^{d_A}\textit{ is a set of states in }\S(\H_E),\quad \{|i\rangle|\}_{i=1}^{d_A}\textit{ is a basic in }\H_R.
$$
\end{property}

\textbf{Note:} By the proof below the inequality (\ref{CHI-LB+}) holds with $\,\varepsilon_{\widehat{\Phi}}=\textstyle\frac{1}{2}\|\widehat{\Phi}\otimes\id_R(\omega)-\vartheta \|_1$, where $\omega$ and $\vartheta$ are any given states in $\S_{\rm m}(\H_{AR})$ and $\C$ correspondingly.\smallskip

\emph{Proof.} Let $\omega$ be a maximally entangled pure states in $\S(\H_{AR})$. Then $\,S(\omega_A)=\ln d_A\,$  and hence relation (\ref{chi-d}) implies that
\begin{equation}\label{tmp-b}
\bar{C}(\Phi)\geq \bar{C}(\Phi,\omega_A)=\ln d_A-D_R(\widehat{\Phi}\otimes\id_R(\omega)).
\end{equation}
By applying the specification to the case $\rho_B=\sigma_B$ of the continuity bound for the quantum discord $D_B$ from Proposition 14 in  \cite[Section 4.3.1]{QC}
and taking into account that $D_R(\vartheta)=0$ for any $\vartheta\in\C$ we obtain
$$
D_R(\widehat{\Phi}\otimes\id_R(\omega))\leq \epsilon \ln \dim \H_R+g(\epsilon)=\epsilon\ln d_A+g(\epsilon),
$$
where $\epsilon=\textstyle\frac{1}{2}\|\widehat{\Phi}\otimes\id_R(\omega)-\vartheta \|_1$. This
inequality and (\ref{tmp-b}) implies (\ref{CHI-LB+}).  $\Box$ \smallskip

The following example shows that there exist nontrivial channels for which $"="$ holds in (\ref{CHI-LB+}).\smallskip

\begin{example}\label{CHI-LB+1} Let $\,\Phi(\rho)=\sum_{k=1}^d \langle \varphi_k|\rho|\varphi_k\rangle|\varphi_k\rangle\langle \varphi_k|\,$ be a channel
from $d$-dimensional quantum system $A$ to $B=A$ determined by a fixed basis $\{\varphi_k\}_{k=1}^{d}$ in $\H_A$.
Then $\widehat{\Phi}=\Phi$ (we may put $E=A$) and it is easy to see that the state $\widehat{\Phi}\otimes\id_R(\omega)$
belongs to the set $\C$ provided that $\,\omega=(1/d)\sum_{k,j=1}|\varphi_k\rangle\langle \varphi_j|\otimes|\psi_k\rangle\langle \psi_j|$,
where $\{\psi_k\}_{k=1}^{d}$ is a basis in $\H_R$. Thus, $\varepsilon_{\widehat{\Phi}}=0$ and hence  the r.h.s. of (\ref{CHI-LB+}) is equal to
$\,\bar{C}(\Phi)=\ln d$.
\end{example}\smallskip

\begin{example}\label{CHI-LB+2} Let $\Omega_p$ be a quantum erasure channel from $d$-dimensional system $A$ to
$(d+1)$-dimensional system $B$ defined in (\ref{era-ch}). Then $\widehat{\Omega}_p=\Omega_{1-p}$ (we may put $E=B$) and hence
$\,\widehat{\Omega}_p\otimes\id_R(\omega)=p\omega+(1-p)|\tau_0\rangle\langle\tau_0|\otimes\omega_R$, $\omega\in\S(\H_{AR})$.
Since the state $\,\vartheta=p(1/d^2)I_{AR}+(1-p)(1/d)|\tau_0\rangle\langle\tau_0|\otimes I_{R}\,$ belongs to the set $\C$, we have
$\,\varepsilon_{\widehat{\Omega}_p}\leq (p/2)\|\omega_m-(1/d^2)I_{AR}\|_1=p(1-1/d^2)$,
where $\omega_m$ is any maximally entangled pure states in $\S(\H_{AR})$. Thus, Proposition \ref{CHI-LB} gives the lower bound
$$
\bar{C}(\Omega_p)\geq (1-p+p/d^2)\ln d-g(p(1-1/d^2))
$$
which is close to $\,\bar{C}(\Omega_p)=(1-p)\ln d\,$ for small $p$. This example  shows that the
term $-g(\varepsilon_{\widehat{\Phi}})$  in (\ref{CHI-LB+}) can not be removed.
\end{example}

\subsection{Local continuity conditions for some characteristics of a quantum channel}

\subsubsection{The function $(\Phi,\rho)\mapsto\bar{C}(\Phi,\rho)$}

In this subsection we apply the generalized Koashi-Winter relation (in the form (\ref{CHI-rep})) to local continuity
analysis of the function $(\Phi,\rho)\mapsto\bar{C}(\Phi,\rho)$ defined by formula (\ref{CHI-QC-def})
on the set $\F(A,B)\times\S(\H_A)$, where  $\F(A,B)$ is the set of all channels from $A$ to any given system $B$ equipped
with the strong convergence topology  (the notion of strong convergence of quantum channels is described in Section 2).

By using representation (\ref{MI-rep}), the lower semicontinuity of the QMI and the basic results of purification theory it is easy to
show that the function $(\Phi,\rho)\mapsto I(\Phi,\rho)$ is lower semicontinuous
on the set $\,\F(A,B)\times\S(\H_A)$. Representation (\ref{CHI-rep}) and (\ref{D-rep}) allow us to prove the same property for the functions $\,(\Phi,\rho)\mapsto\bar{C}(\Phi,\rho)\,$ and $\,(\Phi,\rho)\mapsto I(\Phi,\rho)-\bar{C}(\Phi,\rho)$.\smallskip

\begin{property}\label{CHIQC} A) \emph{The function $(\Phi,\rho)\mapsto\bar{C}(\Phi,\rho)$ is lower semicontinuous
on the set $\,\F(A,B)\times\S(\H_A)$. The  function $(\Phi,\rho)\mapsto \Delta(\Phi,\rho)\doteq I(\Phi,\rho)-\bar{C}(\Phi,\rho)$
is lower semicontinuous
on the set}
$\left\{(\Phi,\rho)\in\F(A,B)\times\S(\H_A)\,|\,\bar{C}(\Phi,\rho)<+\infty\right\}$. \smallskip

B) \emph{Let $\{\rho_n\}$ be a sequence of states in $\S(\H_A)$ converging to a state $\rho_0$
and $\{\Phi_n\}$ a sequence of channels from $A$ to $B$ strongly converging to a channel $\Phi_0$. If
\begin{equation}\label{MIQC-conv}
 \lim_{n\to+\infty }I(\Phi_n,\rho_n)=I(\Phi_0,\rho_0)<+\infty
\end{equation}
then}
\begin{equation}\label{CHIQC-conv}
 \lim_{n\to+\infty }\bar{C}(\Phi_n,\rho_n)=\bar{C}(\Phi_0,\rho_0)<+\infty.
\end{equation}
\emph{Relations (\ref{MIQC-conv}) and (\ref{CHIQC-conv}) hold if}
$$
\textit{either}\quad\lim\limits_{n\to+\infty }S(\rho_n)=S(\rho_0)<+\infty\quad\textit{ or }\quad\lim\limits_{n\to+\infty }S(\Phi_n(\rho_n))=S(\Phi_0(\rho_0))<+\infty.
$$
\end{property}\smallskip

\emph{Proof.} A) Let $\{\rho_n\}$ be a sequence of states in $\S(\H_A)$ converging to a state $\rho_0$
and $\{\Phi_n\}$ a sequence of channels from $A$ to $B$ strongly converging to a channel $\Phi_0$. Let $\{\hat{\rho}_n\}$ be a sequence of pure states in $\S(\H_{AR})$ converging to a pure state $\hat{\rho}_0$ such that $\Tr_R\hat{\rho}_n=\rho_n$ for all $n\geq 0$. Then the sequence $\{\Phi_n\otimes\id_R(\hat{\rho}_n)\}$
converges to the state $\,\Phi_0\otimes\id_R(\hat{\rho}_0)$. So, the lower semicontinuity of the one-way classical correlation and the representation (\ref{CHI-rep}) imply that
$$
\liminf_{n\to+\infty }\bar{C}(\Phi_n,\rho_n)\geq \bar{C}(\Phi_0,\rho_0),
$$
while the lower semicontinuity of the quantum discord (Proposition 2 in Section 5.1) and representation (\ref{D-rep}) show that
$$
 \liminf_{n\to+\infty }\Delta(\Phi_n,\rho_n)\geq \Delta(\Phi_0,\rho_0)
$$
provided that $\bar{C}(\Phi_n,\rho_n)<+\infty$ for all $n\geq0$.

B) The main claim of B follows from the claims of part A. The last claim of B
follows from Proposition 10 in  \cite{CMI}. $\Box$

\subsubsection{The function $(\Phi,\mu)\mapsto\chi(\Phi(\mu))$}

In this subsection we apply the  monotonicity property (\ref{D-m})  to local continuity
analysis of the function $(\Phi,\mu)\mapsto\chi(\Phi(\mu))$ defined by formula (\ref{H-Q})
on the set $\F(A,B)\times\P(\H_A)$, where $\P(\H_A)$ is the set of all generalized ensembles
of states in $\S(\H_A)$ equipped with the weak convergence topology (see Section 6.2.1) and $\F(A,B)$ is the set of all channels from $A$ to any given system $B$ equipped
with the strong convergence topology (see Section 2).\smallskip

\begin{property}\label{CHIQC+} A) \emph{The function $(\Phi,\mu)\mapsto\chi(\Phi(\mu))$
is lower semicontinuous on the set $\F(A,B)\times\P(\H_A)$. The function $(\Phi,\mu)\mapsto \Upsilon(\Phi,\mu)\doteq I(\Phi,\bar{\rho}(\mu))-\chi(\Phi(\mu))$
is lower semicontinuous
on the set}
$\left\{(\Phi,\mu)\in\F(A,B)\times\P(\H_A)\,|\,\chi(\Phi(\mu))<+\infty\right\}$.

\smallskip

B) \emph{Let $\{\mu_n\}$ be a sequence of ensembles in $\P(\H_A)$ weakly converging to an ensemble $\mu_0$
and $\{\Phi_n\}$ a sequence of channels from $A$ to $B$ strongly converging to a channel $\Phi_0$. If
$$
\lim\limits_{n\to+\infty }I(\Phi_n,\bar{\rho}(\mu_n))=I(\Phi_0,\bar{\rho}(\mu_0))<+\infty
$$
then}
\begin{equation}\label{CHIQC-conv+}
 \lim_{n\to+\infty }\chi(\Phi_n(\mu_n))=\chi(\Phi_0(\mu_0))<+\infty.
\end{equation}
\end{property}

\emph{Proof.} A) The lower semicontinuity of the function $(\Phi,\mu)\mapsto\chi(\Phi(\mu))$
on the set $\F(A,B)\times\P(\H_A)$ follows from the lower semicontinuity of the function $\nu\mapsto\chi(\nu)$
on the set $\P(\H_B)$ (cf.\cite[Proposition 1]{H-Sh-2}), since the arguments from the proof of Lemma 1 in \cite{AQC} show
that for any sequence $\{\mu_n\}$ in $\P(\H_A)$ weakly converging to an ensemble $\mu_0$
and any sequence $\{\Phi_n\}$ in $\F(A,B)$ strongly converging to a channel $\Phi_0$ the sequence of ensembles $\{\Phi_n(\mu_n)\}$ weakly converges to the ensemble $\Phi_0(\mu_0)$.

Let $\{\Pi_m\}$ be a sequence of channels from $B$ to $B$ with a finite-dimensional output
strongly converging to the identity channel $\id_B$ as $\,m\to+\infty$. Then
the function $f_m(\Phi,\mu)\doteq I(\Pi_m\circ\Phi,\bar{\rho}(\mu))-\chi(\Pi_m\circ\Phi(\mu))$
is continuous on the set $\,\F(A,B)\times\P(\H_A)\,$ for each $m$. Thus, to prove the second claim of part A it suffices to show that
\begin{equation}\label{chi-eq}
I(\Phi,\bar{\rho}(\mu))-\chi(\Phi(\mu))=\sup_mf_m(\Phi,\mu)
\end{equation}
for any $(\Phi,\mu)$ such that $\chi(\Phi(\mu))<+\infty$. By Proposition \ref{D-m+}
we have $"\geq"$ in (\ref{chi-eq}). To prove the converse inequality note that
$$
 \lim_{m\to+\infty }I(\Pi_m\circ\Phi,\bar{\rho}(\mu))=I(\Phi,\bar{\rho}(\mu))\leq+\infty\quad\textrm{and} \;\;\lim_{m\to+\infty }\chi(\Pi_m\circ\Phi(\mu))= \chi(\Phi(\mu))<+\infty.
$$
These relations follow from the  monotonicity of the mutual information and the output Holevo information
under a concatenation (described at the beginning of Section 6.3) and their lower semicontinuity.

The claim of part B follows directly from the claims of part A. $\Box$\smallskip

\begin{remark}\label{D-m+r}
Propositions \ref{CHIQC} and \ref{CHIQC+} show that relation (\ref{CHIQC-conv+}) holds provided that either
$$
\lim_{n\to+\infty }S(\bar{\rho}(\mu_n))=S(\bar{\rho}(\mu_0))<+\infty\quad \textup{or}\quad \lim_{n\to+\infty }S(\Phi_n(\bar{\rho}(\mu_n)))=S(\Phi_0(\bar{\rho}(\mu_0)))<+\infty.
$$
The same assertion can be proved by using the lower semicontinuity of the entropic disturbance as a function of a pair (channel, input ensemble) \cite[Section VII-A]{CSR}).
\end{remark}\smallskip

It is essential that Proposition \ref{CHIQC+} gives \emph{more strong condition} of local continuity of the function $\mu\mapsto\chi(\Phi(\mu))$.
To show this take a state $\rho$ and a channel $\Phi$ such that $S(\rho)=S(\Phi(\rho))=+\infty$, but $I(\Phi,\rho)<+\infty$.\footnote{An example of such a state and a channel
can be constructed by taking any state $\omega$ in $\S(\H_{BC})$ such that $S(\omega_B)=S(\omega_C)=+\infty$, but $I(B\!:\!C)_{\omega}<+\infty$.
Let $\hat{\omega}$ be a purification of $\omega$ in $\S(\H_{ABC})$. Then the channel $\Phi(\sigma)=\Tr_A \sigma$, $\sigma\in\S(\H_{AB})$
and the state $\rho=\hat{\omega}_{AB}$ have the required properties.}
Proposition \ref{CHIQC+} implies that relation (\ref{CHIQC-conv+}) holds with  $\Phi_n=\Phi$, $n\geq0$, for any sequence $\{\mu_n\}$
of ensembles in $\P(\H_A)$ weakly converging to an ensemble $\mu_0$ such that $\bar{\rho}(\mu_n)=\rho$ for all $n$.
This claim can not be proved by using the conditions stated in Remark \ref{D-m+r}.

\section{Conclusion}

In this article it is shown that the interconnections between correlation measures and information characteristics
of quantum channels (namely, the  generalized versions of Koashi-Winter and Xi-Lu-Wang-Li relations and their "unoptimized" versions) can be  used for solving different tasks concerning characteristics of both types.  In particular, we apply the interconnection technique to obtain advanced continuity bounds for the  Holevo information at the outputs of a channel and its complementary channel, for the Holevo capacity and the unregularized private capacity of a quantum channel depending on the input dimension/energy (which essentially improve the corresponding results obtained earlier \cite{CID}).  We also use this technique to prove  other results concerning the quantum discord in infinite-dimensional bipartite systems and the information characteristics of quantum channels.

Two open questions have been formulated (Conjectures 1 and 2 at the ends of Sections 5-2 and 5-3). It is also not clear whether it is possible to use the interconnection technique for the study of the (regularized) classical and private capacities of quantum channels, while the efficiency of this  technique for the quantitative continuity analysis of the regularization of the one-way classical correlation has been shown in \cite[Section 4.3]{QC}.

\section{Appendix: On closedness of the set of quantum-classical states}

The following "almost obvious" observation is essentially used in the article.\smallskip

\begin{lemma}\label{Q-C-closed}  \emph{The set of all q-c states of an infinite-dimensional bipartite quantum system $AB$ (i.e. states having the form (\ref{qc-def}) with any basis $\{|k\rangle\}$) is a closed subset of $\S(\H_{AB})$.}
\end{lemma}
\smallskip

\emph{Proof.} Let $\,\{\omega_n\}$ be a sequence of q-c states in $\S(\H_{AB})$ converging to a state $\omega$. Then
$\omega_n=\sum_{i}p_i^n\rho^n_i\otimes |\varphi_i^n\rangle\langle\varphi_i^n|$, where $\{p^n_i,\rho^n_i\}_{i}$ is an ensemble of states in $\S(\H_A)$ and $\{\varphi_i^n\}_{i}$ is an orthonormal basis in $\H_B$  for each $n$. We may assume that $\{p^n_i\}_i$ is a non-increasing sequence for each $n$.
Then $[\omega_n]_B=\sum_{i}p_i^n|\varphi_i^n\rangle\langle\varphi_i^n|$
tends to $\omega_B$ as $n\to+\infty$. By Lemma \ref{SP-conv} below there exist
\begin{itemize}
  \item an orthonormal basis $\{\varphi_i^0\}_{i}$ in $\H_B$
and a probability distribution  $\{p^0_i\}_{i}$ such that $\,\omega_B=\sum_{i}p_i^0|\varphi_i^0\rangle\langle\varphi_i^0|$;
  \item a subsequence $\{n_k\}$ of natural numbers such that $\,p_i^{n_k}|\varphi_i^{n_k}\rangle\langle\varphi_i^{n_k}|\,$ tends to $\,p_i^0|\varphi_i^0\rangle\langle\varphi_i^0|\,$ as $\,k\to+\infty\,$ for all $i$.
\end{itemize}
Using this, the convergence of the sequence $\,\{\omega_n\}$ to the state $\omega$ and taking Lemma \ref{D-A} in Section 2 into account
it is easy to show that $\,\omega=\sum_{i}(I_A\otimes|\varphi_i^0\rangle\langle\varphi_i^0|)\shs\omega\shs(I_A\otimes|\varphi_i^0\rangle\langle\varphi_i^0|)$. So, $\omega$ is a q-c state. $\Box$\smallskip

\begin{lemma}\label{SP-conv}  \emph{Let $\{\rho_n\}$ be a sequence in $\S(\H)$ converging to a state $\rho_0$. Let $\rho_n=\sum_{i=1}^{+\infty} \lambda^n_iP^n_i$ be a spectral
decomposition of $\rho_n$ for each $n\neq0$ such that the sequence $\{\lambda^n_i\}_i$ is non-increasing ($P_i^n$ is a one-rank projector, $\lambda^n_i\geq0$). Then there exist
a spectral
decomposition $\rho_0=\sum_{i=1}^{+\infty} \lambda^0_iP^{\shs0}_i$ of $\rho_0$ and an increasing sequence
$\{n_k\}_{k\in\mathbb{N}}$ of natural numbers such that
\begin{equation}\label{SP-conv+}
\lim_{k\to+\infty}\lambda_i^{n_k}P_i^{n_k}=\lambda_i^{0}P_i^{\shs0}\quad \forall i,
\end{equation}
where the limit in the trace norm topology.}
\end{lemma}\smallskip

\emph{Proof.} If the state $\rho_0$ has no multipled eigenvalues then the claim of this lemma
can be  deduced from Theorem VIII.23 in  \cite{R&S} by using the Mirsky inequality (cf.\cite{Mirsky,Mirsky-rr})
\begin{equation}\label{Mirsky-ineq}
  \sum_{i=1}^{+\infty}|\lambda^{n}_i-\lambda^{0}_i|\leq \|\rho_n-\rho_0\|_1\quad \forall n,
\end{equation}
where $\{\lambda^0_i\}_i$ is the sequence
of eigenvalues of $\rho_0$ in the non-increasing order. In this case one can take $n_k=k$.

To prove the lemma in the general case assume first that all the states $\rho_n$ and $\rho_0$ have infinite rank.
Let $\{\mu_k\}_{k=1}^{+\infty}$ be a sequence of \emph{different} eigenvalues of $\rho_0$ arranged in the non-increasing order (i.e. a sequence of eigenvalues of $\rho_0$ excluding multiplicity).
Let $Q^0_k$ be the spectral projector of $\rho_0$ corresponding to $\mu_k$ and $m_k=\rank Q^0_k$  the multiplicity of $\mu_k$.
Let $Q_k^n=\sum_{i=p_{k-1}+1}^{p_k} P_i^n$, where $p_0=0$ and $p_k=m_1+...+m_k$ for $k\in\mathbb{N}$. By using Theorem VIII.23 in \cite{R&S}, inequality (\ref{Mirsky-ineq}) and Lemma \ref{D-A} in Section 2 it is easy to show that $Q_k^n$ tends to $Q_k^0$ in the trace norm for all $k\in\mathbb{N}$.

Thus, by applying Lemma \ref{vul} below to the sequences $\{P_i^{n}\}_{n\in\mathbb{N}}$ for $i=\overline{1,p_1}$ we get an increasing sequence $\mathfrak{A}_1$ of natural
numbers such that the sequence $\{P_i^{n}\}_{n\in\mathfrak{A}_1}$ tends to a one-rank projector $P_i^{0}$ for all $i=\overline{1,p_1}$ and  $\,\sum_{i=1}^{p_1} P_i^0=Q^0_1$.
Now we may apply Lemma \ref{vul} to the sequences $\{P_i^{n}\}_{n\in\mathfrak{A}_1}$ for $\,i=\overline{p_1+1,p_2}$. This gives
an increasing sequence $\mathfrak{A}_2\subseteq\mathfrak{A}_1$ of natural
numbers such that the sequence $\,\{P_i^{n}\}_{n\in\mathfrak{A}_2}$ tends to a one-rank projector $P_i^{0}$ for all  $i=\overline{p_1+1,p_2}$ and  $\,\sum_{i=p_1+1}^{p_2} P_i^0=Q^0_2$. Since $\mathfrak{A}_2\subseteq\mathfrak{A}_1$, we also have  $\,\{P_i^{n}\}_{n\in\mathfrak{A}_2}\to P_i^{0}$ for $\,i=\overline{1,p_1}$ by the construction of $\mathfrak{A}_1$.

By repeating this process we obtain a set $\{P_i^{0}\}_{i=1}^{+\infty}$ of one-rank projectors and a  set $\{\mathfrak{A}_k\}_{k=1}^{+\infty}$ of increasing subsequences  of $\mathbb{N}$ such that  $\,\mathfrak{A}_{k+1}\subseteq\mathfrak{A}_k$ for all $k$, the sequence
$\{P_i^{n}\}_{n\in\mathfrak{A}_k}$ tends to $P_i^{0}$ for $i=\overline{1,p_k}\,$ and $\,\sum_{i=p_{k-1}+1}^{p_k} P_i^0=Q^0_k$ for each $k$.

Now we apply the "diagonal method": for each natural $k$ take $n_k$ in $\mathfrak{A}_k$ in such a way that $\,n_{k+1}>n_k$ for all $k$. Then it is easy to see that
the sequence $\{P_i^{n_k}\}_{k\in N}$ tends to $P_i^{0}$ for all $i$. Since $P_i^{n_k}P_j^{n_k}=0$ for all $i\neq j$  and all $k$, the set $\{P_i^{0}\}_{i=1}^{+\infty}$ consists
of mutually orthogonal projectors. Thus,
$$
\rho_0=\sum_{k=1}^{+\infty}\mu_kQ^0_k=\sum_{k=1}^{+\infty}\mu_k(P_{p_{k-1}+1}^0+...+P_{p_{k}}^0)=\sum_{i=1}^{+\infty}\lambda^0_iP_i^0
$$
is a spectral decomposition of $\rho_0$. Since $\,\lambda^{n}_i\to\lambda^{0}_i\,$ as $\,n\to+\infty\,$ for any $i$ due to inequality (\ref{Mirsky-ineq}), the sequence $\{n_k\}$ has the required properties.

Assume now that $\rho_n$ and $\rho_0$ are states of arbitrary rank. Take an auxiliary infinite-dimensional Hilbert space $\H'$
and a full rank state $\sigma$ in $\S(\H')$ such that the intersection of the spectra of $\sigma$, $\rho_0$ and all the states $\rho_n$ does not contain positive numbers.
Consider the sequence of states $\tilde{\rho}_n=\frac{1}{2}\rho_n\oplus\frac{1}{2}\sigma$ in $\S(\H\oplus\H')$ converging to the state $\tilde{\rho}_0=\frac{1}{2}\rho_0\oplus\frac{1}{2}\sigma$.
If $\sigma=\sum_{i=1}^{+\infty} \nu_i Q_i$ is a spectral
decomposition of $\sigma$ then
\begin{equation}\label{sp-tmp}
\tilde{\rho}_n=\textstyle\frac{1}{2}\displaystyle\sum_{i=1}^{+\infty} \lambda^n_i P^n_i+\textstyle\frac{1}{2}\displaystyle\sum_{i=1}^{+\infty} \nu_i Q_i
\end{equation}
is a spectral decomposition of $\tilde{\rho}_n$ for all $n\neq0$. Since all the states $\tilde{\rho}_n$ and $\tilde{\rho}_0$ have infinite rank, by using the above part of the proof and inequality (\ref{Mirsky-ineq}), it is easy to show the existence of a spectral decomposition of $\tilde{\rho}_0$ having the form (\ref{sp-tmp}) with $n=0$ and a sequence $\{n_k\}$ such that the relations in (\ref{SP-conv+}) hold. $\Box$\smallskip

\begin{lemma}\label{vul}  \emph{Let $\{\{P_i^n\}_{i=1}^m\}_n$, $m<+\infty$, be a sequence of $m$-tuples of mutually orthogonal one-rank projectors on $\H$ such that
$\sum_{i=1}^m P_i^n$ tends to an $\,m$-rank projector $Q$ in the trace norm. Then there exist an  $m$-tuple $\{P_i^0\}_{i=1}^m$ of mutually orthogonal one-rank projectors and  an increasing sequence $\mathfrak{A}\subset\mathbb{N}$
such that the sequence $\{P_i^n\}_{n\in\mathfrak{A}}$ tends to the  projector $P_i^0$ for all $\,i=\overline{1,m}\,$ in the trace norm and $\,\sum_{i=1}^m P_i^0=Q$.}
\end{lemma}\smallskip

{Proof.} The trace norm convergence of $\sum_{i=1}^m P_i^n$ to $Q$ and the compactness criterion for subsets of trace class operators (Proposition 11 in \cite{AQC}) allow us to show that for each $i$
the sequence $\{P_i^n\}_n$ of pure states is relatively compact in $\S(\H)$ and hence any its subsequence has partial limits which are pure states (one-rank projectors).
Using this and the finiteness of $m$ it is easy to prove the claim of the lemma. $\Box$

\bigskip

I am grateful to A.S.Holevo, A.V.Bulinski and E.R.Loubenets for the help and useful discussion.
I am also grateful to L.Lami for the communication concerning characterization of quantum states with zero discord. Special thanks to M.Caleffi for the relevant reference. This work
was funded by Russian Federation represented by the Ministry of Science and
Higher Education (grant number 075-15-2020-788).

\end{document}